\documentclass[%
 reprint,
superscriptaddress,
nofootinbib,
 amsmath,amssymb,
 aps,
]{revtex4-2}

\usepackage{graphicx}
\usepackage{dcolumn}
\usepackage{bm}

\usepackage[utf8]{inputenc}
\usepackage{physics}
\usepackage{hyperref}
\usepackage[version=3]{mhchem}
\usepackage{siunitx}

\usepackage[caption=false]{subfig}

\newcommand{\mr}[1]{\mathrm{#1}}
\newcommand{\mc}[1]{\mathcal{#1}}
\newcommand{\psiin}{{\psi_{\rm in}}}
\newcommand{\psiout}{{\psi_{\rm out}}}
\newcommand{\Nbitstring}{{\{0, 1\}^{N_q}}}

\begin{document}

\title{
Quantum-Selected Configuration Interaction:\\
classical diagonalization of Hamiltonians in subspaces selected by quantum computers}

\newcommand{\qunasys}{QunaSys Inc., Aqua Hakusan Building 9F, 1-13-7 Hakusan, Bunkyo, Tokyo 113-0001, Japan}
\newcommand{\handai}{Graduate School of Engineering Science, Osaka University, 1-3 Machikaneyama, Toyonaka, Osaka 560-8531, Japan}
\newcommand{\qiqb}{Center for Quantum Information and Quantum Biology, Osaka University, Japan}

\author{Keita Kanno}
\email{kanno@qunasys.com}
\affiliation{\qunasys}

\author{Masaya Kohda}
\email{kohda@qunasys.com}
\affiliation{\qunasys}

\author{Ryosuke Imai}
\affiliation{\qunasys}

\author{Sho Koh}
\affiliation{\qunasys}

\author{Kosuke Mitarai}
\affiliation{\handai}
\affiliation{\qiqb}

\author{Wataru Mizukami}
\affiliation{\handai}
\affiliation{\qiqb}

\author{Yuya O. Nakagawa}
\affiliation{\qunasys}

\date{\today}

\begin{abstract}
We propose quantum-selected configuration interaction (QSCI), a class of hybrid quantum-classical algorithms for calculating the ground- and excited-state energies of many-electron Hamiltonians on noisy quantum devices. Suppose that an approximate ground state can be prepared on a quantum computer either by variational quantum eigensolver or by some other method. Then, by sampling the state in the computational basis, which is hard for classical computation in general, one can identify the electron configurations that are important for reproducing the ground state. The Hamiltonian in the subspace spanned by those important configurations is diagonalized on classical computers to output the ground-state energy and the corresponding eigenvector. The excited-state energies can be obtained similarly.
The result is robust against statistical and physical errors because the noisy quantum devices are used only to define the subspace, and the resulting ground-state energy strictly satisfies the variational principle even in the presence of such errors. The expectation values of various other operators can also be estimated for obtained eigenstates with no additional quantum cost, since the explicit eigenvectors in the subspaces are known.
We verified our proposal by numerical simulations, and demonstrated it on a quantum device for an 8-qubit molecular Hamiltonian.
The proposed algorithms are potentially feasible to tackle some challenging molecules by exploiting quantum devices with several tens of qubits, assisted by high-performance classical computing resources for diagonalization.
\end{abstract}

\maketitle
\section{Introduction}
Recent years have seen a rapid development of quantum computers towards their practical use. Although current quantum devices are prone to errors due to physical noise, ways to achieve \textit{quantum advantage} over classical computations have been explored experimentally~\cite{arute2019quantum, zhong2020quantum, madsen2022quantum},
and such noisy intermediate-scale quantum (NISQ) devices are believed to become useful in the near future~\cite{preskill2018quantum}.
Quantum chemistry is at the top of the list of such useful applications (see, e.g., Refs.~\cite{cao2019quantum, mcardle2020quantum, cerezo2021variational, bharti2022noisy, tilly2022variational}): for instance, energy eigenvalues of a molecular Hamiltonian can be calculated by quantum algorithms developed for NISQ devices, where the most notable is the variational quantum eigensolver (VQE)~\cite{peruzzo2014variational} to find the ground-state energy.

However, VQE faces several challenges to be overcome for practical use.
The major obstacle comes from errors caused by statistical fluctuation and physical noise inherent in the noisy devices. Suppressing the statistical error to a practically acceptable level needs a prohibitively large number of samples~\cite{gonthier2020identifying}, and error mitigation techniques~\cite{viola1999dynamical, temme2017error, li2017efficient,endo2018practical,koczor2021exponential,huggins2021virtual,mcardle2019error,bonet2018low,maciejewski2020mitigation,endo2021hybrid} for reducing physical noise require even more samples to compensate the additional statistical error they introduce~\cite{wang2021can,takagi2022fundamental, tsubouchi2022universal,takagi2022universal}.
In particular, the effect of the errors can spoil the \textit{variational} nature of VQE: that is, the energy estimated by quantum devices is not guaranteed to give an upper bound on the exact ground-state energy.
This is problematic because lowering the resulting energy of VQE does not necessarily mean approaching to the exact ground state.
Besides, there are other challenges for VQE such as the barren plateau problem, which can interrupt the optimization~\cite{mcclean2018barren}.

In this paper, we propose a class of hybrid quantum-classical methods, which we call quantum-selected configuration interaction (QSCI), to find low-lying eigenvalues and eigenstates of a many-electron Hamiltonian.\footnote{We focus on applications to quantum chemistry in this paper. However, the proposed methods can be applied to a variety of many-body Hamiltonians, including many-electron and spin problems in condensed matter physics.} QSCI is noise resilient
and, in principle, free of costly optimization of parametrized quantum circuits.
In particular, QSCI sets rigorous upper bounds on the ground-state energy\footnote{QSCI can also set rigorous upper bounds on the excited-state energies, depending on its algorithmic implementation.} even under the effect of physical and statistical errors.
Here we outline a version of QSCI for finding a ground state:
suppose that an approximate ground state, which we call an \textit{input state} in this paper, can be prepared on a quantum computer; one then repeats a measurement of the state to identify the computational basis states, or electron configurations, that are important to express the ground state~\cite{kohda2022quantum};
one then diagonalizes, on classical computers, the truncated Hamiltonian matrix in the subspace spanned by the identified configurations to obtain the smallest eigenvalue and eigenvector. 
The resulting eigenvalue approximates the ground-state energy.
The diagonalization is classically tractable unless the number of selected configurations is exponentially large in the system size.
The algorithm can be extended to find excited states by enlarging the subspace or by repeating the procedure for each energy eigenstate.

Since the matrix elements of the Hamiltonian in the computational basis can be exactly calculated on classical computers, the diagonalization results in an energy that gives a definite upper bound on the exact ground-state energy regardless of the quality of the subspace spanned by the identified configurations; the quality only affects how tight the bound is. The states need to be measured only in the computational basis, and thus no additional gate operation is required for the measurement. In the presence of symmetries with conserved quantities such as the particle number, the post-selection of the computational basis states in the sampling outcome allows one to mitigate the bit-flip errors. We experimentally demonstrate the effectiveness of the post-selection in this paper.
The algorithm may take any quantum states as the input states, if they roughly approximate the desired eigenstates and can be prepared on quantum devices. Such input states can be prepared, e.g., by parametrized quantum circuits moderately optimized via
VQE and its variants~\cite{tilly2022variational}, and other preparation schemes are discussed in Sec.~\ref{subsec:state-preparation}. Sampling from such quantum states can be hard for classical computers~\cite{arute2019quantum}, and thereby providing a potential quantum speed-up in QSCI.

QSCI can also be advantageous as a technique for \textit{eigenstate} tomography in that it can (classically) estimate the expectation values of a variety of observables at no additional quantum cost: as we already have the classical representation of the state, one can efficiently compute the expectation values using that representation.
Unlike QSCI, other efficient tomography techniques such as classical shadows~\cite{huang2020, zhao2021}, neural network tomography~\cite{Torlai_2018}, and tensor network tomography~\cite{Cramer_2010} do not exploit the fact that the states of our interest are eigenstates of some problem Hamiltonian.

As the name suggests, QSCI can be viewed as a configuration interaction (CI), where the many-body basis set is determined by quantum computers via sampling of an input state. 
There are established techniques~\cite{helgaker2014molecular} that choose fixed basis sets. 
A common approach in electronic structure theory is to select only one- and two-particle excitations from a reference wavefunction. When the reference wavefunction is chosen to be Hartree-Fock, the resulting method is known as CI with singles and doubles (CISD). If the reference wavefunction is a correlated wavefunction beyond the mean-field approximation, the method is called multi-reference CISD (MR-CISD). In the context of quantum computing, MR-CISD has sometimes been called as quantum subspace expansion (QSE)~\cite{mcclean2017hybrid,takeshita2020increasing,urbanek2020jctc}.

Another approach is the adaptive selection of a suitable basis set for a target system. In quantum chemistry, a systematic selection of important bases has a long history~\cite{bender1969pr,whitten1969jcp,huron1973jcp,buenker1974tca,buenker1975tca,nakatsuji1983cluster,cimiraglia1987jcc,harrison1991jcp,greer1995jcp,greer1998jcpss}. And there are recently active studies along with such a systematic selected CI~\cite{evangelista2014jcp,holmes2016jctc,schriber2016jcp,holmes2016jctc2,tubman2016deterministic,ohtsuka2017jcp,schriber2017jctc,sharma2017semistochastic,chakraborty2018ijqc,coe2018jctc,coe2019jctc,abraham202jctc,tubman2020modern,zhang2020jctc,zhang2021jctc,chilkuri2021jcc,chilkuri2021jctc,goings2021jctc,pineda2021jctc,jeong2021jctc,coe2023jctc,seth2023jctc}. Thanks to such developments, systematic selected CIs are now gradually being considered as a promising approach for large-scale quantum chemical simulations. The likely reason for this revival is that selected CI is an algorithm that can be adapted to current classical computer architectures with sufficient memory. QSCI may be seen as a new systematic selected CI that utilizes quantum computers. 

Our methods are capable of selecting electron configurations which are necessary to describe the eigenstates to some accuracy but are missed in the conventional methods with a fixed basis set. Note that our methods call the diagonalization procedure at most only once for each eigenstate, while the adaptive methods iteratively repeat the diagonalization to search for a configuration to be added in the basis set; our methods require much less classical computational time compared to those adaptive methods.

The classical diagonalization is already utilized in various hybrid quantum-classical algorithms to find energy eigenstates. Most notable is QSE, which spans the subspace by states built upon the reference VQE state,
and is widely used for various applications, e.g., excited state calculations~\cite{mcclean2017hybrid}, band structure calculations~\cite{yoshioka2022variational}, and noise reduction~\cite{mcclean2017hybrid,bonet2018low, takeshita2020increasing,mcclean2020decoding,yoshioka2022generalized, epperly2022theory}. 
More generally, one can span the subspace by various methods~\cite{huggins2020non,motta2020determining, parrish2019quantumfilter, stair2020multireference,parrish2019quantum, seki2021quantum,baek2022say, kirby2022exact}, which are sometimes collectively called as the quantum subspace diagonalization. In those methods, however, the matrix elements of the subspace Hamiltonian are calculated on quantum computers, and thus are subject to the physical and statistical errors. There is a proposal~\cite{radin2021classically} where some of the matrix elements are classically calculated, but the method still requires some matrix elements which are efficiently computable only by quantum computers for a possible quantum speed-up. In QSCI, on the other hand, all the matrix elements are classically computed, giving up the use of more complex and physically-motivated states as basis states that define the subspace.

The rest of the paper is organized as follows. The proposed methods are introduced in Sec.~\ref{sec:methods}, and numerically tested in Sec.~\ref{sec:numerical}. A demonstration on a quantum device is presented in Sec.~\ref{sec:noisy-simulation-experiment}, along with a noisy simulation as a preparatory study. We discuss aspects of the proposed methods in Sec.~\ref{sec:discussion}, and finally conclude in Sec.~\ref{sec:conclusion}. Details of the algorithms, numerical simulations and experiment, as well as supplemental numerical results are given in the appendices.

\section{Methods}
\label{sec:methods}

In this section, we present the methods of QSCI.
Two ways of implementation are introduced: single diagonalization scheme in Sec.~\ref{sssec:method-single} and sequential diagonalization scheme in Sec.~\ref{sssec:method-sequantial}.
They are designed for finding multiple energy eigenstates, and reduce to the same simplified method when used for finding the ground state alone.
After introducing necessary ingredients, we begin with the algorithm specific to finding the ground state, which is simple and illustrative, and then proceed to the two methods which can also find excited states.

\subsection{
Preliminary
}

We consider electronic structure problems of molecules in the second-quantization formalism with the Born-Oppenheimer approximation.
A Hamiltonian and wave functions for electrons, in this setup, can be mapped onto $N_q$ qubits such that the Slater determinants\footnote{Instead, linear combinations of Slater determinants such as configuration state functions may be mapped to the computational basis states. QSCI can work with such a mapping, if the matrix elements of the Hamiltonian in the computational basis can be efficiently computed by classical computation.} for the Hartree-Fock state and its excitations are associated with the computational basis states $\ket{x}$, where $x\in\Nbitstring$ is an $N_q$-bit string (see, e.g., Ref.~\cite{cao2019quantum,mcardle2020quantum}).
In the Jordan-Wigner mapping, which we adopt in the numerical study, $N_q$ corresponds to the number of spin orbitals, and ``1'' or ``0'' represents whether each spin orbital is occupied or not.
The methods can work with other mapping schemes such as the Bravyi-Kitaev mapping~\cite{bravyi2002fermionic}, although the fermion-qubit correspondence is less intuitive and the error mitigation (discussed later) is less effective.
We denote the qubit Hamiltonian by $\hat{H}$.
A linear combination of all the computational basis states,
\begin{align}
\ket{\psi}
=\sum_{x\in\Nbitstring} \alpha_x \ket{x},
\label{eq:general_state}
\end{align}
encompasses the full-CI wave function.
Note that for a fixed number of electrons only a subset of the computational basis states is needed.

In the full-CI method, sets of the CI coefficients \{$\alpha_x$\} that correspond to energy eigenstates are found by diagonalizing the Hamiltonian in the full Fock space.
The method is costly due to the combinatorial growth of the Fock-space dimension as the number of spin-orbitals increases.
For reducing the computational cost, there exist various classical approaches which truncate the Fock space and approximate the sum in Eq.~\eqref{eq:general_state} using a fixed or adaptively selected basis set, as mentioned in the previous section.
In line with these efforts, but from a different viewpoint, we propose methods which harness quantum computers to identify important computational basis states, or electron configurations, for truncating the Fock space.

\subsection{QSCI for ground state
\label{subsec:ground-state}
}

We now describe the explicit algorithms.
We begin with the algorithm for finding the lowest eigenvalue and the corresponding eigenstate (ground state) of an electronic Hamiltonian $\hat{H}$ on $N_q$ qubits.
For simplicity, we assume the ground state is unique.
When the degeneracy exists, the algorithms given in the next subsection, which is aimed at finding multiple eigenstates, can be straightforwardly applied.
Indeed, the algorithm introduced in this subsection is a special case of each of the two algorithms in the next subsection.

Let $\ket{\psiin}$ be an input state, which roughly approximates the ground state, and suppose $\ket{\psiin}$ can be prepared by a quantum circuit with $N_q$ qubits.
Then, one prepares the input state on a quantum computer and measures the state in the computational basis, which results in an outcome bit string $x\in\Nbitstring$.
Repeating such a sampling procedure (or shot) for $N_{\rm shot}$ times, one counts how many times each $x$ appears.
Based on the total sampling result, the most frequent $R$ computational basis states are selected to define the set
\begin{align}
\mc{S}_R = \{ \ket{x} | x\in\Nbitstring, R~{\rm most~frequent} \},
\label{eq:set_GS}
\end{align}
where $R$ is a positive integer manually determined.
This is to truncate the Fock space.
One may in principle include all the computational basis states appeared in the measurements, while choosing an appropriately small $R$ can reduce the computational cost for diagonalization.

One then solves the eigenvalue problem in the subspace spanned by $\mathcal{S}_R$:
\begin{align}
    \bm{H}_R\bm{c} = E_R\bm{c},
\end{align}
where $\bm{H}_R$ is the $R\times R$ Hermitian matrix defined by
\begin{align}
 (\bm{H}_R)_{xy}= \mel{x}{\hat{H}}{y}~{\rm for}~\ket{x}, \ket{y} \in \mathcal{S}_R,
\end{align}
and $\bm{c}$ is an eigenvector with eigenvalue $E_R$, satisfying $\bm{c}^\dagger \bm{c}=1$.
This step of the algorithm proceeds via classical computations: calculations of the matrix elements $\mel{x}{\hat{H}}{y}$ and the diagonalization of $\bm{H}_R$.
The former calculations can be efficiently done by some classical method, e.g., by the Slater-Condon rules in the fermionic basis.
The latter diagonalization is performed to obtain the smallest eigenvalue $E_R$ and the eigenvector $\bm{c}$, which are output of the algorithm.
See Sec.~\ref{ssec:discussion-computational-cost} for further discussion on costs of these classical computations. 
Here, $E_R$ approximates the exact ground-state energy of $\hat{H}$, while $\bm{c}$ approximately gives the (normalized) CI coefficients, or the vector representation of the ground state, respectively.
The corresponding quantum state, which we call the \textit{output state}, is constructed as
\begin{align}
    \ket{\psiout}=\sum_{\ket{x}\in \mathcal{S}_R} c_x \ket{x},
    \label{eq:output-state-gs}
\end{align}
where $c_x$ is an element of the eigenvector $\bm{c}$.
The output state $\ket{\psiout}$ approximates the true ground state of $\hat{H}$.
We remark that one does not need to realize the output state on quantum computers. Retaining the eigenvector $\bm{c}$ as classical data is enough for the application explained below.

The output state can be used to estimate the expectation values of observables other than the Hamiltonian for the ground state, solely based on classical computations.
Specifically, suppose that an observable in question is represented by a qubit operator $\hat{O}$. If the matrix elements $\mel{x}{\hat{O}}{y}$ can be efficiently computed on classical computers, so does the expectation value $\ev{\hat{O}}{\psiout}$, which is expected to give an approximation to the expectation value for the true ground state.
In particular, if $\hat{O}$ can be expressed as a linear combination of $\text{poly}(N_q)$ Pauli strings, 
which is the case in many physical quantities, its expectation value can be efficiently computed on classical computers.

Comments are in order for technical details.
We identify the set $\mathcal{S}_R$ to span the subspace by sampling the input state.
In this way, we expect that important computational basis states, or Slater determinants, to describe the ground state wave function can be selected.
This is because in the sampling procedure a bit string $x$ occurs with the probability $\abs{\bra{x}\ket{\psiin}}^2$, while $\bra{x}\ket{\psiin}$ gives the CI coefficient of the corresponding Slater determinant in the input wave function $\ket{\psiin}$.\footnote{See, e.g., Eq.~\eqref{eq:general_state}. There, the CI coefficients can be expressed as $\alpha_x = \bra{x}\ket{\psi}$.}
Indeed, $\mathcal{S}_R$ gives the $R$ Slater determinants with the largest coefficients $\abs{\bra{x}\ket{\psiin}}$ in $\ket{\psiin}$, under the ideal situation where physical noise can be ignored and the sampling is performed with an infinite number of shots.
In passing, we sometimes adopt a method equivalent to this ideal situation to define $\mc{S}_R$ in the numerical study: that is, we just pick up the $R$ Slater determinants with the largest absolute values of the CI coefficients in the input state, instead of performing actual sampling procedures.
We call this method as the \textit{idealized sampling} in this paper.
Note that we assume the input state roughly approximates the true ground state.
This is just to ensure the two states share the important computational basis states, and there is no need for a precise agreement between the CI coefficients of the two states.
Such an input state can be prepared, e.g., by a parametrized quantum circuit moderately optimized via VQE.
In Sec.~\ref{subsec:state-preparation}, we discuss methods to prepare the input state, including non-VQE based ways.

The set $\mathcal{S}_R$ is defined in Eq.~\eqref{eq:set_GS} by specifying $R$, the number of the computational basis states retained in the subspace.
But this is not the unique choice.
For instance, one may define the set by taking all the computational basis states in the measurement outcome, as already mentioned.
Or, one may instead set a threshold on the rate of occurrence $f_x$ for an outcome $x$ in the total sampling result, and then define an alternative set $\mathcal{S}_\epsilon = \{ \ket{x} | f_x \geq \epsilon \}$ with a threshold parameter $\epsilon$, 
For a proof-of-principle demonstration, we adopt Eq.~\eqref{eq:set_GS} to define the subspace for diagonalizing the Hamiltonian in the rest of the paper.

In reality, physical noise and statistical fluctuation, the latter due to a finite number of shots, cannot be ignored, causing some errors in the output.
However, the effect is only indirect and the method is robust against those errors: that is, the errors can degrade the quality of the selected subspace by missing important configurations or by picking up irrelevant configurations in the sampling procedures, but the lowest eigenvalue and eigenvectors are exact within the subspace. 
The latter point, the exactness within the subspace, results from the use of diagonalization for the matrix $\bm{H}_R$, whose elements are exactly computed.
Consequently, the obtained energy $E_R$ sets an upper bound on $E_{\rm exact}$, the true ground-state energy of $\hat{H}$:
\begin{align}
E_{\rm exact}\leq E_R.
\label{eq:variational-inequality}
\end{align}
Note that this variational inequality holds even under statistical fluctuation and physical noise.
The situation is in contrast with VQE, where such an inequality is not guaranteed as the energy is directly measured on quantum computers and hence is susceptible to the errors.\footnote{In VQE, physical noise in the state preparation can hardly lead to the violation of the variational inequality, but it may be possible that an error during the measurement procedure causes it. The use of error mitigation techniques can also lead to the breakdown of the inequality.
}
It is also worth mentioning that, for a given sampling outcome, increasing $R$, the subspace size, always leads to a better approximation of the ground-state energy: $E_{\rm exact} \leq E_{R_a} \leq E_{R_b}$ for $R_a > R_b$.
This can be used to see if the calculation converges.
On the other hand, smaller $R$ can reduce the classical computational cost.
Such a trade-off between the accuracy and cost is discussed in Sec.~\ref{subsec:scaling}.

The algorithm finds the lowest energy state in the subspace $\mc{S}_R$, which gives an approximation to the ground state in the {\it full} Fock space. 
When there exists symmetry in the Hamiltonian, there are associated conserved quantities, e.g., the total electron number $N_e$ (or the charge of molecule) and the $z$-component of total electron spin $S_z$.
Given this, one may wish to find the lowest energy state in a specific symmetry sector. In such a case, the method can be similarly applied but by relying on the subspace with fixed conserved quantities. For $N_e$ and $S_z$, this can be easily achieved as follows since each computational basis state corresponds to a Slater determinant with definite $N_e$ and $S_z$: one prepares an input state with the desired values of $(N_e, S_z)$, for which the sampling results in configurations each with the desired $(N_e, S_z)$; or, if such an input state cannot be prepared, one may post-select the sampling outcome, where one discards an outcome $x\in\Nbitstring$ if it conflicts with the desired $(N_e,S_z)$. It is worth noting that the variational inequality~\eqref{eq:variational-inequality} still holds in each sector of Fock space specified by $(N_e, S_z)$.

\begin{figure*}
    \includegraphics[width=\textwidth]{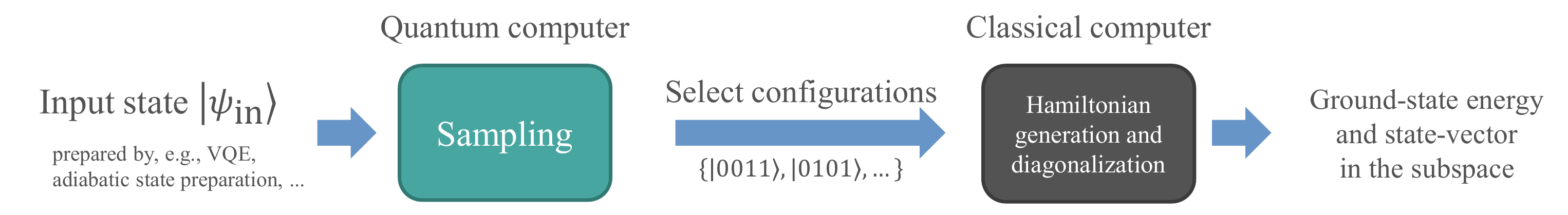}
    \caption{Schematic description of the QSCI algorithm for finding the ground state. When selecting the configurations, one may post-select the configurations by using conserved quantities such as the electron number or spin $S_z$ to mitigate the errors.}
    \label{fig:algorithm-gs}
\end{figure*}

Physical noise can cause a contamination of symmetry sectors: for an input state with fixed $(N_e, S_z)$, sampling on a noisy device can result in electron configurations with unwanted values of $(N_e, S_z)$, due to the bit-flip noise\footnote{Note that an error that corresponds to a phase-flip error occurring at the end of a circuit does not affect the probability distribution $\abs{\bra{x}\ket{\psiin}}^2$ and hence the sampling outcome.} or readout error.
Nevertheless, one can mitigate such errors by post-selecting the sampling outcome according to the conserved quantities, as described above.  
One then diagonalizes the Hamiltonian in the post-selected subspace.
We find that the post-selection is particularly effective to mitigate the readout error in the Jordan-Wigner mapping, while it is also applicable to other fermion-qubit mapping schemes (see Appendix~\ref{subsec: post-selection} for discussions). 

The algorithm is schematically summarized in Fig.~\ref{fig:algorithm-gs}.

\subsection{
QSCI for multiple energy eigenstates
\label{subsec:excited-states}
}

\begin{figure*}
  \begin{minipage}{\textwidth}
    \subfloat[][Single diagonalization]{
         \includegraphics[width=\textwidth]{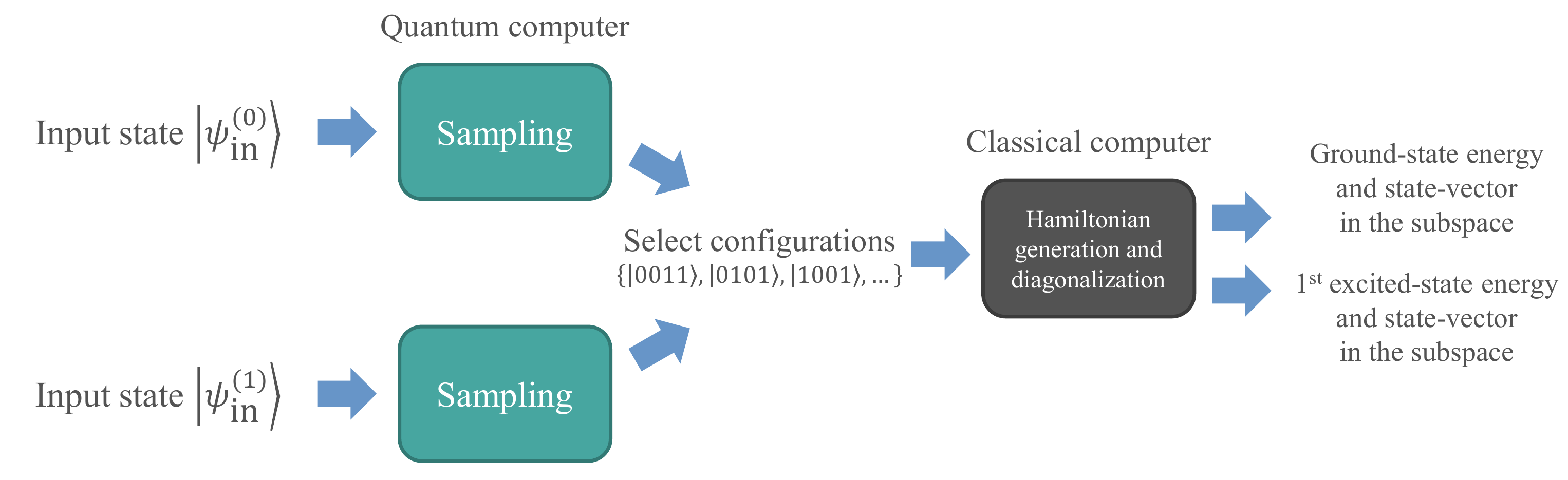}}
        
    \subfloat[][Sequential diagonalization]{
         \includegraphics[width=\textwidth]{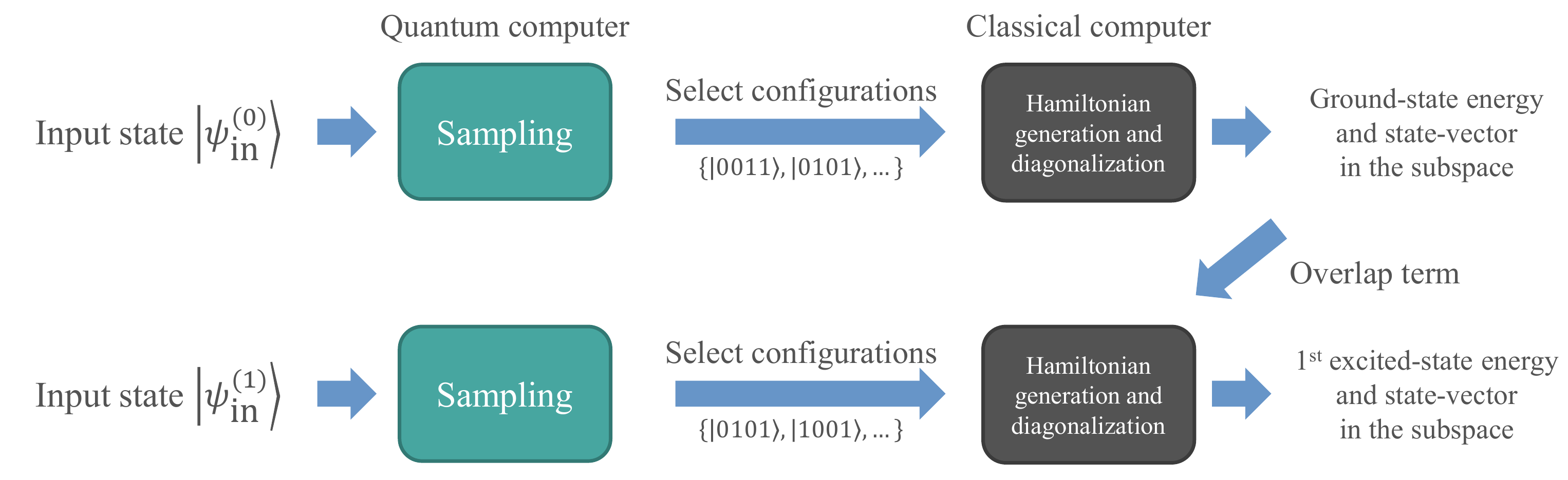}}
    \end{minipage}
    \caption{Schematic descriptions of the QSCI algorithms for finding the ground state and the first excited state: (a) single diagonalization scheme, and (b) sequential diagonalization scheme. In both panels, $|\psi_{\rm in}^{(0)}\rangle$ ($|\psi_{\rm in}^{(1)}\rangle$) is the input state for the ground (first excited) state. In the panel (b), the overlap term is constructed from the preobtained output state $|\psi_{\rm out}^{(0)}\rangle$ to define the effective Hamiltonian for the first excited state, $\hat{H}^{(1)} =\hat{H}+\beta_0|\psi_{\rm out}^{(0)}\rangle\langle \psi_{\rm out}^{(0)}|$.}
    \label{fig:algorithm-es}
\end{figure*}

We now extend the algorithm to find multiple energy eigenstates, including low-lying excited states.
For this, we note that the previous algorithm can output multiple eigenvectors, which can be taken to approximate excited states as well as the ground state.
Yet, the quality of the approximation would not be satisfactory for the excited states, as the subspace is tailored for the ground state.
Hence, we introduce extra input states to construct subspace(s) which can capture the excited states.
In the following, we present two distinct algorithms to find multiple energy eigenstates and energies, schematically shown in Fig.~\ref{fig:algorithm-es}.
The first algorithm, which we call the single diagonalization scheme, constructs a common subspace for both ground and excited states of interest, and performs the diagonalization in the subspace to simultaneously obtain all the desired eigenstates and energies.
On the other hand, the second algorithm, dubbed as the sequential diagonalization scheme, constructs multiple subspaces, each tailored for each energy eigenstate, and sequentially diagonalizes the Hamiltonian in each subspace.
Both of the algorithms contain the algorithm specific to the ground state, introduced in the preceding subsection, as a special case.

\subsubsection{
Single diagonalization scheme
\label{sssec:method-single}
}

Here we describe the single diagonalization scheme.
Suppose one seeks for $N_s$ low-lying eigenstates of $\hat{H}$, which consist of the ground state(s) and subsequent excited states.
In this case, one prepares multiple input states $|\psi_{\rm in}^{(i)}\rangle$ ($i=0,1,\cdots,N_{\rm in}-1$), which correspond to the low-lying energy eigenstates.
Here, we allow $N_{\rm in}\leq N_s$, although the natural choice would be $N_{\rm in}=N_s$.
For each of the input states, one repeats the sampling procedure as in the previous subsection.
One then obtains the set of important configurations $\mc{S}_{R_i}^{(i)}$, formed by most frequent $R_i$ bit strings in the total sampling outcome for the $i$-th input state.
Combining all the sets $\mc{S}_{R_i}^{(i)}$, one constructs the common subspace\footnote{Note that the set $\mc{S}_R$ defined here agrees with the definition~\eqref{eq:set_GS} in the preceding subsection when $N_{\rm in}=1$.}:
\begin{align}
\mc{S}_R =
\mc{S}_{R_0}^{(0)} \cup \mc{S}_{R_1}^{(1)} \cup \cdots \cup \mc{S}_{R_{N_{\rm in}-1}}^{(N_{\rm in}-1)}.
\label{eq:set_single-step}
\end{align}
In this case, the parameters $R_i$ may be eigenstate dependent, while $R$ is the number of the elements in the common subspace $\mc{S}_R$.
$R\geq N_s$ is required to yield at least $N_s$ eigenvectors in the diagonalization procedure shortly explained.

One may treat all $R_i$ as free parameters, which determine $R$ in turn.
Or, one may first choose a value for $R$ and, then, decide each $R_i$ following some strategy.
There are various ways for the latter strategy, depending on the purpose of using the algorithm. 
For example, if one prioritizes the ground state in terms of accuracy, a possible choice would be $R_0=R$ and $R_i=0$ for $i\neq 0$, albeit extreme.
Or, if one wishes to treat all the input states on equal footing, each of $R_i$ can be chosen as equal as possible.\footnote{One can make each of $R_i$ as equal as possible by the following cycle of procedures, starting from an empty set $\mathcal{S}_R$, for a given $R$: in the first cycle, for each of the $N_\text{in}$ input states, the most frequent bit string is selected from the sampling outcome and then added to $\mathcal{S}_R$;
this procedure is executed from the 0-th input state to $(N_{\rm in}-1)$-th input state, where one skips the state if the selected bit string already exists in $\mathcal{S}_R$;
in the second cycle, the second frequent bit string is added to $\mathcal{S}_R$ for each input state according to the same rule;
one repeats such a cycle until $\mathcal{S}_R$ is filled with $R$ distinct bit strings.
Suppose such procedures finished after completing $R'$ cycles.
Then, one can ensure that at least $R'$ most frequent bit strings for each input state are included in $\mathcal{S}_R$.
This implies $R'$ or $R'+1$ most important configurations in each input state are included in the common subspace~\eqref{eq:set_single-step}, in the ideal situation where statistical fluctuation and physical noise can be ignored.}

With the common subspace $\mc{S}_R$ constructed, one then diagonalizes the Hamiltonian in $\mc{S}_R$ as in the previous subsection: one constructs the $R\times R$ Hermitian matrix $\bm{H}_R$, solves the eigenvalue equation $\bm{H}_R\bm{c} = E_R\bm{c}$, and then picks up $N_s$ low-lying eigenvectors and eigenvalues, $(\bm{c}^{(0)}, E_R^{(0)}), (\bm{c}^{(1)}, E_R^{(1)}), \cdots, (\bm{c}^{(N_s-1)}, E_R^{(N_s-1)})$, where $\bm{c}^{(i)\dagger} \bm{c}^{(j)}=\delta_{ij}$.
Here, $E_R^{(i)}$ ($E_R^{(0)}$) approximates the true energy of the $i$-th excited state (ground state), when the ground state is unique, for instance.
The corresponding output states can be constructed as
\begin{align}
    |\psi_{\rm out}^{(i)}\rangle =\sum_{\ket{x}\in \mathcal{S}_R} c_x^{(i)} \ket{x},
    \label{eq:output-state_single-step}
\end{align}
for $i=0,1,\cdots, N_s-1$.

Note that the algorithm in the previous subsection is a special case of the single diagonalization scheme with a single input state ($N_{\rm in}=1$).
In this method, one can apply the same error mitigation technique by the post-selection as described in the previous subsection.
The variational inequality now holds for each of energy eigenstates by Cauchy's interlace theorem~\cite{helgaker2014molecular} (see also Refs.~\cite{hylleraas1930numerical, macdonald1933successive}):
\begin{align}
 E^{(i)}_{\rm exact} \leq E_R^{(i)},
 \label{eq:variational-inequality-single}
\end{align}
for $i=0,1,\cdots, N_s-1$, where $E^{(i)}_{\rm exact}$ is the $i$-th eigenvalue (in ascending order) by the exact diagonalization.
We remark that QSE~\cite{mcclean2017hybrid} and multistate-contracted VQE (MCVQE)~\cite{parrish2019quantum}, which also rely on the subspace diagonalization to obtain excited states, need to measure matrix elements, while the current method exactly calculates the matrix elements. 
Hence, we expect our method to be more noise-robust with the guarantee of the variational inequality.

\subsubsection{
Sequential diagonalization scheme
}
\label{sssec:method-sequantial}

We now give another scheme of QSCI to find excited states.
The sequential diagonalization finds the ground state(s) and subsequent excited states by sequential diagonalization procedures of the Hamiltonian $\hat{H}$ in distinct subspaces.
The algorithm is similar to the variational quantum deflation (VQD)~\cite{higgott2019variational}, a variant of VQE for excited states.

Suppose one seeks for the $k$-th excited state\footnote{We implicitly assume the ground state is unique for ease of illustration. One can straightforwardly translate the description here to cases of degenerate ground (and possibly excited) states.} given that preceding $(k-1)$ excited states and ground state are already obtained by this method with the output states $|\psi_{\rm out}^{(i)}\rangle$ ($i=0,1,\cdots, k-1$).
As in the previous methods, one repeats the preparation and measurement of the input state $|\psi_{\rm in}^{(k)}\rangle$ to obtain the set of important configurations:
\begin{align}
\mc{S}_{R_k}^{(k)} = \{ \ket{x} | x\in\Nbitstring, R_k~{\rm most~frequent} \}.
\label{eq:set_multi-step}
\end{align}
One then has to find the lowest energy state of $\hat{H}$ in this subspace, under the restriction that this state is orthogonal to the states already found, $|\psi_{\rm out}^{(i)}\rangle$ ($i=0,1,\cdots, k-1$).
This can be achieved by diagonalizing the following effective Hamiltonian\footnote{This is not the unique choice of the effective Hamiltonian. For instance, the orthogonality can be imposed without introducing extra parameters though the implementation would be less suitable for NISQ devices~\cite{lee2018generalized}.} in the subspace spanned by $\mc{S}_{R_k}^{(k)}$:
\begin{align}
 \hat{H}^{(k)} =\hat{H}+
 \sum_{i=0}^{k-1}\beta_i
 |\psi_{\rm out}^{(i)}\rangle
 \langle \psi_{\rm out}^{(i)} |,
 \label{eq:sequential-Heff}
\end{align}
where $\beta_i$ are real parameters, which need to be sufficiently large for ensuring the orthogonality. 
The additional terms correspond to the overlap terms in VQD. 
This is equivalent to solving the eigenvalue equation
\begin{align}
    \bm{H}_{R_k}^{(k)}\bm{c}^{(k)} = E_{R_k}^{(k)}\bm{c}^{(k)},
\label{eq:eigenvalue-eq}
\end{align}
and then pick up the smallest eigenvalue $E_{R_k}^{(k)}$ and eigenvector $\bm{c}^{(k)}$, normalized by $\bm{c}^{(k)\dagger} \bm{c}^{(k)}=1$.
Here, $\bm{H}_{R_k}^{(k)}$ is the $R_k \times R_k$ Hermitian matrix defined by
\begin{align}
 (\bm{H}_{R_k}^{(k)})_{xy}= \mel{x}{\hat{H}^{(k)}}{y}~{\rm for}~\ket{x}, \ket{y} \in \mathcal{S}^{(k)}_{R_k},
 \label{eq:sequential-matrix}
\end{align}
whose matrix elements can be efficiently calculated by classical computations based on the expression
\begin{align}
(\bm{H}_R^{(k)})_{xy}
= \mel{x}{\hat{H}}{y} +\sum_{i=0}^{k-1}\beta_i c_x^{(i)}c_y^{(i)*}.
\end{align}
One then constructs the output state
\begin{align}
    |\psi_{\rm out}^{(k)}\rangle =\sum_{\ket{x}\in \mathcal{S}_{R_k}^{(k)}} c_x^{(k)} \ket{x},
    \label{eq:output-state_multi-step}
\end{align}
which approximates the $k$-th excited state.

Note that the expressions are specific to the $k$-th excited state.
In order to find entire (low-lying) spectrum, one has to repeat the above procedure sequentially, starting from $k=0$, the ground state, which can be found by the ground-state algorithm already explained.
This is similar to VQD, but the QSCI method does not require extra circuits to calculate the overlap terms.

The coefficients $\beta_i$ can be chosen in the same manner as VQD.
We want the smallest eigenvalue of $\bm{H}_{R_k}^{(k)}$ to approximate $E^{(k)}_{\rm exact}$, the $k$-th eigenvalue of $\hat{H}$.
Following the discussion in Ref.~\cite{higgott2019variational}, it suffices to choose $\beta_i > E^{(k)}_{\rm exact}-E^{(i)}_{\rm exact}$ for $i=0,\cdots, k-1$; 
or, one may apply the looser condition of $\beta_i > 2\sum_j \abs{c_j}$, where $c_j$ are coefficients in the qubit Hamiltonian $\hat{H}=\sum_j c_j P_j$, expressed by the Pauli strings $P_j$ (see Appendix~\ref{subsec:details-sequential} for details).
In practice, the condition $\beta_i > E^{(k)}_{\rm exact} - E^{(i)}_{\rm exact}$ can be utilized if one has prior knowledge on the energy spectrum, e.g., based on variational quantum algorithms.
Even if such information is not available, one may still rely on the looser condition $\beta_i > 2\sum_j \abs{c_j}$.
Note that in the sequential diagonalization scheme, the variational inequality like Eq.~\eqref{eq:variational-inequality-single} is not guaranteed due to the inexactness of the effective Hamiltonian, i.e., as Eq.~\eqref{eq:sequential-Heff} would be constructed only by approximate eigenstates in practice (see Appendix~\ref{subsec:details-sequential} for further discussion).

\section{Benchmark of QSCI with noiseless simulations
\label{sec:numerical}
}

In this section, we test various aspects of QSCI for small molecules by noiseless numerical simulations, where the effects of physical noise are not included. In Secs.~\ref{subsec:ground-state-simulation-with-noiseless-vqe} and \ref{subsec:simulation-excited-h2o}, QSCI calculations are performed for ground states and excited states, using input states prepared by VQE and VQD~\cite{higgott2019variational}, a variant of VQE for excited states. Then the scalability of QSCI is examined in Sec.~\ref{subsec:scaling}, and finally the effect of the statistical error in QSCI is studied in Sec.~\ref{ssec:sampling-simulation}. For the numerical simulations, a quantum-circuit simulation library Qulacs~\cite{suzuki2021qulacs} is used with the help of QURI Parts~\cite{quri_parts}, a library for developing quantum algorithms. The simulations in Sec.~\ref{ssec:sampling-simulation} are carried out by the sampling simulator which takes into account the statistical error, while all the other simulations are performed by the state-vector simulator, where the expectation values are exactly calculated without errors. 

For each simulation and experiment in this paper, the molecular Hamiltonian is first prepared as the second-quantized electronic Hamiltonian using the Born-Oppenheimer approximation with Hartree-Fock orbitals using the STO-3G basis unless otherwise stated, and converted to the qubit one by the Jordan-Wigner mapping. Active spaces are explicitly specified when employed, otherwise the full-space Hamiltonians are used.
The electronic Hamiltonians are generated by OpenFermion~\cite{mcclean2020openfermion} interfaced with PySCF~\cite{sun2018pyscf}. The molecular geometries and other details are shown in Appendix~\ref{sec:appendix-details-of-sim-and-exp}. Stable geometries are chosen for all the molecules except for the hydrogen chains, and a potential impact of unstable geometry is briefly analyzed in Appendix~\ref{ssec:appendix-bond-length}.

\subsection{QSCI for ground state}
\label{subsec:ground-state-simulation-with-noiseless-vqe}

\begin{figure}
    \includegraphics[width=.45\textwidth]{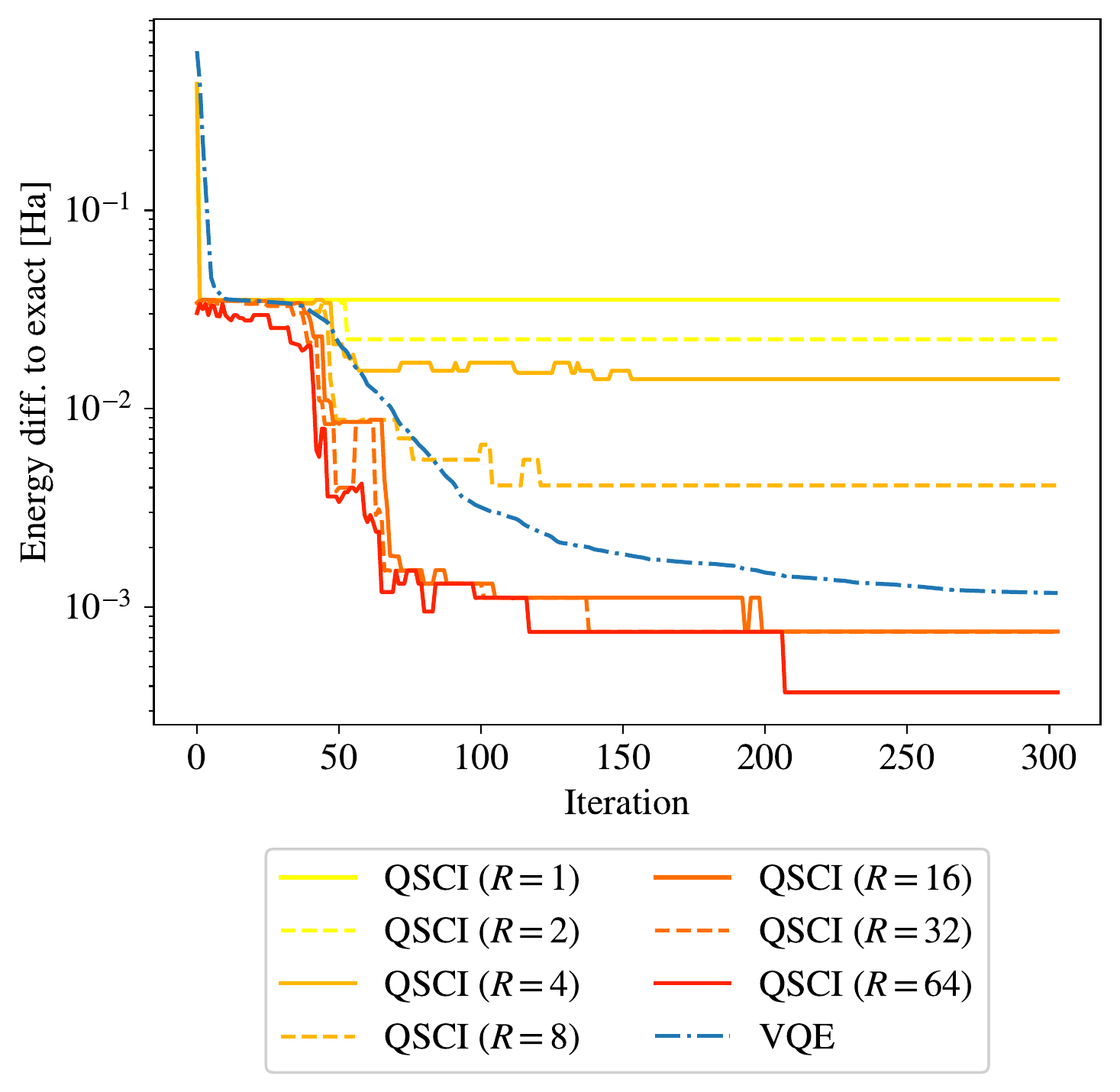}
    \caption{
    The result of QSCI, the proposed method, for the ground state of \ce{H2O} molecule by noiseless simulation, shown with optimization history of VQE, which is used to prepare the input states of QSCI. 
    Each of the resulting energies is shown as the difference to the CASCI result $E_{\rm exact}$ in Hartree. The dash-dotted line shows the result by the state-vector simulation of VQE. The lines specified by the parameter $R$ show the results of QSCI, $E_R - E_{\rm exact}$, for the given value of $R$, using the parametrized state at each iteration of VQE as the input state. The parameter $R$ determines the classical computational cost for QSCI, as explained in the main text.}
    \label{fig:noiseless-vqe}
\end{figure}

We first show the result of numerical simulation for ground state with input states prepared by noiseless VQE. We choose \ce{H2O} molecule with five active spatial orbitals and six active electrons as our problem, which leads to a 10-qubit Hamiltonian after the Jordan-Wigner mapping. 
In the VQE calculation, the parametrized quantum circuit is constructed by the real-valued symmetry-preserving ansatz~\cite{gard2020efficient, ibe2022calculating} with the depth 10, and is optimized by the Broyden–Fletcher–Goldfarb–Shanno (BFGS) optimizer in the scientific library SciPy~\cite{virtanen2020scipy}. 
See Appendix~\ref{subsec:setup-noiseless-vqe} for details.

The QSCI calculation is performed for each iteration of the VQE optimization: given the values of ansatz parameters obtained at the iteration, the input state is prepared by the parametrized quantum circuit with those values assigned; then, the QSCI calculation with the idealized sampling introduced in Sec.~\ref{subsec:ground-state} is performed to estimate the ground-state energy $E_R$ for a given $R$, the number of configurations in the subspace $\mc{S}_R$.
This calculation is repeated for all the iterations of VQE with different values of $R$.
The effect of the uncertainty due to the finiteness of the number of shots is addressed later in Secs.~\ref{ssec:sampling-simulation}, \ref{sec:noisy-simulation-experiment}, and Appendix~\ref{ssec:appendix-sampling}.

In Fig.~\ref{fig:noiseless-vqe}, the result is shown along with the optimization history of VQE: $E_R - E_{\rm exact}$ is plotted (in Hartree) for each optimization step of VQE, where $E_{\rm exact}$ is the ground-state energy obtained by the exact diagonalization in the active space, called the complete active space configuration interaction (CASCI).
The energies obtained by VQE are shown in the same way. 

Comparing the results at the last iteration in the plot, one can see that QSCI gives a lower energy than VQE for $R\gtrsim 16$. 
This shows that the method is able to improve the results of VQE even in the noiseless setting, 
where the effect of error mitigation is not present. 
We emphasize that, as discussed in Sec.~\ref{subsec:ground-state}, a lower energy by QSCI means that the energy is closer to the exact ground-state energy, which is manifested in the plot where $E_R - E_{\rm exact}$ is always positive.
It is notable that we can already achieve the chemical accuracy\footnote{In this paper, we define the chemical accuracy 
by $\SI{1}{kcal/mol} \simeq \SI{1.6e-3}{Hartree}$ for the deviation of the calculated energy from the one obtained by the exact diagonalization of the Hamiltonian.
} of $\SI{1.6e-3}{Hartree}$ with $R\sim 16$ while the CASCI in this case uses 100 determinants to express the ground state.\footnote{
For the active space restriction of five active orbitals and six active electrons with $S_z=0$, the number of the Slater determinants
is $\binom{5}{3}\cdot \binom{5}{3}=100$.
If one does not know
the number of electrons and $S_z$ of the ground state before the calculation, then one would need to deal with the full Hamiltonian in the Fock space of $2^{10}=1024$ dimensions.}

A similar tendency is observed for iterations of $\gtrsim 200$. Note, in this case, the VQE results already achieve the chemical accuracy. 
On the other hand, for intermediate iterations of 70--200, the VQE results do not reach the chemical accuracy, while QSCI can improve them to meet the chemical accuracy if $R \gtrsim 16$.
This suggests that an intermediate result of VQE, which is not seeing convergence in the optimization yet, is already useful as an input state of QSCI, 
and that one can reduce the number of optimization steps for VQE by employing QSCI as a post-processing.
We note that the QSCI results do not monotonically decrease, 
as a QSCI calculation for an input state with a lower energy does not necessarily result in a lower output energy.

\subsection{QSCI for excited states}
\label{subsec:simulation-excited-h2o}

\begin{figure}
  \begin{minipage}{0.5\textwidth}
     \subfloat[][$T_1$ state]{
         \includegraphics[width=\textwidth]{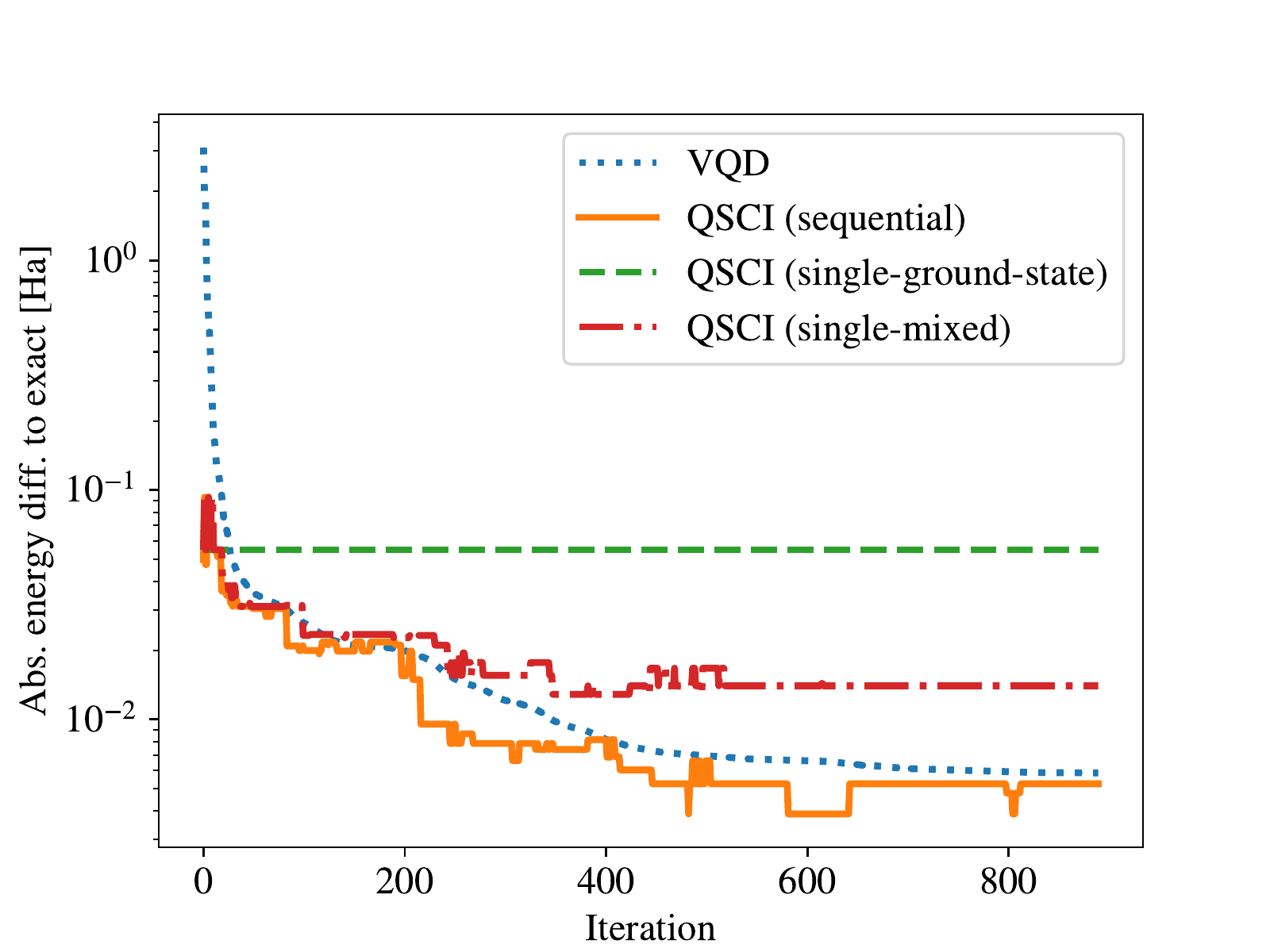}}
    
     \subfloat[][$S_1$ state]{
         \includegraphics[width=\textwidth]{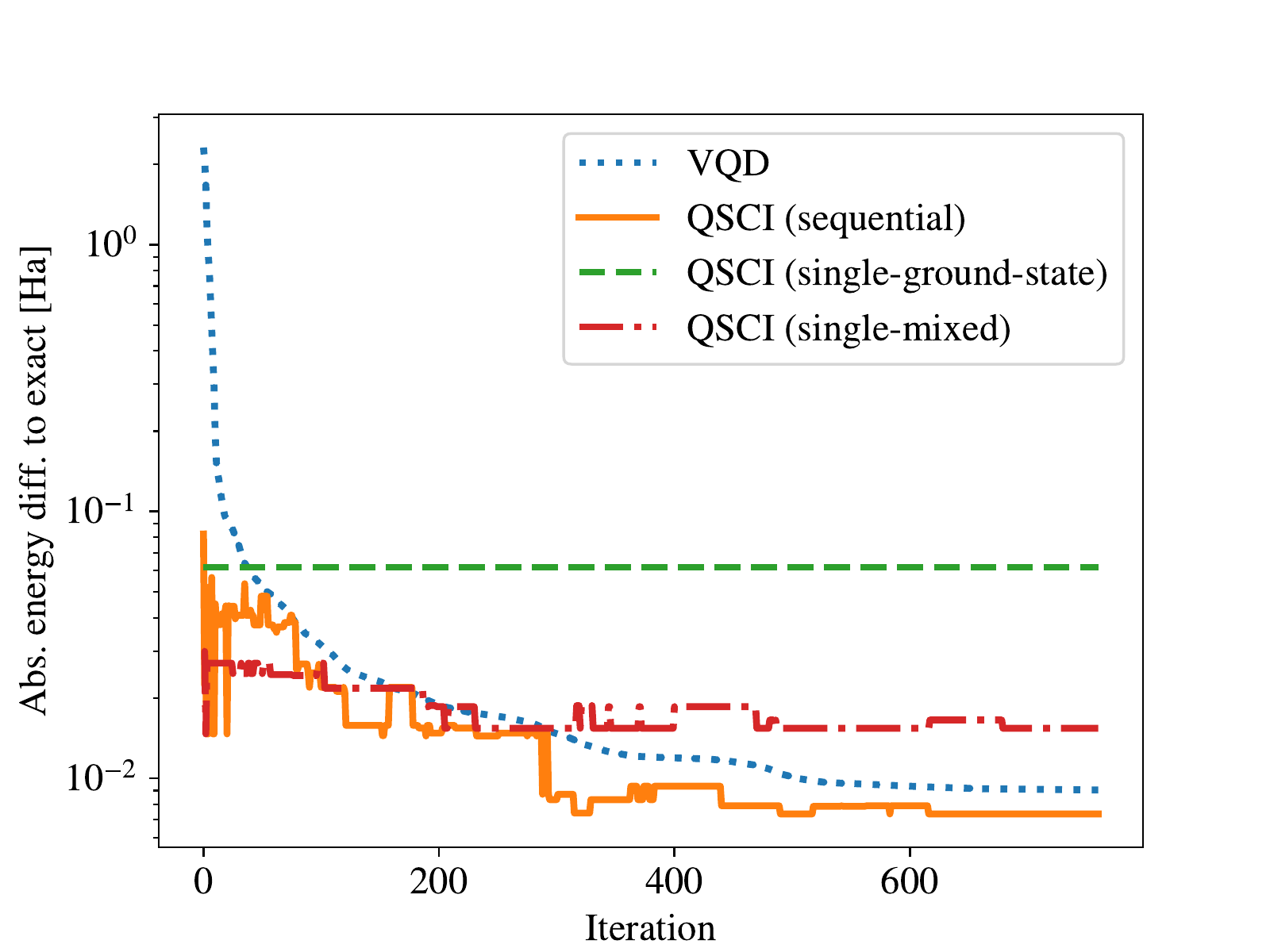}}
  \end{minipage}
    \caption[]{
    Same as Fig.~\ref{fig:noiseless-vqe} but for the first ($T_1$) and second ($S_1$) excited states of \ce{H2O} with $S_z = 0$, along with optimization histories of VQD, shown by dotted lines, for input-state preparation; the energy differences are plotted by the absolute values.
    For the QSCI calculation of the $T_1$ ($S_1$) state, the input state(s) corresponding to the lower energy states, i.e., $S_0$ ($S_0$ and $T_1$) state(s), are prepared by converged sets of parameters of VQD.
    QSCI results are shown for the sequential diagonalization and the single diagonalization with two types of configuration selection, as described in the main text. The QSCI calculations with sequential diagonalization are done with $R_i=16$ for $i=0,1,2$, while the value of $R$ is set to $R=16$ for single diagonalization.}
    \label{fig:noiseless-vqd}
\end{figure}

We next show the results of noiseless simulations for excited states of \ce{H2O} using the two distinct implementations of QSCI presented in Sec.~\ref{subsec:excited-states}, namely the single diagonalization and sequential diagonalization schemes, which take the input states for excited states as well as the ground state.
The numerical setup is similar to the previous subsection, with some differences explained below.
In the input-state preparation, we employ VQE for the ground state and VQD for excited states, each with the same ansatz and optimizer as in the previous subsection;  
we use the same 10-qubit Hamiltonian, but with the overlap terms and penalty terms~\cite{mcclean2016theory,ryabinkin2018constrained,kuroiwa2021penalty} in VQD, for orthogonality between the eigenstates and for symmetry restrictions (charge neutrality for the molecule and $S_z=0$ for the total electron spin) on the excited states.
Under the same symmetry restrictions, the first excited state is a triplet state ($T_1$) and the second excited state is a singlet state ($S_1$), according to the exact diagonalization.
The VQD calculation for $T_1$ requires information of the ground state ($S_0$) to generate the overlap term, for which the ansatz state is used with the converged parameters in VQE.
A similar procedure is applied for $S_1$, but with the extra overlap term for $T_1$ added. 
With the input states for $S_0$, $T_1$ and $S_1$, we perform QSCI calculations to find $T_1$ and $S_1$, where the idealized sampling is used. 
See Appendix~\ref{subsec:setup-noiseless-vqd} for details.

The results are shown in Fig.~\ref{fig:noiseless-vqd} along with the optimization history of VQD, in the similar way as Fig.~\ref{fig:noiseless-vqe}, but for $|E_R^{(i)} - E_{\rm exact}|$ ($i=1$ for $T_1$ and $i=2$ for $S_1$).
At each iteration, three types of QSCI calculations are performed: sequential, single-ground-state, and single-mixed. Sequential diagonalization uses the $T_1$ ($S_1$) state at the iteration, and one (two) lower energy state(s) at their final iterations as input states. Two single diagonalization methods use different input states: ``single-ground-state'' uses the ground state prepared by the converged VQE calculation, and that is why they are constant in the plot; on the other hand, ``single-mixed'' uses the two (three) states as input states, and selects $R$ configurations so that each of the two (three) states contributes as equally as possible, as explained in Sec.~\ref{sssec:method-single}.
Note that $R$ is the dimension of the common subspace $\mc{S}_R$ in Eq.~\eqref{eq:set_single-step}.
The coefficient(s) $\beta_0$ (and $\beta_1$ for $S_1$ state) of the overlap term(s) for orthogonality is set to $\beta_0=\beta_1=\SI{1}{Hartree}$, which is sufficiently larger than the energy gaps between the states in question.
For sequential diagonalization, the values of $R_i$ are fixed to $R_i=16$ for $i=0,1$ and 2, corresponding to $S_0$, $T_1$, and $S_1$ states, respectively;
for single diagonalization, the value of $R$ is set to $R=16$,
so that the sizes of the subspace Hamiltonian matrices to be diagonalized are the same among all the setups.

Comparing the three QSCI results for excited states, the sequential diagonalization performs the best 
except for the initial steps of iterations where the quality of the input state is significantly low.
Moreover, the sequential diagonalization outperforms the VQD calculation, even with a moderate value of $R_i=16$.
For some larger $R$, the single diagonalization is also expected to improve and eventually outperform the VQD result at the same iteration as it can achieve the same representability as the sequential one\footnote{To show this explicitly, assume $N_s=2$ for simplicity. The single diagonalization with the subspace $\mc{S}_{R}=\mc{S}_{R_0}^{(0}\cup\mc{S}_{R_1}^{(1)}$, where the subspaces on the right-hand side denote those of the sequential diagonalization, have at least the same representability as the sequential diagonalization calculation with $\mc{S}_{R_1}^{(1)}$.}.
Although the sequential diagonalization seems to be better in terms of performance, it should be noted that there is no guarantee for the variational inequality in the sequential diagonalization. The inequality for excited states holds in the single diagonalization, as explained in Sec.~\ref{subsec:excited-states}.

\subsection{Scaling of computational costs}
\label{subsec:scaling}

\begin{figure}
     \begin{minipage}{0.45\textwidth}
     \subfloat[][\ce{Cr2}]{
         \includegraphics[width=\textwidth]{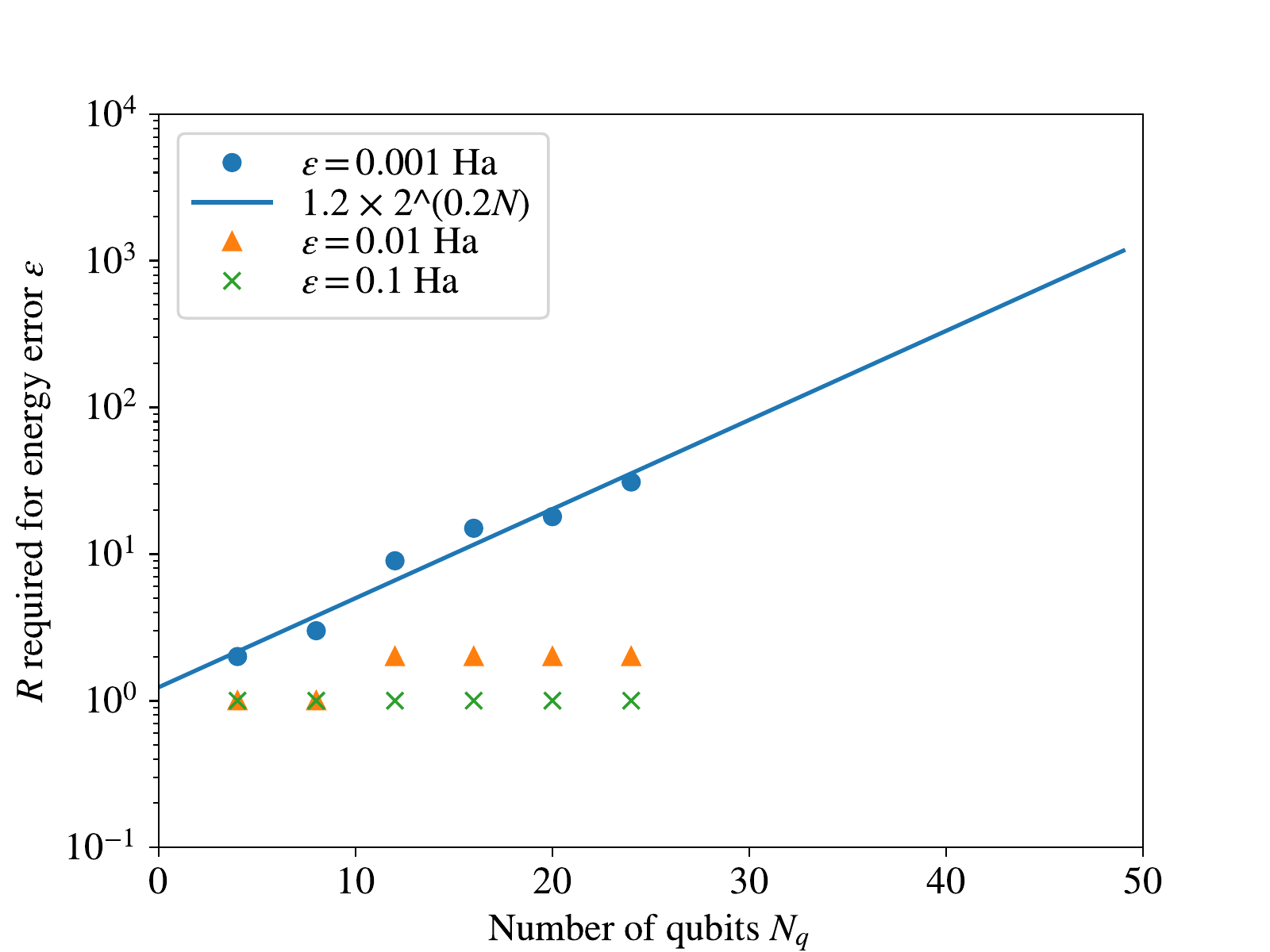}
        }
        
     \subfloat[][Hydrogen chain]{
         \includegraphics[width=\textwidth]{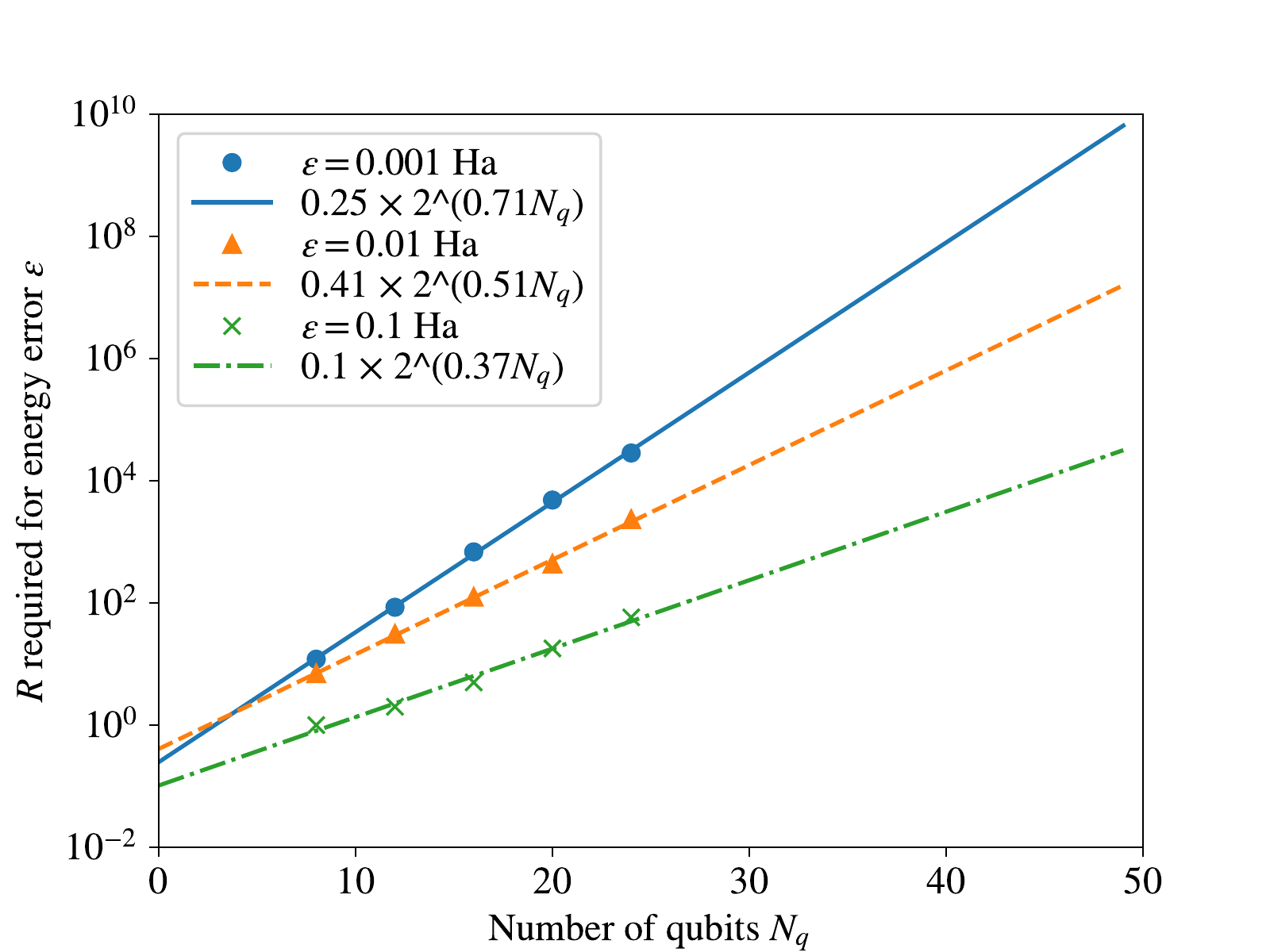}
        }
     \end{minipage}
    \caption{Estimated $R$ required for a given energy error $\epsilon$. Results with (a) expanding active spaces (\ce{Cr2}) or (b) various numbers of atoms (hydrogen chain) are shown by markers, along with the linear fit of each plot.}
    \label{fig:scaling-a}
\end{figure}

We now investigate the scalability of the proposed method by estimating the classical and quantum computational costs 
to calculate the ground states for molecular Hamiltonians of various sizes. More concretely, we estimate the minimum value for $R$ and the required number of shots $N_{\rm shot}$ to obtain the ground-state energy within an error $\epsilon$ for those Hamiltonians.

For this sake, we employ the chromium dimer \ce{Cr2} with various active spaces 
and the linear hydrogen chains with different numbers of atoms. Both \ce{Cr2} and hydrogen chains are known to be challenging molecules in quantum chemistry (see, e.g., Refs.~\cite{larsson2022chromium, motta2020ground} and references therein), while the hydrogen chains are also expected to show a clear scaling in the number of atoms. For \ce{Cr2}, the cc-pVQZ basis set is used with $n$ active orbitals and $n$ active electrons with $n=2,4,\dots,12$; 
the Jordan-Wigner mapping produces $4,8,\cdots,24$-qubit Hamiltonians, respectively. For the linear hydrogen chains, we consider 
$4,6,\cdots,12$ hydrogen atoms equally separated by a distance \SI{1.0}{\AA}; we use the STO-3G basis set without specifying the active space, corresponding to full-space Hamiltonians of $8,12,\cdots, 24$-qubit after the Jordan-Wigner mapping, respectively.

For each setup, the exact ground state of the Hamiltonian is prepared as the input state, and the QSCI calculation is performed by the idealized sampling introduced in Sec.~\ref{subsec:ground-state}, which picks up the $R$ Slater determinants with the largest absolute values of CI coefficients in the input-state wavefunction.
Then, for a given accuracy $\epsilon$, the minimal $R$ that satisfies $\abs{E_R -E_{\rm exact}} \leq \epsilon$ is determined, where $E_R$ is the energy obtained by QSCI with the $R$ configurations and $E_{\rm exact}$ by the exact diagonalization. In Fig.~\ref{fig:scaling-a}, the results are plotted for each molecule by varying the number of qubits, for $\epsilon=0.1, 0.01$ and $0.001$~Hartree; they are extrapolated by fitting (shown by lines) to discuss the feasibility for larger system sizes.

As detailed in Sec.~\ref{ssec:discussion-computational-cost}, 
we infer that the diagonalization with $R\simeq \SI{5e7}{}$ configurations is achievable by the current state-of-the-art classical computing according to the reports~\cite{stampfuss2003improved, garniron2019quantum}.
The result for \ce{Cr2} suggests that $R$ is expected to be manageable even when we require $\epsilon=\SI{0.001}{Hartree}$ for a system larger than 50 qubits, where the exact diagonalization in the whole Fock space, i.e., CASCI, is challenging for classical computers. In the case of the hydrogen chains, on the other hand, the exponential growth of $R$ 
is more clearly observed, and it may become hard to achieve $\epsilon=\SI{0.001}{Hartree}$ for a system much larger than 50 qubits due to the limitation of classical computing. Note that the two scalings have slightly different meanings: the active space is enlarged for \ce{Cr2} while fixing the molecule, i.e., the system size, while the system size itself is enlarged for the hydrogen chains. The results may suggest that our method is more suited to a localized system with many electrons involved, rather than a spatially extended system. For similar studies and results for several diatomic and aromatic molecules, see Appendix~\ref{ssec:more-results-scaling}.

\begin{figure}
     \begin{minipage}{0.5\textwidth}
     \subfloat[][\ce{Cr2}]{
         \includegraphics[width=.9\textwidth]{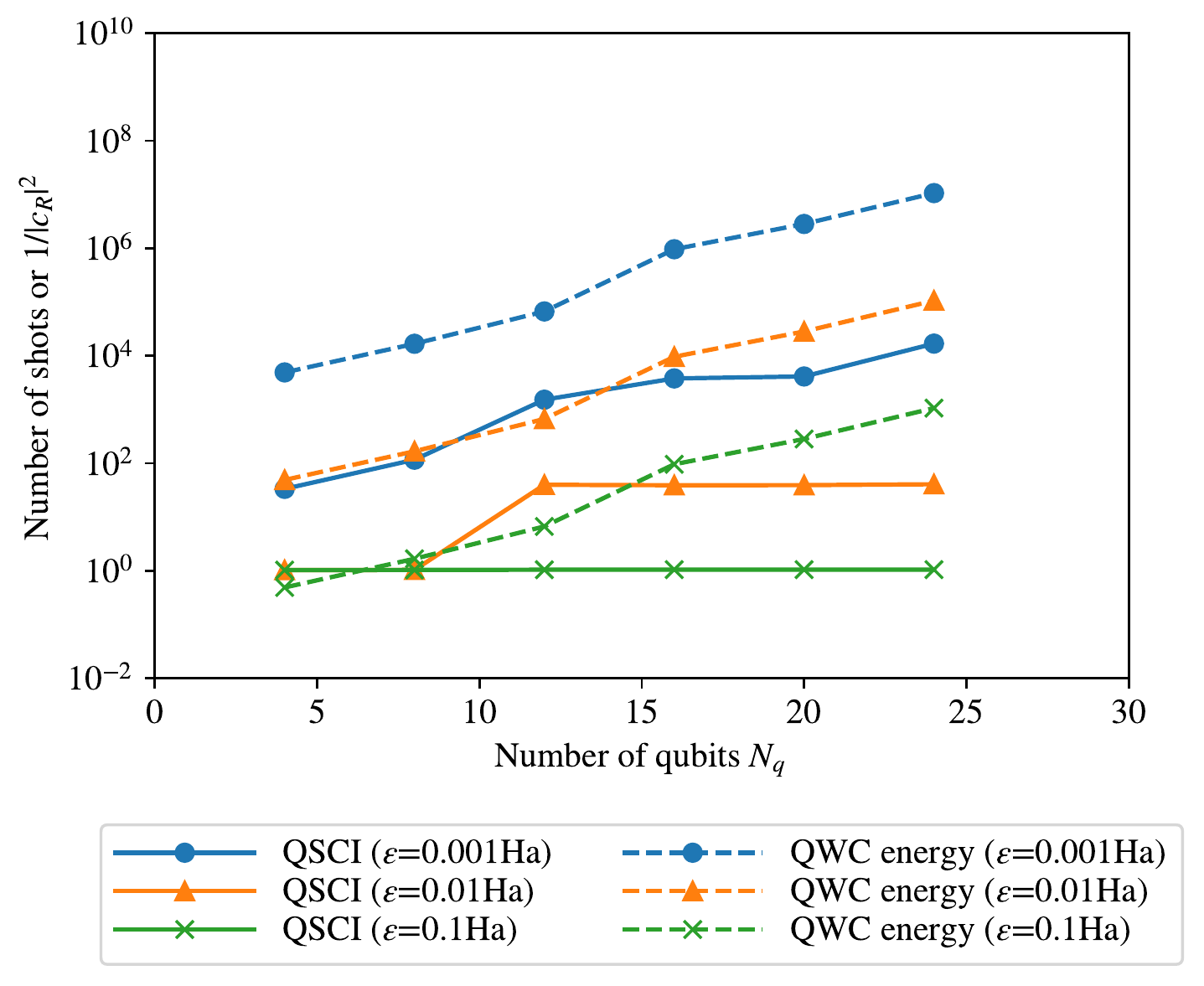}
        }
        
     \subfloat[][Hydrogen chain]{
         \includegraphics[width=\textwidth]{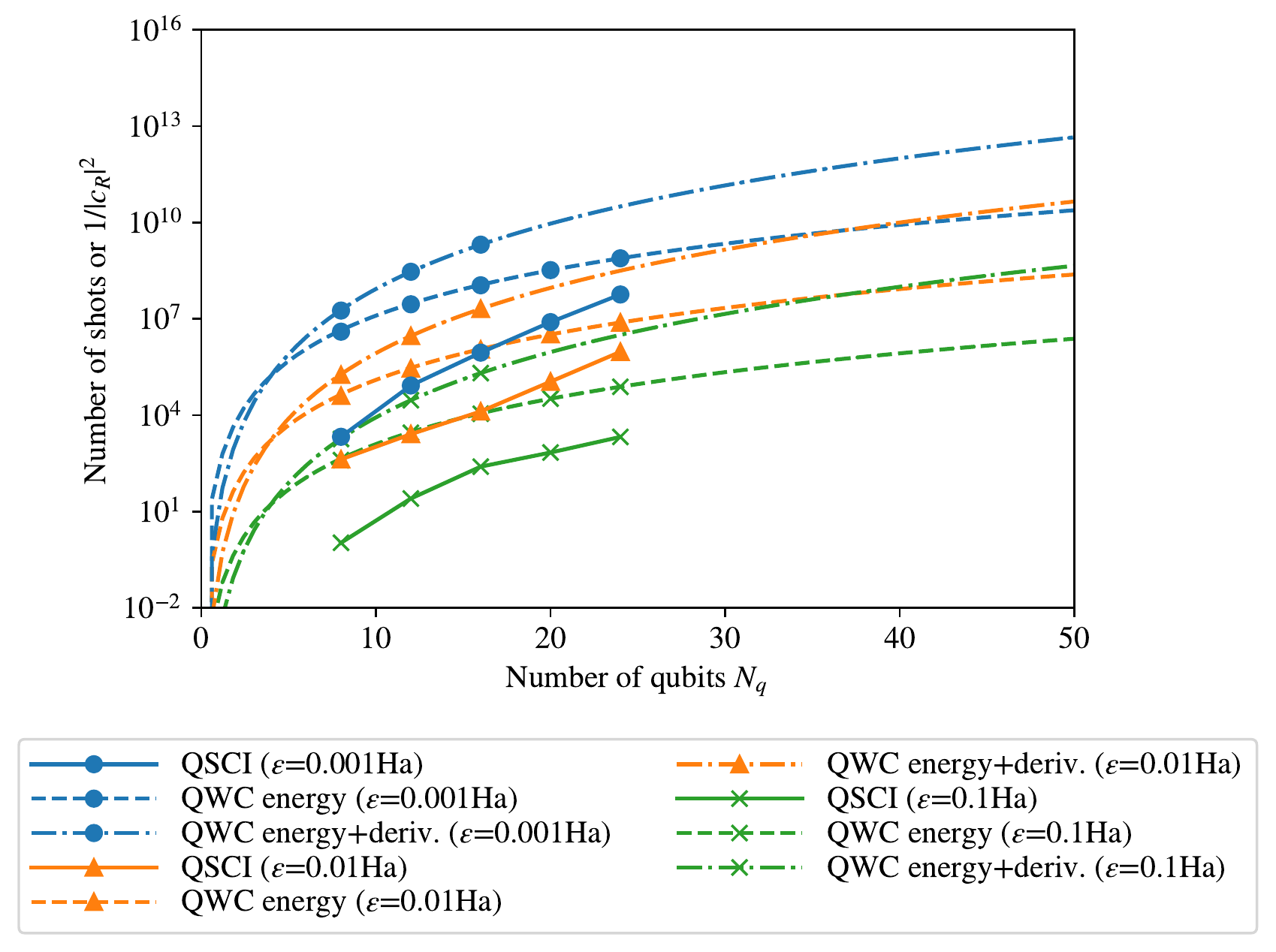}
        }
     \end{minipage}
    \caption{Estimated number of shots for a given energy error $\epsilon$. For QSCI, the number of shots are approximated by $1/\abs{c_R}^2$, where $R$ for each $\epsilon$ is obtained in Fig.~\ref{fig:scaling-a} and $c_R$ is the CI coefficient of the input state with $R$-th largest absolute value. The reasoning for this approximation is explained in the main text.
    For comparison,
    the results of the conventional expectation-value estimation with QWC grouping are plotted for each $\epsilon$, fitted by a logarithmic function in the plot: the required numbers of shots for evaluating the energy are shown by dashed curves; for hydrogen chains, the ones for evaluating the gradients and Hessians in addition to the energy are shown by dash-dotted curves; the precision of gradients and Hessians are set to be $\epsilon/\text{\AA}$ and $\epsilon/\text{\AA}^2$, respectively.}
    \label{fig:scaling-b}
\end{figure}

We next estimate the number of shots for sampling required 
to achieve an error of $\epsilon$ by using the value of $1/\abs{c_R}^2$ for each setup (Fig.~\ref{fig:scaling-b}). Here, $c_R$ is the CI coefficient that has the $R$-th largest absolute value in the input state, where $R$ is taken to be the values shown in Fig.~\ref{fig:scaling-a}. When the state is sampled $1/\abs{c_R}^2$ times, the probability of obtaining the $R$-th most significant configuration is $O(1)$, and in that sense, $1/\abs{c_R}^2$
gives a rough estimator for the number of shots required to sample $R$ most significant configurations. 
We see in the next section, especially in Fig.~\ref{fig:conventional}, 
that this gives a ballpark estimate of the required number of shots for a given accuracy.

For comparison, the total number of shots required in a conventional expectation-value estimation 
is also estimated. More precisely, we analytically estimate the number of shots for which the standard deviation of the expectation-value estimations equals $\epsilon$ for the exact ground state (see, e.g., Ref.~\cite{kohda2022quantum}). In the conventional methods, the expectation value of the Hamiltonian, which is expressed as a linear combination of Pauli strings, is estimated by directly measuring the quantum state in the basis of the Pauli strings multiple times and taking the average of the measurement outcome. To reduce the number of measurements, we employ the qubit-wise commuting (QWC) grouping~\cite{mcclean2016theory} with the sorted insertion algorithm~\cite{crawford2021efficient}.
The total shot is distributed to each of the groups 
with the shot allocation optimized for the exact ground state\footnote{This shot allocation may not be possible in practice without prior knowledge of the exact ground state, but 
this estimation gives the lower-bound on the required number of total shots among possible shot allocation strategies, for a given error tolerance with the given grouping method and the state.
}~\cite{wecker2015progress, rubin2018application}.
Note that, although there are methods that are capable of reducing the number of measurements better than QWC, they require more gate operations for measurements than QWC does; QWC requires a layer of single-qubit rotations after the state preparation, which is minimal for methods that measure the Pauli strings directly, while QSCI requires no gate operation. Most of the other grouping methods are thus expected to be more vulnerable to noise, and QWC is chosen for a fair comparison in this study. 

Figure~\ref{fig:scaling-b} shows the values of $1/\abs{c_R}^2$ in QSCI 
for various numbers of qubits, along with the estimated numbers of shots in QWC. For the hydrogen chains, the results of QWC are fitted by a function $a(N_q)^b$ with parameters $a$ and $b$ as they are expected to be polynomial in the number of qubits\footnote{More precisely, the fit was performed by a function
\begin{equation}
    \log (N_{\text{shot}}(N_q))= B\log(N_q)+A,
\end{equation}
where $A$ and $B$ are the free parameters. 
Similarly in Fig.~\ref{fig:scaling-a}, a linear function $c N_q+d$ was used to fit the data for $\log(R)$, rather than $D\times 2^{C N_q}$ for $R$.
}, 
while the scaling of QSCI is unclear and fitting is not performed.
In the case of \ce{Cr2}, the number of shots for the proposed method seems to be consistently smaller than that of QWC, while the advantage of QSCI, in terms of reducing the effect of statistical fluctuation, is more non-trivial in hydrogen chains with the numbers of qubits $N_q\gtrsim 30$. 

The more operators are evaluated with the same output state, the more advantageous QSCI becomes; as we already noted in the previous section, QSCI does not require any additional quantum computation to evaluate additional observables, because QSCI 
outputs the classical vector representation of the state, and the expectation values are evaluated classically.
On the other hand, in the conventional methods, quantum computational cost becomes more expensive, e.g., to measure additional Pauli strings introduced by the extra operators.
To exploit this feature, we explore a scenario where the nuclear gradient and Hessian are evaluated along with the energy in the case of the hydrogen chains.
For the shot allocation in the QWC grouping, we developed a method that is optimized for measuring multiple operators at once and is used in the simulation; see Appendix~\ref{subsec:appendix-scaling-multiple-operators} for details.
The result, shown also in Fig.~\ref{fig:scaling-b} (b), implies that such a scenario makes QSCI much more advantageous\footnote{It is numerically shown in Appendix~\ref{ssec:appendix-multiple-observable-accuracy} that the accuracy of the gradients and Hessians in QSCI are of the same order as $\epsilon$ when expressed in the units of Hartree$/\mathrm{\AA}$ and Hartree$/\mathrm{\AA}^2$, respectively.}, as the number of shots for QWC significantly increases.

QSCI generally outperforms QWC in terms of the sampling cost within the range of the system size that we studied.
Although the scaling of QSCI seems to be worse than that of QWC in hydrogen chains, we should emphasize here that, even if QWC outperforms QSCI at, say, 50 qubits, it does not mean that QSCI is not useful for Hamiltonians with more than 50 qubits: QSCI has various features, such as error mitigation and the explicit representation of the output state, over the conventional methods, in addition to the reduction of the number of shots. The result should be interpreted as an implication that QSCI can be advantageous in moderately smaller but still classically-challenging systems, even when we only consider the effect of reduction of the number of shots.

\subsection{Sampling simulation}
\label{ssec:sampling-simulation}
For assessing the effect of the statistical error during the sampling in QSCI, sampling simulation with different numbers of shots is performed. The result for a linear \ce{H6} molecule (12 qubits) is shown in Fig.~\ref{fig:conventional}, and results for other molecules are in Appendix~\ref{ssec:appendix-sampling}. For this simulation, the exact ground state is used as the input state, and we include all the configurations obtained in the sampling into the basis set $\mathcal{S}_R$ and we do not specify $R$ beforehand.
For comparison, we also performed a conventional sampling estimation for the exact ground state with QWC grouping and a shot allocation optimized for Haar random states.

For both QSCI and QWC, we perform 10 trials of sampling simulation for each number of shots, and the average of the absolute differences to the exact ground-state energy is plotted along with the standard deviation of the 10 trials. The absolute differences to the exact value are much smaller in QSCI compared to the conventional sampling with QWC grouping. 

It is worth noting that the standard deviation of QSCI energy is smaller than its average difference, while those of QWC sampling are roughly equal. Energy values obtained by QSCI are biased estimators for the exact expectation values even when using the exact ground states as input states. Thus, the absolute difference can roughly be calculated as a sum of the intrinsic bias existing in QSCI and the standard deviation which comes from statistical fluctuation of the subspace $\mathcal{S}_R$. In QWC, on the other hand, the statistical error is the only source of error. One can say that the QSCI result is much less affected by the statistical error compared to the conventional method.

Furthermore, as one can see in Fig.~\ref{fig:conventional}, $1/\abs{c_R}^2$ calculated in the previous simulation gives a relatively accurate estimation of the total shots that gives an average error close to $\epsilon$. Thus the plots in Fig.~\ref{fig:scaling-b} for both QWC and QSCI give reasonable estimations of the number of shots with expected average error $\epsilon$, and the comparison is fair in this sense.
\begin{figure}
     \includegraphics[width=.45\textwidth]{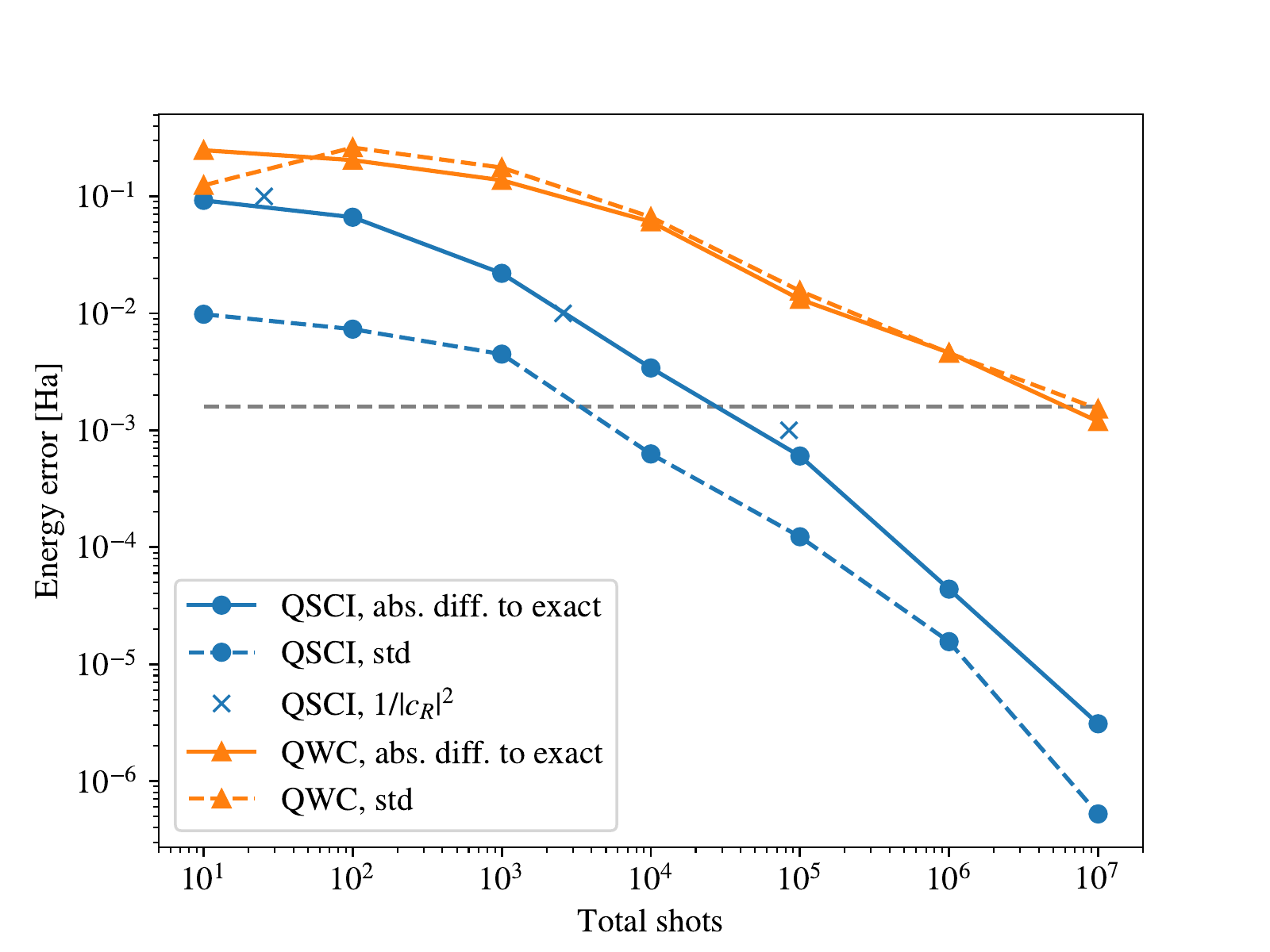}
    \caption{Energy error results for both QSCI and conventional QWC in sampling simulations. For each set of 10 trials for each method, the standard deviation and the average of the absolute error to the exact value obtained by exact diagonalization are shown. QSCI energy error using $1/\abs{c_R}^2$ shots obtained in Fig.~\ref{fig:scaling-b} for $\epsilon=0.1,0.01,0.001$ Ha are also plotted for reference. The horizontal line indicates the chemical accuracy, \SI{1.6}{mHa}.}
    \label{fig:conventional}
\end{figure}

\section{Benchmark of QSCI with noisy simulation and experiment}
\label{sec:noisy-simulation-experiment}

\begin{figure*}
     \begin{minipage}{\textwidth}
     \subfloat[][Noisy simulator w/o post-selection]{
         \includegraphics[width=.45\textwidth]{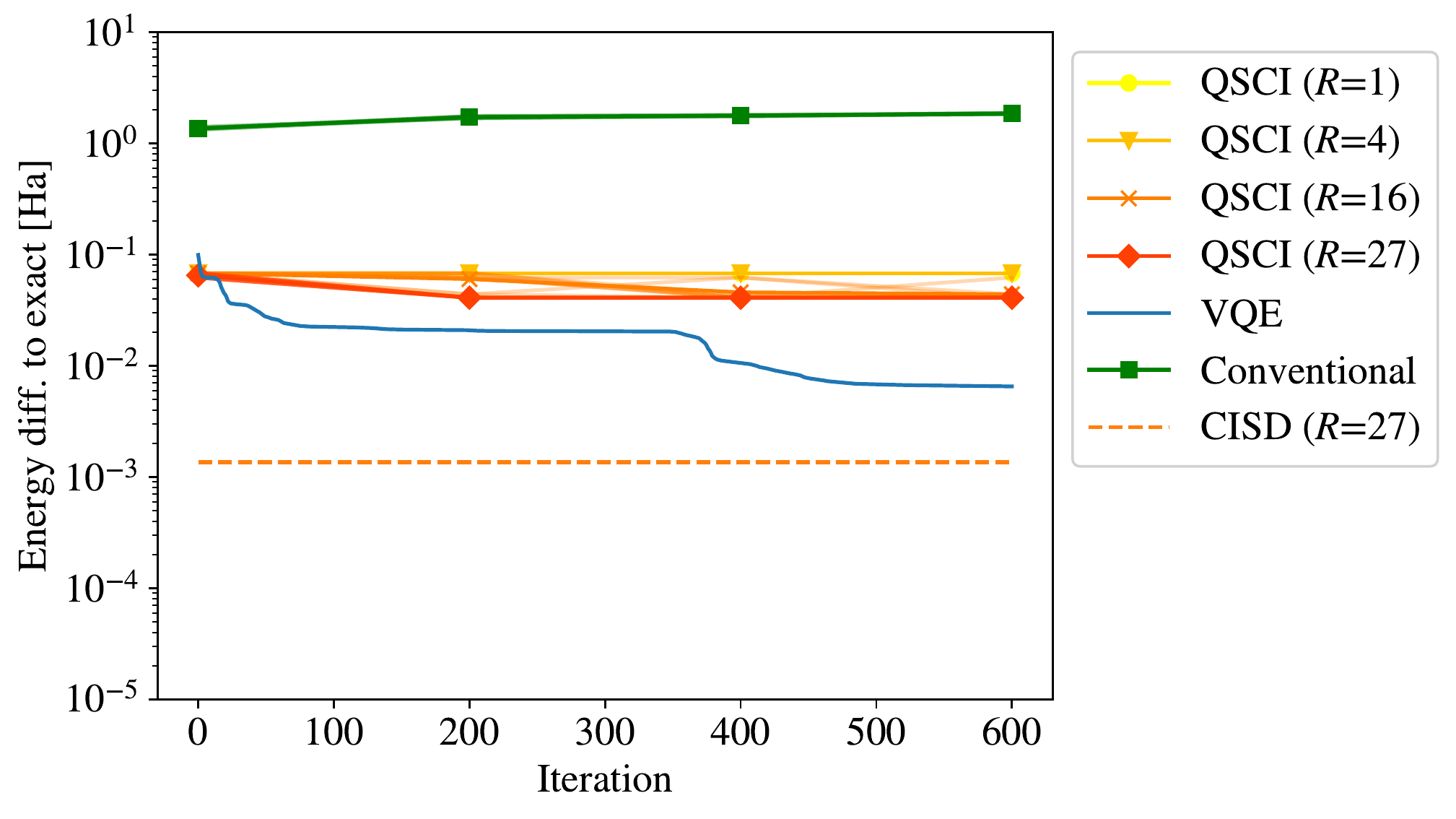}
        }
     \subfloat[][Noisy simulator w/ post-selection]{
         \includegraphics[width=.45\textwidth]{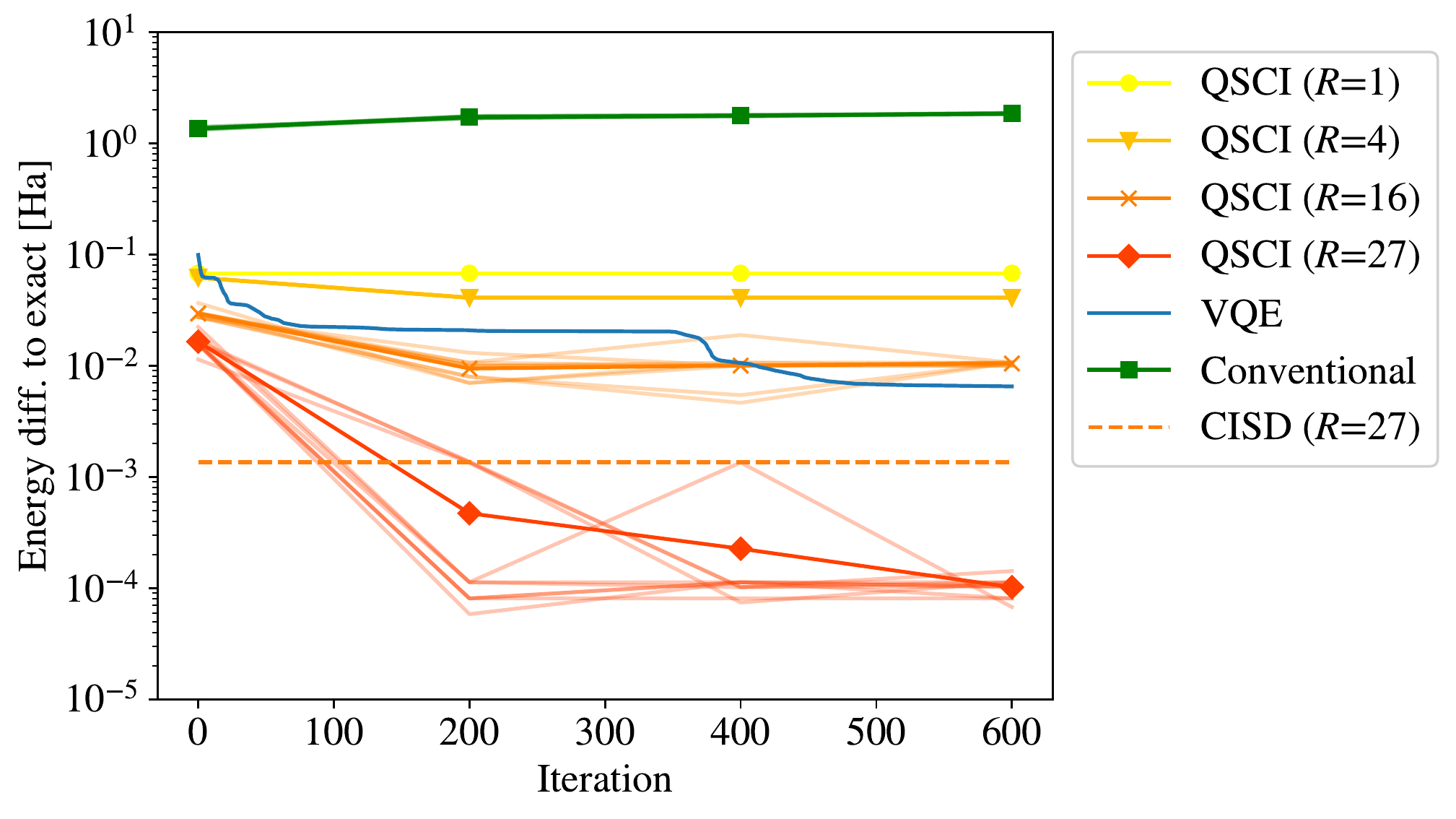}
        }
        
     \subfloat[][IonQ device w/o post-selection]{
         \includegraphics[width=.45\textwidth]{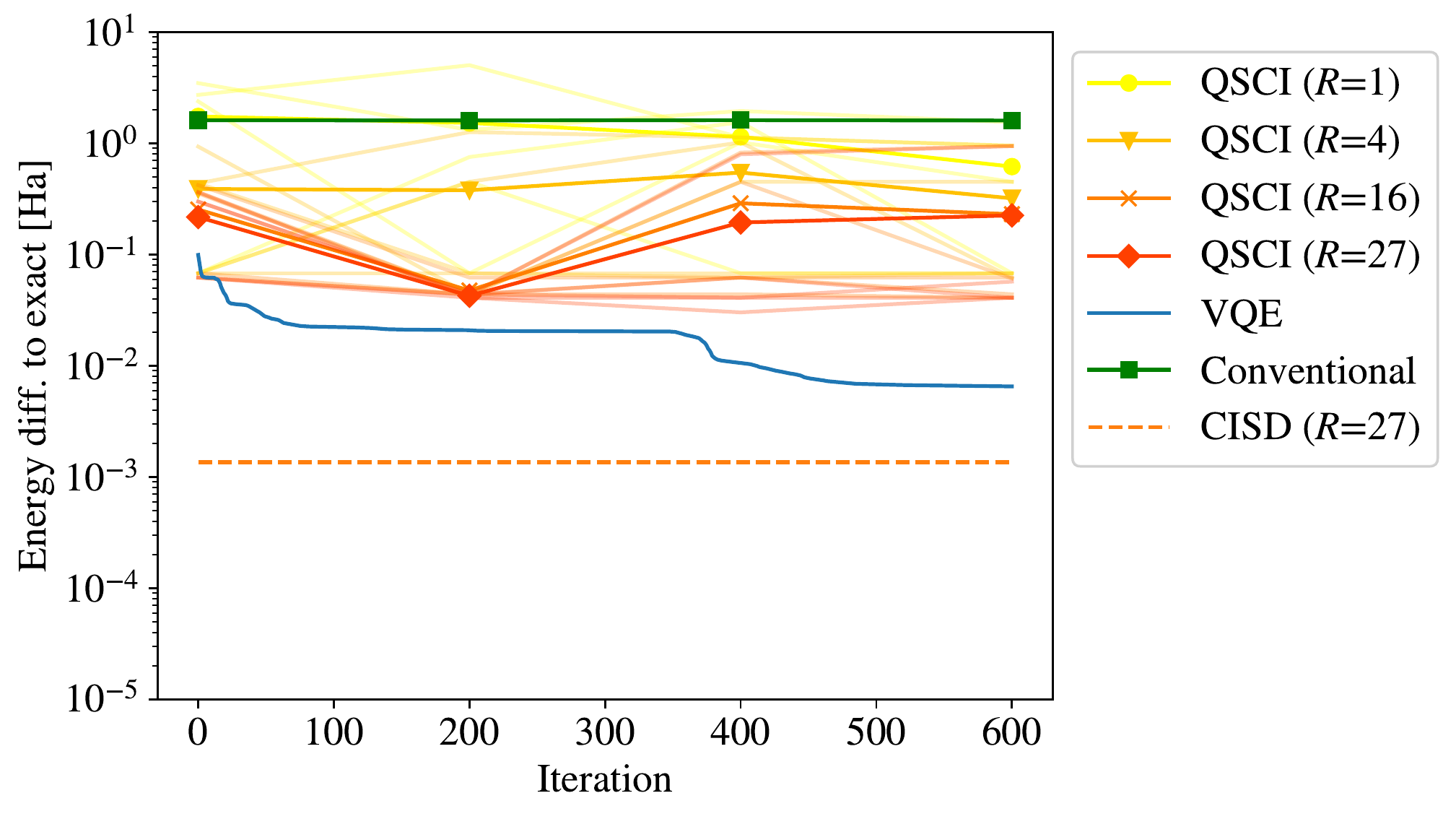}
        }
     \subfloat[][IonQ device w/ post-selection]{
         \includegraphics[width=.45\textwidth]{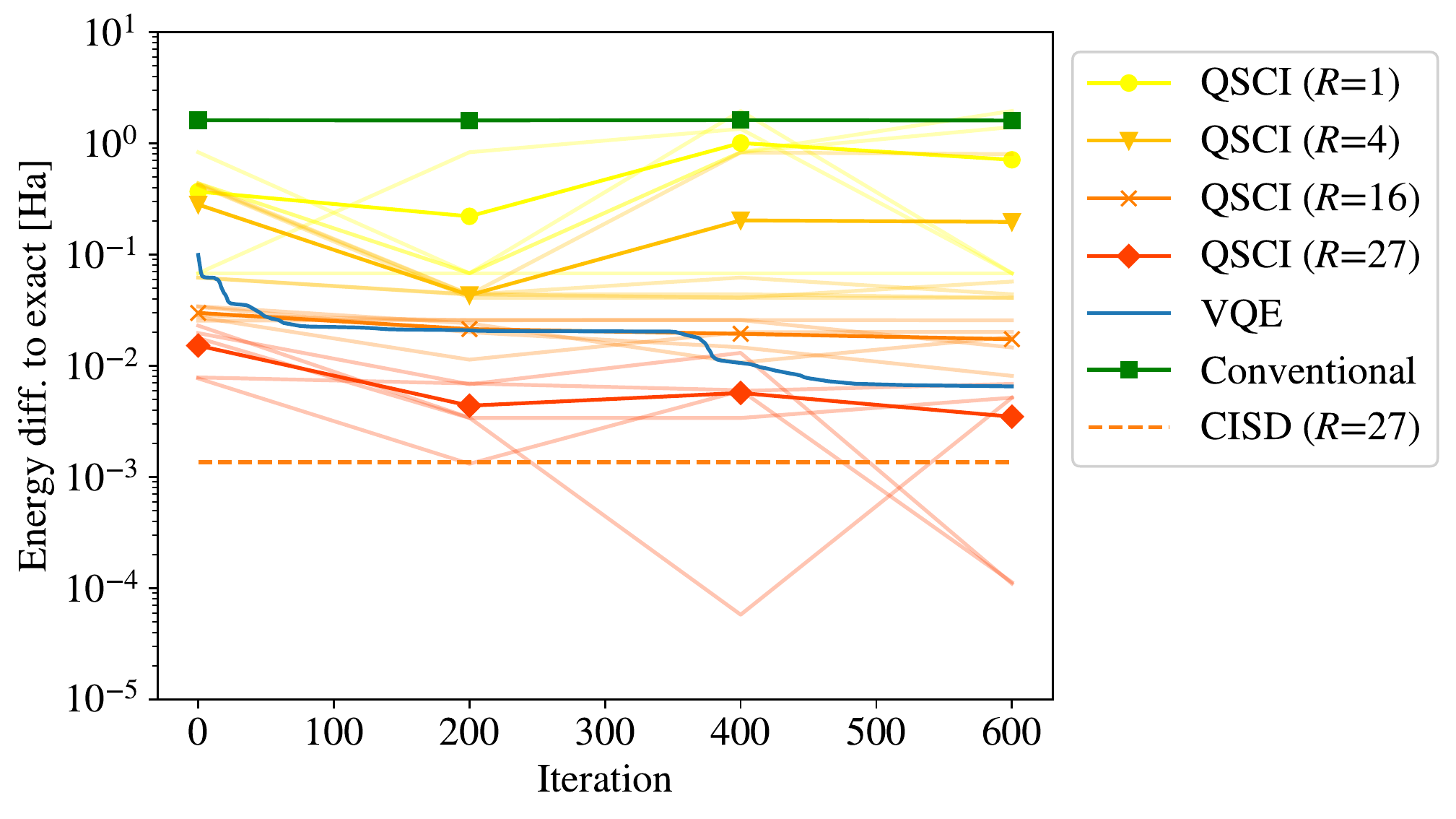}}
     \end{minipage}
    \caption{
    QSCI results for the ground state of the linear hydrogen chain $\ce{H4}$ on 8 qubits by the noisy simulator [(a), (b)] and by the IonQ device [(c), (d)], with and without post-selection, compared with the conventional method of quantum-expectation value estimation and CISD, which uses 27 Slater determinants. The resulting energies are plotted in Hartree as deviations from the one obtained by the exact diagonalization (CASCI).
    Lines specified by ``VQE'' show the exact energy value of the parametrized state at each iteration, and the states at four selected iterations are used as the input states for the QSCI calculations, shown by ``QSCI'' with four different values of $R$. 
    The conventional method uses the QWC grouping in the energy estimation.
    The markers on the solid lines show the average value of some trials while each line without markers shows the result of one of the trials. The number of trials is ten for both QSCI and the conventional method on the noisy simulator, one for the conventional method and five for QSCI on the IonQ device. 
    }
    \label{fig:vqe-experiment}
\end{figure*}

In this section, we describe the result of the experiment for the ground state of the hydrogen chain \ce{H4} (8 qubits), conducted on 
the IonQ 11-qubit device through Amazon Braket service, along with the result of noisy sampling simulation using Qulacs with the identical setup. We first run a VQE calculation of a linear \ce{H4} molecule 
with bond lengths \SI{1.0}{\AA} on a noiseless state-vector simulator. We use the STO-3G basis set without freezing any orbitals, and thus the problem Hamiltonian is 8-qubit. The so-called Ry ansatz with depth 8 is employed for the VQE calculation. See Appendix~\ref{ssec:setup-experiment} for details, including the circuit diagram of the ansatz. Then, we perform QSCI calculations on the quantum hardware and the noisy simulator using four sets of parameters at four distinct iterations of the VQE calculation. We use 10,000 shots for each sampling, and the most frequent $R$ configurations are selected to define the subspace, with and without the post-selection.
The post-selection of the sampling result is performed using the number of electrons $N_e=4$ and the spin $S_z=0$.
For noisy simulation, to simulate the physical noises on the device, single-qubit depolarizing noise is added after each gate and bit-flip noises are added at the end of the circuit to mimic the measurement error. The level of each type of the noise is determined by the single-qubit and two-qubit gate fidelities, and the measurement fidelity of the actual device: 99.61\%, 96.868\%, and 99.824\%, respectively\footnote{More precisely, the error rate of the single-qubit depolarizing noise for each single-qubit gate is set to $p_1$, where $p_1$ is the single-qubit gate infidelity. For the two-qubit gates, single-qubit depolarizing noise is applied to each of the two qubits with probability $1-\sqrt{1-p_2}$ for a two-qubit gate infidelity $p_2$. The bit-flip noise is applied to each qubit with probability $p_{\text{ro}}$, the measurement infidelity.}. 

For comparison, on the quantum device and the noisy simulator, the calculation of the expectation value of the energy using a conventional method is performed with 10,000 shots. The QWC grouping and the shot allocation optimized for Haar random states are employed.
Error mitigation techniques, which may improve the result at the cost of additional quantum resources, are not employed in this study.

The results are presented in Fig.~\ref{fig:vqe-experiment}. By comparing the results from the noisy simulator and the quantum device, one can see that they have a reasonable agreement, although the result from the quantum device seems to be more affected by the errors. Moreover, it is clear that the post-selection is powerful in both simulation and experiment. It is particularly worth noting that even on the physical device, some of the QSCI calculations with $R=27$ do outperform the result of CISD, which also diagonalizes the subspace Hamiltonian with 27 Slater determinants, and achieve the chemical accuracy on the 8-qubit system.

Some minor comments are in order: firstly, at the earlier iterations, the number of sampled (and post-selected) configurations was sometimes less than the given $R$, because the state is concentrated in some computational basis states. In that case, we only used the sampled configurations for the QSCI calculation; secondly, CASCI result, i.e., the exact diagonalization result, corresponds to\footnote{The number of Slater determinants which have the required particle number and $S_z$ is $\binom{4}{2}\cdot \binom{4}{2}=36$.} $R=36$; this number may seem to be comparable to $R=27$, but it is still a non-trivial task to choose 27 configurations out of 36 possibilities.

\section{Discussion}
\label{sec:discussion}

In this section, we discuss various aspects of QSCI. We start from its classical and quantum computational costs in Sec.~\ref{ssec:discussion-computational-cost}, and then discuss its benefits for refining VQE results in Sec.~\ref{ssec:discussion-qsci-refine-vqe}. In Sec.~\ref{subsec:state-preparation}, several ideas for preparing input states are introduced. The aspect of QSCI as a selected CI is discussed in Sec.~\ref{ssec:discussion-qsci-as-selected-ci}, and 
ideas for future directions are finally introduced.

\subsection{Computational costs}
\label{ssec:discussion-computational-cost}

Here classical and quantum computational costs are examined.
In QSCI, classical computing is used for generating the truncated Hamiltonian matrix $\bm{H}_R$ and diagonalizing it. 
Exploiting the Slater-Condon rules, one can generate the sparse matrix $\bm{H}_R$
efficiently in both $R$ and the number of orbitals (see, e.g., Ref.~\cite{tubman2020modern} for details). For diagonalizing $\bm{H}_R$, one can employ algorithms
to diagonalize a sparse matrix, such as the Lanczos method or the Davidson method. 
The generation and diagonalization of the Hamiltonian matrix are common procedures in the selected CI methods, and it is reported~\cite{garniron2019quantum} that $R\simeq \SI{5e7}{}$ of Slater determinants are manageable when a state-of-the-art high-performance computing resource is available, even for the method that repeats the Hamiltonian generation and diagonalization.
In our method, such a repetition is not needed, and thus the computational cost should be smaller.
As already discussed in Sec.~\ref{subsec:scaling}, 
Fig.~\ref{fig:scaling-a} suggests that, for some challenging molecules of $\sim$50 qubits, the QSCI calculation is feasible in terms of the classical cost by the current state-of-the-art classical computing, while meeting the accuracy requirement of $\epsilon \lesssim \SI{0.001}{Hartree}$. 
Note such a system size would be beyond the reach of the exact diagonalization.

The quantum computational time is $t_Q=N_{\text{shot}}\times t_{\text{prepare}}$, 
where $N_{\text{shot}}$ is the number of shots for the sampling, i.e., the repetitions of the input-state preparation and measurement, and $t_{\text{prepare}}$ is the time needed for a single shot.
Note that the total computational time can be reduced if 
multiple quantum computers are available, since the sampling procedures are completely parallelizable. $t_{\text{prepare}}$ highly depends on the type of quantum device to be used and the way to prepare the input state. For example, the Sycamore processor used in the Google's quantum supremacy experiment~\cite{arute2019quantum} can achieve $N_{\text{shot}}=\SI{1e6}{}$ in 200 seconds for a quantum circuit with 53 qubits and 20 repetitions of entangling operations, which corresponds to $N_{\text{shot}}\sim\SI{4e8}{}$ in a day. 
Hence, Fig.~\ref{fig:scaling-b} implies that the sampling cost is affordable for \ce{Cr2} with several tens of qubits, while it may be challenging at the moment to achieve $\epsilon=\SI{0.001}{Hartree}$ for a hydrogen chain with, say, 50 qubits.

We remark that the sampling cost can be significantly reduced if one can 
prepare a state $\ket{\Delta\psi}$ that is orthogonal to a classically tractable state $\ket{\psi_c}$ such that, for some complex numbers $\alpha$ and $\beta$,
$\ket{\psi_{\text{GS}}}=\alpha \ket{\psi_c} + \beta \ket{\Delta\psi}$
approximates the ground state, and can sample from $\ket{\Delta\psi}$ on a quantum computer. The state $\ket{\psi_c}$ can be the Hartree-Fock state or more intricate states such as the CISD state. For example, if $\abs{\alpha}^2=0.9$, then the sampling cost for a given precision can be reduced by a factor of ten. On the other hand,  $\ket{\Delta\psi}$ can be prepared, e.g., by the method of Ref.~\cite{radin2021classically}.

\subsection{
Use of QSCI to refine VQE results
}
\label{ssec:discussion-qsci-refine-vqe}
QSCI can be viewed as a post-processing technique for VQE and its variants, when they are used to prepare the input states. 
Our methods have the following advantages:
\begin{description}
    \item[Error reduction]
    By virtue of the classical diagonalization of a Hamiltonian matrix generated classically, the proposed methods can refine the VQE results, 
    as demonstrated in Sec.~\ref{sec:numerical} and Sec.~\ref{sec:noisy-simulation-experiment}. 
    Although results of noiseless VQE simulations are used to prepare the input states in our numerical and experimental studies, our results suggest that QSCI is also effective to refine \textit{dirty} VQE results subject to the statistical and physical errors.
    Figure~\ref{fig:vqe-experiment} also shows the effectiveness of the post-selection: the rate of the readout error, which is one of the major sources of physical errors, can be reduced from $O(p)$ to $O(p^2)$ with the Jordan-Wigner mapping, as discussed in Appendix~\ref{subsec: post-selection}. 
    Note that QSCI does not require extra gate operations for the measurement, unlike expectation-value estimations in VQE.
    As already shown in Fig.~\ref{fig:noiseless-vqe}, our method is also effective to improve the quality of the input state even in the absence of physical and statistical errors. This feature may enable one to use ansatzes with shallower circuits, or to reduce the number of optimization steps in VQE, by employing QSCI to improve the final result.
    \item[Reliability] Our method is free of errors in the sense that the resulting ground-state energy is exact within the subspace spanned by the quantum-selected configurations.
    This means that the obtained energy is a definite upper bound for the exact ground-state energy, which is not the case in conventional VQE because of physical and statistical errors, as discussed in Sec.~\ref{subsec:ground-state}.
    This is advantageous for comparing the QSCI result obtained on noisy quantum devices with the results of classical variational methods such as CISD or density matrix renormalization group (DMRG)~\cite{white1992,white1993,Schollwock2005}: the variational nature of these methods guarantees that the method that gives the lowest energy is the most accurate one. 
    Similar variational inequalities hold for excited states in the single diagonalization scheme of QSCI, while there is no such guarantee in the sequential diagonalization scheme. Although the latter appeared to be more accurate in our numerical simulation, the former is of great use if one is interested in giving rigorous upper bounds on excited-state energies.
    \item[Handiness] 
    As one has the classical representations of the eigenstates as output, 
    one can compute the expectation values of a large class of observables with no additional quantum computation.  
    Our method becomes more valuable 
    when more observables are to be evaluated, as exemplified in Fig.~\ref{fig:scaling-b}. Moreover, one can also analyze the classical vectors themselves, which may be useful to study the significance of each Slater determinant.
\end{description}

\subsection{
Use of QSCI with more general input states
}
\label{subsec:state-preparation}
As discussed in the previous sections, input states for ground state can be prepared by VQE, and those for excited states by its variants, but the proposed methods are applicable to more general input states.

Our method can in principle be applied to any kind of input states 
that can be prepared and sampled on a quantum computer.
We give an incomplete list of possible preparation schemes for input states in the following: 
the adiabatic state preparation~\cite{farhi2000quantum, aspuru2005simulated}, the imaginary time evolution~\cite{williams2004probabilistic, terashima2005nonunitary, mcardle2019variational,mao2022measurementbased}, classically-boosted VQE~\cite{radin2021classically}, classically-optimized shallow ansatz circuits~\cite{okada2022identification}, unitary coupled-cluster ansatz circuits with classically-optimized parameters~\cite{mcclean2016theory, romero2018strategies, kuroiwa2023clifford+, hirsbrunner2023beyond}, and parametrized states classically optimized by Clifford circuits~\cite{mitarai2022quadratic, ravi2022cafqa}.
Note that the performance of QSCI depends on the quality of the input state and also on the form of the exact eigenstate. For example, if the exact eigenstate is the equal superposition of all the computational basis states, then our algorithm will not perform well.

The algorithm can also be useful for a Hamiltonian that has an exactly-known ground state. For example, one can calculate an exact ground-state energy of a system that is solvable with the Bethe ansatz, but there are quantities, such as a class of correlation functions, that cannot be computed efficiently~\cite{verstraete2009quantum}. Our method provides 
the classical representation of an approximate eigenstate, 
which means that one can evaluate various physical quantities without additional quantum resource, as we already discussed for states prepared by VQE. The preparation of the Bethe ansatz states on quantum computers is addressed in Refs.~\cite{van2021preparing, sopena2022algebraic}.

Moreover, although we proposed the method as a hybrid quantum-classical algorithm, one can apply the method to input states that can be sampled efficiently on classical computers. 
This is shortly discussed in Sec.~\ref{ssec:outlook}.

\subsection{
QSCI as selected CI
}
\label{ssec:discussion-qsci-as-selected-ci}
As selected CI methods, the novelty of QSCI comes simply from how to define the subspace on which we construct the subspace Hamiltonian. Quantum computers are used to sample important configurations from the input state, and there is a quantum speed-up when the input state is hard to sample classically. In selected CI methods, the subspace of the Fock space for the diagonalization is either fixed by the method, e.g., CISD, or adaptively chosen according to the algorithm. We have shown experimentally that CISD performs worse even when compared to the QSCI result on the current NISQ device (Fig.~\ref{fig:vqe-experiment}).

One of the most advanced methods for sampling dynamically important bases is the adaptive sampling configuration interaction (ASCI) algorithm developed by Tubman and co-workers~\cite{tubman2016deterministic,tubman2020modern}. The idea of systematically selecting important bases based on perturbation theory was developed about 50 years ago~\cite{bender1969pr,whitten1969jcp,huron1973jcp,buenker1974tca,buenker1975tca}, and a selection scheme based on Monte Carlo methods was proposed in the 1990s~\cite{greer1995jcp,greer1998jcpss}. However, systematically selected CI was not widely used in quantum chemistry calculations for many years. Recently, it has undergone rapid development and is now becoming applicable to large-scale quantum chemical simulationst~\cite{evangelista2014jcp,holmes2016jctc,schriber2016jcp,holmes2016jctc2,tubman2016deterministic,ohtsuka2017jcp,schriber2017jctc,sharma2017semistochastic,chakraborty2018ijqc,coe2018jctc,coe2019jctc,abraham202jctc,tubman2020modern,zhang2020jctc,zhang2021jctc,chilkuri2021jcc,chilkuri2021jctc,goings2021jctc,pineda2021jctc,jeong2021jctc,coe2023jctc,seth2023jctc}. Indeed, Tubman \textit{et al.} showed that ASCI is capable of handling 34 electrons in 152 spatial orbitals~\cite{tubman2020modern}.

ASCI has hyperparameters that define the size of the search space to adaptively select the configurations, and we will see in Appendix~\ref{ssec:comparison-to-asci} that, with some set of hyperparameters, QSCI can perform better than ASCI.

\subsection{Outlook}
\label{ssec:outlook}
QSCI is applicable to diverse systems, and has many directions for generalizations.
\begin{itemize}
    \item It would be possible to consider a hybrid of the proposed method and another adaptive selected CI method, such as ASCI, by combining the configurations suggested by QSCI with those of the other method. In this way, one could improve the results of the state-of-the-art selected CI methods by using quantum computers.
    \item QSCI is essentially a selected CI where the configurations are randomly selected according to a probability distribution $p(x) = \abs{\braket{x}{\psi_{\mathrm{in}}}}^2$. A classical counterpart of this approach is called Monte-Carlo configuration interaction (MCCI)~\cite{greer1995jcp,greer1998jcpss}. MCCI does not seem to have been extensively studied since the first proposal in 1995, and the use of a more sophisticated (classical) probability distribution for MCCI is yet to be explored. It would be an interesting future work to use a classically tractable $p(x)$ for MCCI and compare/combine it with QSCI. For example, some of the tensor network states, such as the matrix product states (MPS) and the multi-scale entanglement renormalization ansatz (MERA) states, can be efficiently sampled on classical computers~\cite{ferris2012perfect}.
    \item Our method, compared to the conventional VQE, has an advantage that it can evaluate physical observables classically with no additional quantum computational cost. One may leverage this feature by using QSCI for the geometry optimization problem of a molecule, or a molecular dynamics calculation. In those applications, one may skip the sampling for some iterations, and continue to run with the same state subspace defined by the $R$ electron configurations, thereby reducing the quantum computational cost further.
\end{itemize}

We remark that the performance of QSCI depends highly on the quality of the input state.
It would be great if there is a way to start from an input state with modest quality, and then improve the quality of the input state by an iterative use of QSCI.

\section{Conclusion}
\label{sec:conclusion}

In this work, we proposed QSCI, a class of hybrid quantum-classical algorithms, to find low-lying eigenvalues and eigenstates of a many-electron Hamiltonian.
Taking rough approximations of such eigenstates as input, QSCI selects important electron configurations to represent the eigenstates by sampling the input states on quantum computers, and then classically diagonalizes the Hamiltonian in the subspace(s) spanned by the selected configurations to yield better approximations for the eigenstates and their energies.
QSCI is robust against noise and statistical fluctuation, as quantum computation is used only to define the subspaces. A quantum speed-up potentially arises in that sampling a quantum state is, in general, classically intractable.

We verified the algorithms for ground and excited states of small molecules by numerical simulations and experiment, where the latter was conducted on the quantum device with the 8-qubit quantum circuits.
We discussed potential utility of QSCI in various aspects: for instance, taking a state obtained by VQE as the input state, QSCI can be used to refine the VQE result, which may not be accurate enough due to statistical fluctuation, physical noise, and poor optimization; 
QSCI can be used as a technique for eigenstate tomography, which enables estimation of a variety of observables with no additional quantum computational cost.
We also argued that QSCI is potentially feasible to tackle challenging molecules such as the chromium dimer by exploiting quantum devices with several tens of qubits, assisted by a high-performance classical computing resource for diagonalization.

\begin{acknowledgements}
The authors thank Amazon Web Services for supporting this work through their Amazon Braket service.
A part of this work is supported by JST PRESTO JPMJPR2019 and JPMJPR191A, MEXT Quantum Leap Flagship Program (MEXTQLEAP) Grant No. JPMXS0118067394 and JPMXS0120319794, and JST COI-NEXT program Grant No. JPMJPF2014.
\end{acknowledgements}

\appendix
\section{The effect of post-selection
\label{subsec: post-selection}
}
In this section, we discuss the effect of post-selection to mitigate errors in QSCI.
To be specific, we consider the post-selection technique introduced in Sec.~\ref{subsec:ground-state}, which exploits the conservation of particle number and spin, targeting the bit-flip noise.
In the following, we consider to measure an $N$-qubit computational basis state.
We assume that the state is prepared without the effects of noise but each bit of the measurement result is flipped with error probability $p$. It is equivalent to the situation where the bit-flip noise is introduced to each qubit independently after the input state is generated. The probability that the $N$-bit string describing the state is measured correctly is $(1-p)^N$.

\subsection{Jordan-Wigner mapping}
Let us now assume that we consider an electronic Hamiltonian converted by the Jordan-Wigner mapping. In this case, the number of 1's in the $N$-bit string, which we denote by $n_1$, corresponds to the number of electrons in the system 
and is 
sometimes known prior to the calculation for a ground state or an excited state.
One can thus perform the post-selection for a measurement outcome that excludes resulting bit strings with the number of 1's not equal to $n_1$.
Although one may still get incorrect results, the probability is reduced.

More concretely, the probability to get a result with correct $n_1$ is
\begin{equation}
    (1-p)^N+n_0n_1p^2(1-p)^{N-2}+O(p^4),
\end{equation}
where we define $n_0:=N-n_1$.
After the post-selection, the probability to get the correct result is thus
\begin{align}
    &\dfrac{(1-p)^N}{(1-p)^N+n_0n_1p^2(1-p)^{N-2}+O(p^4)}\\
    &=\dfrac{1}{1+n_0n_1p^2/(1-p)^2+O(p^4)}\\
    &=\dfrac{1}{1+n_0n_1p^2+O(p^3)}\\
    &=1-n_0n_1p^2+O(p^3).
\end{align}
The error rate of getting an incorrect result is reduced from $1-(1-p)^N\sim pN$ to $n_0n_1p^2$ with the ratio being
\begin{equation}
    \dfrac{n_0n_1p^2}{pN}=\left(\dfrac{n_0}{N}\right)\left(\dfrac{n_1}{N}\right)Np \sim O(pN),
\end{equation}
which is less than one in a sensible situation $Np \ll 1$, where the original success probability $(1-p)^N$ is not too small.

Although we here considered a computational basis state as the input state,
we expect that 
the post-selection similarly works 
for a general input state that is a superposition of computational basis states with some fixed $n_1$.
Note that, if one also knows the total spin $S_z$ of electrons 
prior to the calculation, one can count the number of 1's separately for up- and down-spin electrons, and make the post-selection more efficient.

\subsection{Other mappings}
For most of the other fermion-qubit mappings, it is not expected that the reduction of the error probability from $O(p)$ to $O(p^2)$ happens. For example, in the parity mapping~\cite{bravyi2002fermionic,seeley2012bravyi} and the Bravyi-Kitaev mapping~\cite{bravyi2002fermionic}, the states $\ket{01}$ and $\ket{11}$ are connected by just one bit flip, but both of them are one-electron states. This bit flip, which cannot be detected by the post-selection, occurs with probability $O(p)$, and thus the error rate after the post-selection is still $O(p)$. The same is true for a Hamiltonian with a reduced number of qubits using symmetries, where there is always a bit flip that does not change the total number of electrons.

\section{Details of the algorithms}

In this section, we present several detailed discussions on the QSCI algorithms.

\subsection{Choice of $\beta_i$ parameters and variational inequalities in sequential diagonalization scheme}
\label{subsec:details-sequential}

Here we discuss the sequential diagonalization scheme, introduced in Sec.~\ref{sssec:method-sequantial}, on how to choose the $\beta_i$ parameters and a potential violation of the variational inequality, following the discussion in Ref.~\cite{higgott2019variational}.

Suppose $k$ low-lying eigenstates of $\hat{H}$, $\ket{E_0}, \cdots, \ket{E_{k-1}}$, are known exactly.
Then, the effective Hamiltonian to find the $k$-th eigenstate can be exactly constructed as
\begin{align}
 \hat{H}^{(k)\prime} =\hat{H}+
 \sum_{i=0}^{k-1}\beta_i \ket{E_i}\bra{E_i}.
\end{align}
This can be formally expressed as
\begin{align}
 \hat{H}^{(k)\prime} =
 \sum_{i=0}^{k-1}(E_i + \beta_i) \ket{E_i}\bra{E_i} + \sum_{i\geq k} E_i \ket{E_i}\bra{E_i},
\end{align}
where $E_i$ represents the $i$-th eigenvalue of $\hat{H}$ in this appendix.
For $\beta_i > E_k - E_i$ ($i=0,\cdots, k-1$), the following inequality holds for an arbitrary $\ket{\psi}$ with $\bra{\psi}\ket{\psi}=1$:
\begin{align}
 \bra{\psi} \hat{H}^{(k)\prime} \ket{\psi} \geq E_k,
\end{align}
where the equality holds if and only if $\ket{\psi}=\ket{E_k}$ up to a phase factor.
In the language of the eigenvalue problem of Eq.~\eqref{eq:eigenvalue-eq}, this implies $E_{R_k}^{(k)\prime}\geq E_k$, where $E_{R_k}^{(k)\prime}$ is the smallest eigenvalue of $\bm{H}_{R_k}^{(k)\prime}$, the subspace matrix for $\hat{H}^{(k)\prime}$, defined in the same way as Eq.~\eqref{eq:sequential-matrix}.

In practice, the condition $\beta_i > E_k - E_i$ can be utilized if one has prior knowledge on the energy spectrum, e.g., based on variational quantum algorithms.
But even without such knowledge, one may still rely on the stronger condition of $\beta_i > 2\sum_j \abs{c_j}$~\cite{higgott2019variational}, which is written in terms of the coefficients $c_j$ of the qubit Hamiltonian $\hat{H}=\sum_j c_j P_j$, expressed by the Pauli strings $P_j$.

In reality, the effective Hamiltonian cannot be exactly constructed as the $k$ low-lying eigenstates would be obtained only approximately and, hence, the inequality $E_{R_k}^{(k)}\geq E_k$ is not guaranteed.
For instance, in the problem to find the first excited state, 
the effective Hamiltonian $\hat{H}^{(1)}$ is constructed with $|\psi_{\rm out}^{(0)}\rangle$, the output state for the ground state obtained by the preceding step in sequential diagonalization.
Unless the output state perfectly overlaps with the true ground state $\ket{E_0}$, or $|\langle\psi_{\rm out}^{(0)}| E_0\rangle|=1$, 
there is no guarantee that $\ev{\hat{H}^{(1)}}{\psi}$ is bounded by the exact eigenvalue $E_1$.
Instead, $\min_\psi \ev{\hat{H}^{(1)}}{\psi}$ is only bounded as~\cite{higgott2019variational}:
\begin{align}
E_1 - O((E_1-E_0)\epsilon_0) \leq \min_\psi \ev{\hat{H}^{(1)}}{\psi} \leq E_1 +\beta_0\epsilon_0,
\end{align}
where $\epsilon_0 = 1 - |\langle\psi_{\rm out}^{(0)}| E_0\rangle|^2$ and $\bra{\psi}\ket{\psi}=1$.
A concrete example for breaching the variational inequality is given as follows.
Consider a system with the unique ground state, i.e., $E_0 < E_1$. Suppose that one has a poor output state $|\psi_{\rm out}^{(0)}\rangle$ that is orthogonal to the true ground state $\ket{E_0}$, i.e., $\epsilon_0=1$. 
Then, $\bra{\psi}\hat{H}^{(1)} \ket{\psi} \geq E_0$ for any positive $\beta_0$, where the equality holds if and only if $\ket{\psi}=\ket{E_0}$ up to a phase factor.
This means that the variational inequality $E_{R_1}^{(1)}\geq E_1 (> E_0)$ is violated at least in the limit where the subspace $\mc{S}_{R_1}^{(1)}$ is enlarged to cover the part of the Fock space necessary to express $\ket{E_0}$.
Such a subspace can be constructed, e.g., if the input state is chosen to be $|\psi_{\rm in}^{(1)}\rangle = \ket{E_0}$ with a sufficiently large $R_1$.

\subsection{An optimal shot allocation for evaluating expectation values of multiple observables in conventional method}
\label{subsec:appendix-scaling-multiple-operators}
In this subsection, we describe the details of the numerical estimation of computational cost in Sec.~\ref{subsec:scaling} for evaluating multiple operators. In the numerical simulation, we considered a situation where we want to calculate the expectation values of the nuclear gradient $\left\{\pdv{\hat{H}}{x_i} \mid i=1,\dots,3N_{\text{atom}} \right\}$ and the nuclear Hessian $\left\{\pdv{\hat{H}}{x_i}{x_j} \mid i,j=1,\dots,3N_{\text{atom}} \right\}$ along with the Hamiltonian $\hat{H}(\left\{x_i\right\})$, where $x_i$ are the nuclear coordinates and $N_\mr{atom}$ is the number of atoms in the molecule.
The most naive way of doing it would be to calculate each expectation value completely separately. This is, though, too naive to be considered as the optimal strategy; all the observables are linear combinations of operators $a^\dagger_i a_j$ and $a^\dagger_i a_j a^\dagger_k a_l$ in the fermionic basis, and the expectation values of these operators can be reused among the operators.

Before discussing the optimal strategy for evaluating expectation values of multiple observables, let us review the one for a single observable, following the discussion in Ref.~\cite{rubin2018application}.
Consider a quantum state $\ket{\psi}$ and the expectation value of an operator $\hat{O}$ which can be written as a sum of operators $\hat{O}_l$:
\begin{equation}
    \hat{O}=\sum_{l=1}^{L}\hat{O}_l.
\end{equation}
Each term $\hat{O}_l$ can be either a Pauli string or a sum of Pauli strings that commute with each other, which admits the projective measurement on eigenvalues of each $\hat{O}_l$.
We denote the variance of each term $\hat{O}_l$ per one shot by $\sigma_l^2:=\text{Var}(\hat{O}_l) := \ev{\hat{O}_l^2}{\psi}-\ev{\hat{O}_l}{\psi}^2$. By measuring each term $\hat{O}_l$ with a number of shots $M_l$, the observed expectation value has the variance $\sum_l \sigma_l^2/M_l$. Employing the method of Lagrange multiplier with the Lagrangian
\begin{equation}
    \mathcal{L}=\sum_l M_l +\lambda \left( \sum_l \dfrac{\sigma_l^2}{M_l} -\epsilon^2\right),
\end{equation}
one can get the optimal allocation of the number of shots with the total variance of the expectation value fixed to $\epsilon^2$,
which is
\begin{equation}
    M_l\propto \sigma_l.
\end{equation}
In general, $\sigma_l$ is not exactly known a priori, so one may use $\sigma_l$ for Haar random states to get a reasonable strategy. One may also try to improve the strategy by dividing the shot budget for one evaluation of an expectation value into several iterations: one can simply evaluate the expectation value with a mildly optimized strategy in the first iteration, and then, in the rest of the iterations, one can adjust the strategy by calculating $\sigma_l$ by using the expectation values obtained in the previous iterations.

Generalizing the above discussion, let us consider a situation where one calculates the expectation values of a set of operators $\left\{\hat{O}^{(i)} \mid i=1,\dots,n\right\}$.
We assume that $\hat{O}^{(i)}$ is decomposed as
\begin{equation}
    \hat{O}^{(i)}=\sum_{l=1}^{L} \hat{O}^{(i)}_l,
\end{equation}
where all of $\left\{\hat{O}^{(i)}_l\mid i=1,\dots,n\right\}$ are simultaneously measurable for each $l$, i.e., $[\hat{O}^{(i)}_l, \hat{O}^{(j)}_l] = 0$ for any $i,j$.
In our numerical simulation, the grouping was done by firstly taking the sum of all the observables $\hat{O}^{(i)}$ with each Pauli string with negative coefficient multiplied by $-1$ to make it positive. Then the greedy qubit-wise grouping of Refs.~\cite{mcclean2016theory, crawford2021efficient} was used.
Our aim here is to find a good strategy to estimate the expectation values of all the operators $O^{(i)}$ with statistical error less than $\epsilon$.
Note that one can always rescale the observables so that the required precision is the same for all observables even when one requires different precision for different operators.
To get an analytical solution, we choose the following Lagrangian with slightly modified constraint,
\begin{equation}
    \mathcal{L}=\sum_l M_l +\lambda \left(\sum_l \sum_i \left(\dfrac{{\sigma_l ^{(i)}}^2}{M_l}\right)-\epsilon_{\text{tot}}\right),
\end{equation}
where $\epsilon_{\text{tot}}$ can be $N\times \epsilon$ but it turns out that the choice of $\epsilon_{\text{tot}}$ does not affect the final result.
By solving the extremal condition of this Lagrangian, one can get the best shot allocation that minimizes the total number of shots, while keeping the sum of the variances of all the operators less than $\epsilon_{\text{tot}}$. The result implies that
\begin{equation}
    M_l\propto \sqrt{\sum_i {\sigma_l^{(i)}}^2}.
\end{equation}
By estimating the variance of each operator $\hat{O}^{(i)}$ with this shot allocation, and by adjusting the total number of shots so that the statistical error of each operator is $\epsilon$ at worst, one can obtain the total number of shots $\sum_l M_l$ with desired precision for all the operators. This may not be the optimal shot allocation to achieve the statistical error $\epsilon$ for each operator as we are minimizing the total variance rather than the maximum value of the variances, but this will give a reasonable strategy that is analytically available.

There is one comment to make on the evaluation of nuclear Hessians.
In the following, the expectation value is always taken by an ansatz state $\ket{\psi(\bm{\theta}(\left\{x_i\right\}))}$ parametrized by the ansatz parameters $\bm{\theta}(\left\{x_i\right\})$; we assume that $\bm{\theta}(\left\{x_i\right\})$ is optimized so that $\ket{\psi(\bm{\theta}(\left\{x_i\right\}))}$ has the minimum energy within the ansatz for each $\left\{x_i\right\}$. We denote the energy expectation value of the state by $E(\left\{x_i\right\})$. In the case of the nuclear gradient,
\begin{equation}
    \pdv{E(\left\{x_i\right\})}{x_i}=\expval{\pdv{\hat{H}(\left\{x_i\right\})}{x_i}}
\end{equation}
holds thanks to the Hellmann-Feynman theorem, and it suffices to compute the right-hand side to obtain the nuclear gradient of the energy.
In the case of the nuclear Hessian of the energy~\cite{mitarai2020theory}, on the other hand, it is in general necessary to evaluate the contribution of derivatives acting on the state as well as on the Hamiltonian operator,
\begin{equation}
    \pdv{E(\left\{x_i\right\})}{x_i}{x_j}=\expval{\pdv{\hat{H}(\left\{x_i\right\})}{x_i}{x_j}}+\dots
\end{equation}
In our numerical simulation, we ignored the contribution of the derivatives acting on the state for simplicity.
In the case of QWC, this contribution requires additional quantum resources.
On the other hand, in the case of QSCI, one can generate and diagonalize the Hamiltonians at small finite distance $x_i\to x_i\pm \delta$ to get the derivatives of the state within the same selected subspace of the Fock space with no additional quantum resources. If we take this contribution into account properly, the advantage of QSCI will increase.

It should also be noted that, although we evaluate $O(N_{\text{atom}}^2)$ observables in the numerical simulations for the hydrogen chain in the main text, due to the rich geometrical symmetry of the molecules, many of the observables are zero as an operator.
It is likely that, for more generic molecules, the crossing-point of QSCI and QWC comes at a larger number of qubits.

\section{Details of numerical simulations and experiments}
\label{sec:appendix-details-of-sim-and-exp}
In this section, we explain details of the numerical simulations and the experiment on quantum hardware in the main text.
For all the molecules examined in this study, the second-quantized electronic Hamiltonian under the Born-Oppenheimer approximation is generated by OpenFermion~\cite{mcclean2020openfermion} interfaced with PySCF~\cite{sun2018pyscf} using the Hartree-Fock orbitals with the STO-3G minimal basis set, unless otherwise stated. The electronic Hamiltonians are mapped to qubit ones by the Jordan-Wigner transformation. The molecular geometries used in our study are shown in Table~\ref{tab: geometries}. Stable geometries for diatomic molecules are taken from CCCBDB database~\cite{johnson2022nist} and Ref.~\cite{wang2016relativistic}, while those for the other molecules are taken from PubChem~\cite{kim2023pubchem}, except for the hydrogen chains which are not in their stable geometries.
We list the details specific to each of simulations and experiment in the following.

\begin{table*}[] 
 \caption{Geometries of molecules. ``$(\mr{X}, (x,y,z))$" denotes three dimensional coordinates $x,y,z$ of an atom X in units of \AA. 
 \label{tab: geometries}
 }
 \begin{tabular}{c|p{12cm}}
 \hline \hline
 Molecule & Geometry  \\ \hline
 \ce{H2O} & (O, (0, 0, 0)), (H, (0.2774, 0.8929, 0.2544)), (H, (0.6068, -0.2383, -0.7169)) \\
 \ce{H}$_n$ ($n=4,6,8,10,12)$ & (H, (0, 0, 0)), (H, (0, 0, 1.0), \dots,  (H, (0, 0, $n \times 1.0$)) \\
 \ce{LiH} & (Li, (0, 0, 0)), (H, (0, 0, 1.595))\\
\ce{N2} & (N, (0, 0, 0)), (N, (0, 0, 1.1))\\
\ce{O2} & (O, (0, 0, 0)), (O, (0, 0, 1.2))\\
\ce{F2} & (F, (0, 0, 0)), (F, (0, 0, 1.4))\\
\ce{Cl2} & (Cl, (0, 0, 0)), (Cl, (0, 0, 2.0))\\
\ce{HCl} & (H, (0, 0, 0)), (Cl, (0, 0, 1.3))\\
\ce{CO} & (C, (0, 0, 0)), (O, (0, 0, 1.1))\\
\ce{Cr2} & (Cr, (0, 0, 0)), (Cr, (0, 0, 1.6))\\
 \ce{Benzene} &(C, (-1.2131, -0.6884, 0)), (C, (-1.2028, 0.7064, 0.0001)), (C, (-0.0103, -1.3948, 0)), (C, (0.0104, 1.3948, -0.0001)), (C, (1.2028, -0.7063, 0)), (C, (1.2131, 0.6884, 0)), (H, (-2.1577, -1.2244, 0)), (H, (-2.1393, 1.2564, 0.0001)), (H, (-0.0184, -2.4809, -0.0001)), (H, (0.0184, 2.4808, 0)), (H, (2.1394, -1.2563, 0.0001)), (H, (2.1577, 1.2245, 0))\\
 \ce{Naphthalene} &(C, (0, -0.7076, 0)), (C, (0, 0.7076, 0.0001)), (C, (1.225, -1.3944, 0.0001)), (C, (1.225, 1.3944, 0)), (C, (-1.225, -1.3943, 0)), (C, (-1.225, 1.3943, 0)), (C, (2.4327, -0.6958, 0)), (C, (2.4327, 0.6959, -0.0001)), (C, (-2.4327, -0.6958, -0.0001)), (C, (-2.4327, 0.6958, 0)), (H, (1.2489, -2.4822, 0.0001)), (H, (1.2489, 2.4821, -0.0001)), (H, (-1.2489, -2.4822, -0.0001)), (H, (-1.249, 2.4821, 0.0001)), (H, (3.3733, -1.239, -0.0001)), (H, (3.3732, 1.2391, -0.0001)), (H, (-3.3733, -1.239, -0.0001)), (H, (-3.3732, 1.239, 0))\\
 \ce{Anthracene} &(C, (-1.225, 0.706, 0.0001)), (C, (-1.2251, -0.7061, 0.0001)), (C, (1.2251, 0.7061, 0.0002)), (C, (1.2251, -0.7061, 0.0001)), (C, (0, 1.3937, 0.0001)), (C, (0, -1.3938, 0)), (C, (-2.4504, 1.393, -0.0001)), (C, (-2.4505, -1.393, 0)), (C, (2.4505, 1.3929, 0)), (C, (2.4505, -1.3929, 0)), (C, (-3.6587, 0.6956, -0.0001)), (C, (-3.6588, -0.6955, -0.0001)), (C, (3.6587, 0.6956, -0.0002)), (C, (3.6587, -0.6956, -0.0002)), (H, (0, 2.4838, 0)), (H, (0, -2.4839, -0.0001)), (H, (-2.4742, 2.4808, -0.0001)), (H, (-2.4744, -2.4809, 0)), (H, (2.4742, 2.4808, 0)), (H, (2.4743, -2.4808, 0)), (H, (-4.5989, 1.2394, -0.0003)), (H, (-4.5991, -1.2391, -0.0002)), (H, (4.5989, 1.2393, -0.0003)), (H, (4.5989, -1.2393, -0.0004))\\
 \ce{Tetracene} & (C, (0, 0.7045, -0.0002)), (C, (0, -0.7046, -0.0001)), (C, (-2.451, 0.7058, 0)), (C, (-2.4511, -0.7058, 0.0002)), (C, (2.4511, 0.7057, 0.0001)), (C, (2.4511, -0.7058, -0.0001)), (C, (1.2254, 1.3923, -0.0001)), (C, (1.2254, -1.3924, -0.0003)), (C, (-1.2254, 1.3923, -0.0002)), (C, (-1.2255, -1.3923, 0.0002)), (C, (-3.6764, 1.3928, -0.0001)), (C, (-3.6764, -1.3929, 0.0002)), (C, (3.6764, 1.3929, 0.0003)), (C, (3.6765, -1.3929, -0.0001)), (C, (-4.8846, 0.6957, -0.0001)), (C, (-4.8847, -0.6955, 0.0001)), (C, (4.8846, 0.6957, 0.0004)), (C, (4.8847, -0.6956, -0.0001)), (H, (1.2253, 2.4825, -0.0001)), (H, (1.2254, -2.4825, -0.0003)), (H, (-1.2254, 2.4824, -0.0003)), (H, (-1.2255, -2.4824, 0.0003)), (H, (-3.6999, 2.4807, -0.0002)), (H, (-3.7001, -2.4808, 0.0003)), (H, (3.6999, 2.4807, 0.0004)), (H, (3.7001, -2.4807, -0.0003)), (H, (-5.8248, 1.2393, -0.0002)), (H, (-5.8249, -1.2392, 0.0002)), (H, (5.8248, 1.2394, 0.0005)), (H, (5.8249, -1.2392, -0.0002))\\
\hline \hline
 \end{tabular}
\end{table*}

\subsection{Noiseless simulation for ground state}
\label{subsec:setup-noiseless-vqe}
\begin{figure}[h!]
     \includegraphics[width=.45\textwidth]{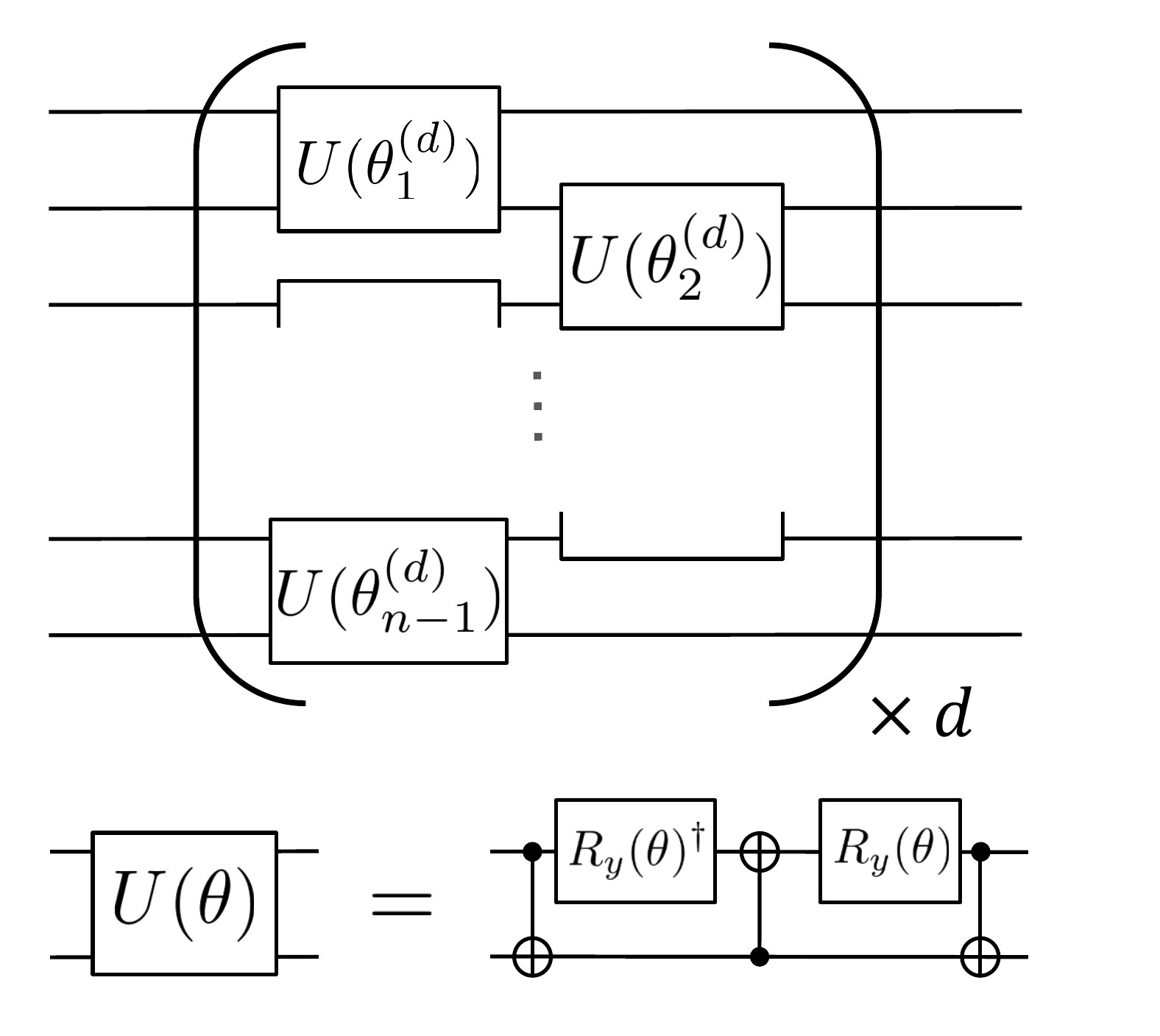}
    \caption{Real-valued symmetry-preserving ansatz with $n$ qubits and depth $d$.}
    \label{fig:rsp-ansatz}
\end{figure}
In Sec.~\ref{subsec:ground-state-simulation-with-noiseless-vqe}, the \ce{H2O} molecule with six active electrons and five active orbitals, is chosen to find the ground state by QSCI. In the VQE calculation for preparing the input states, the BFGS optimizer is employed through the scientific library SciPy~\cite{virtanen2020scipy}, and the real-valued symmetry-preserving ansatz~\cite{ibe2022calculating} is used to construct parametric quantum circuits with depth 10 (Fig.~\ref{fig:rsp-ansatz}).
The initial state of the ansatz circuits is set to be the Hartree-Fock state, and the initial parameters in the optimization are randomly chosen.

\subsection{Noiseless simulations for excited states}
\label{subsec:setup-noiseless-vqd}

In Sec.~\ref{subsec:simulation-excited-h2o}, QSCI is demonstrated for the same \ce{H2O} molecule but to find excited states.
To prepare the input states, the VQD calculations are performed in the same setup as the previous VQE calculation,
but with the penalty terms~\cite{mcclean2016theory,ryabinkin2018constrained,kuroiwa2021penalty} added to the Hamiltonian for constraining the resulting states to have $S_z=0$ and $N_e=6$; specifically, the following operator (in atomic units) is added to the Hamiltonian
\begin{equation}
    3.0 (\hat{S}_z)^2 + 3.0 (\hat{N}_e-6)^2,
\end{equation}
where $\hat{S}_z$ is the operator for the total electron spin in $z$-direction, and $\hat{N}_e$ for the particle number operator of electrons in the active space.
Furthermore, the overlap terms to constrain the state to be orthogonal to lower energy eigenstates~\cite{higgott2019variational} are added with coefficients of unity (in Hartree).
For the sequential diagonalization scheme of QSCI, the coefficients $\beta_i$ for ensuring orthogonality are also set to unity.

\begin{figure}[h!]
     \includegraphics[width=.25\textwidth]{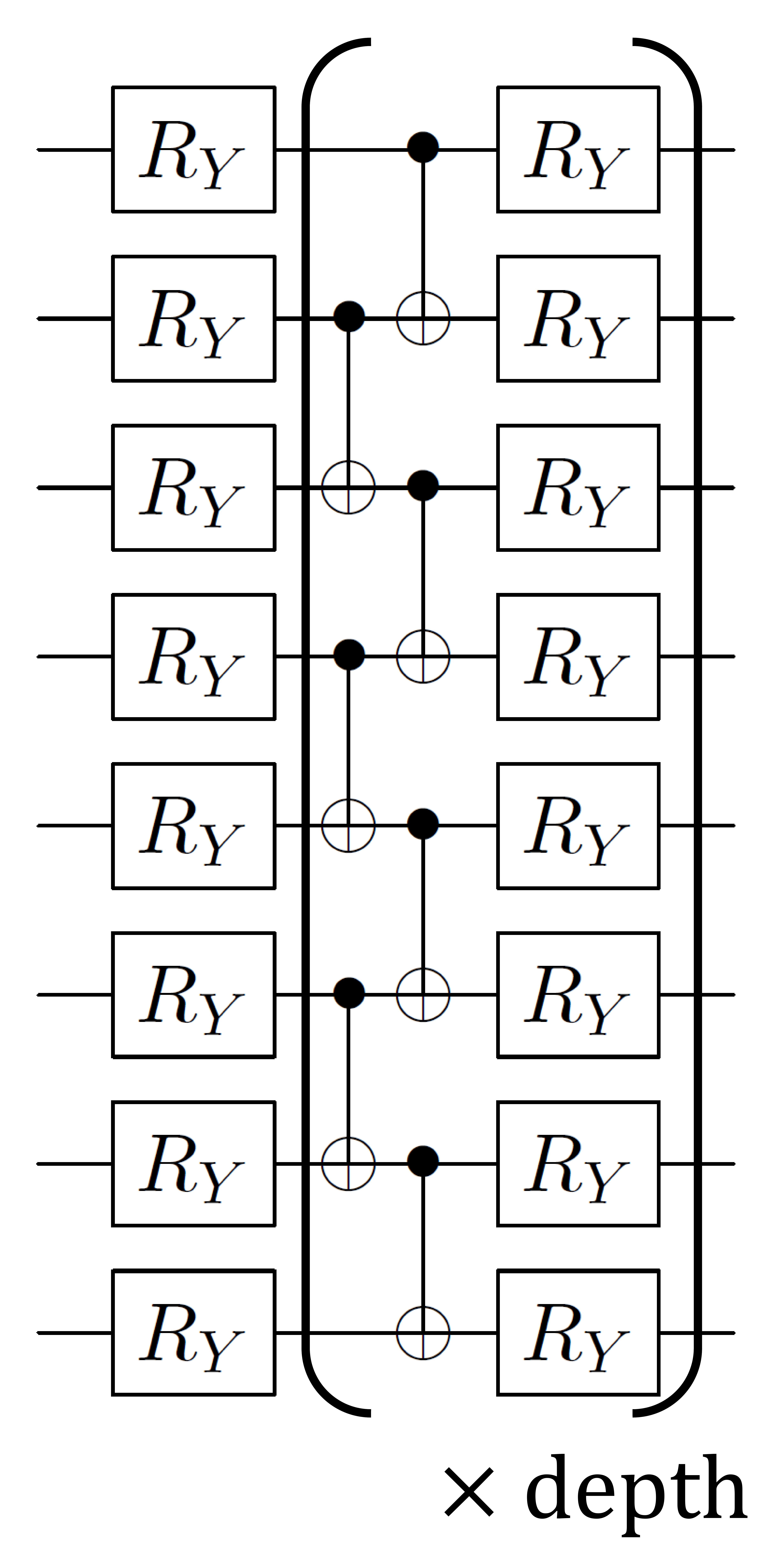}
    \caption{Ry ansatz with 8 qubits. All the rotational gates have independent parameters. The depth is set to be 8 in our experiment.}
    \label{fig:ryansatz}
\end{figure}

\subsection{Noisy simulation and experiment}
\label{ssec:setup-experiment}

For the noisy simulation and experiment in Sec.~\ref{sec:noisy-simulation-experiment}, the input states are prepared by noiseless VQE simulations. The VQE calculations are performed with the BFGS optimizer and Ry ansatz (Fig.~\ref{fig:ryansatz}) with depth 8. Other details are described in the main text.

\section{Supplemental numerical results}
In this section, we provide additional numerical results to supplement the contents in Sec.~\ref{sec:numerical}.
\subsection{Scaling of computational costs with various molecules}
\label{ssec:more-results-scaling}
\begin{figure*}
     \begin{minipage}{0.9\textwidth}
     \subfloat[][$R$ required for energy error $\epsilon=\SI{0.001}{Hartree}$.]{
         \includegraphics[width=.45\textwidth]{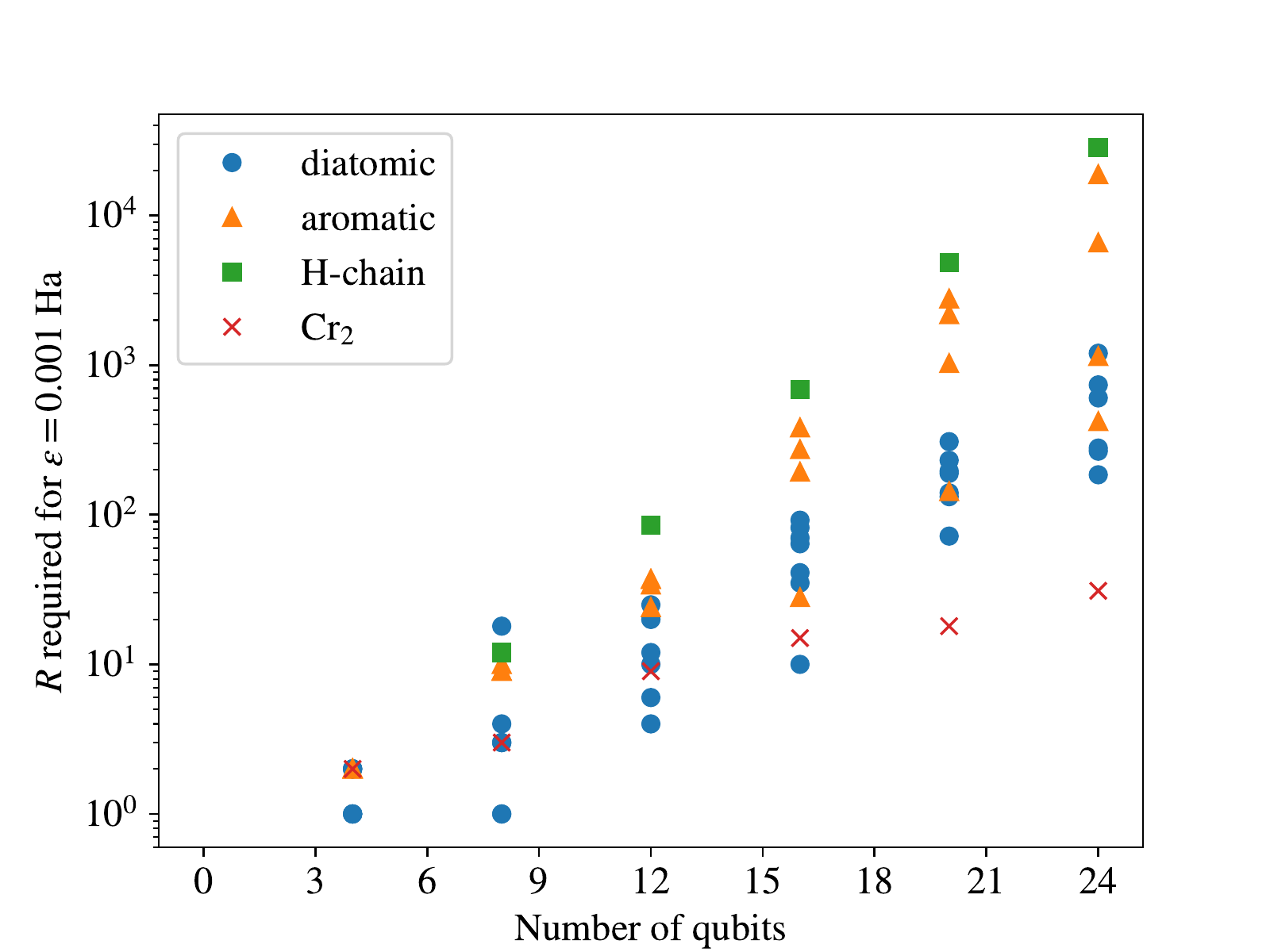}
     }
     \subfloat[][$1/\abs{c_R}^2$ for energy error $\epsilon=\SI{0.001}{Hartree}$.]{
         \centering
         \includegraphics[width=.45\textwidth]{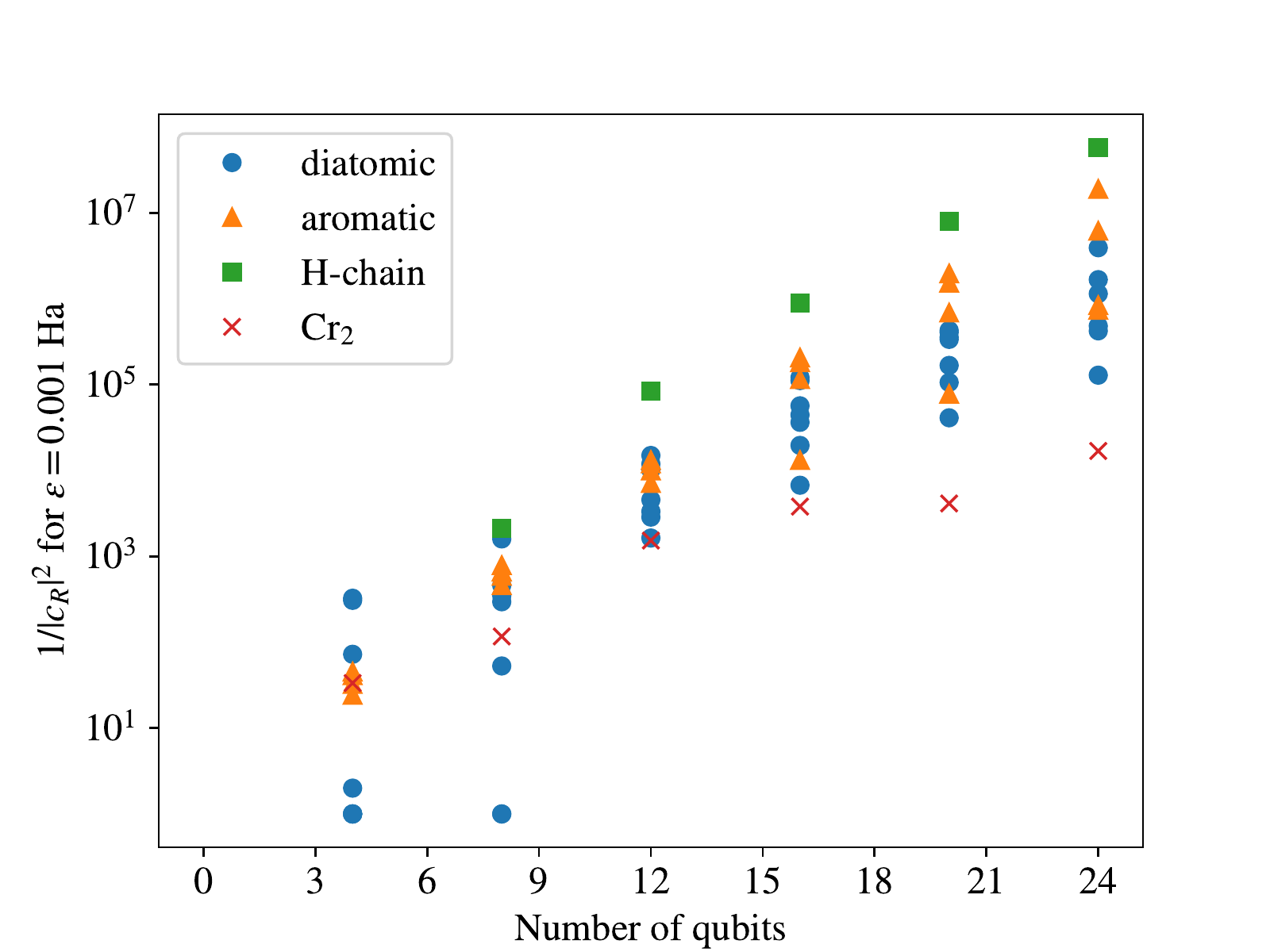}
     }
     
     \subfloat[][$R$ required for energy error $\epsilon=\SI{0.01}{Hartree}$.]{
         \centering
         \includegraphics[width=.45\textwidth]{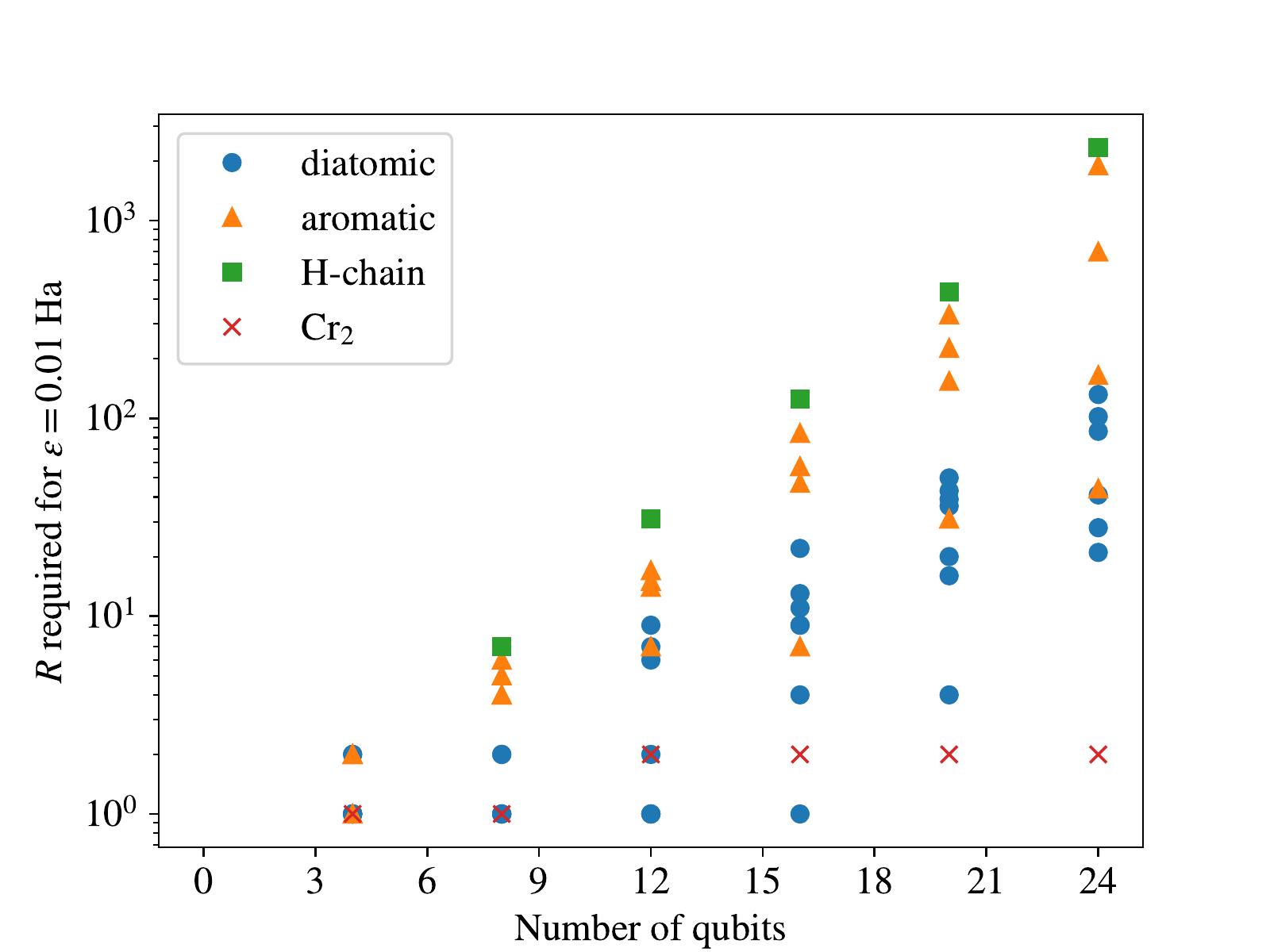}
     }
     \subfloat[][$1/\abs{c_R}^2$ for energy error $\epsilon=\SI{0.01}{Hartree}$.]{
         \includegraphics[width=.45\textwidth]{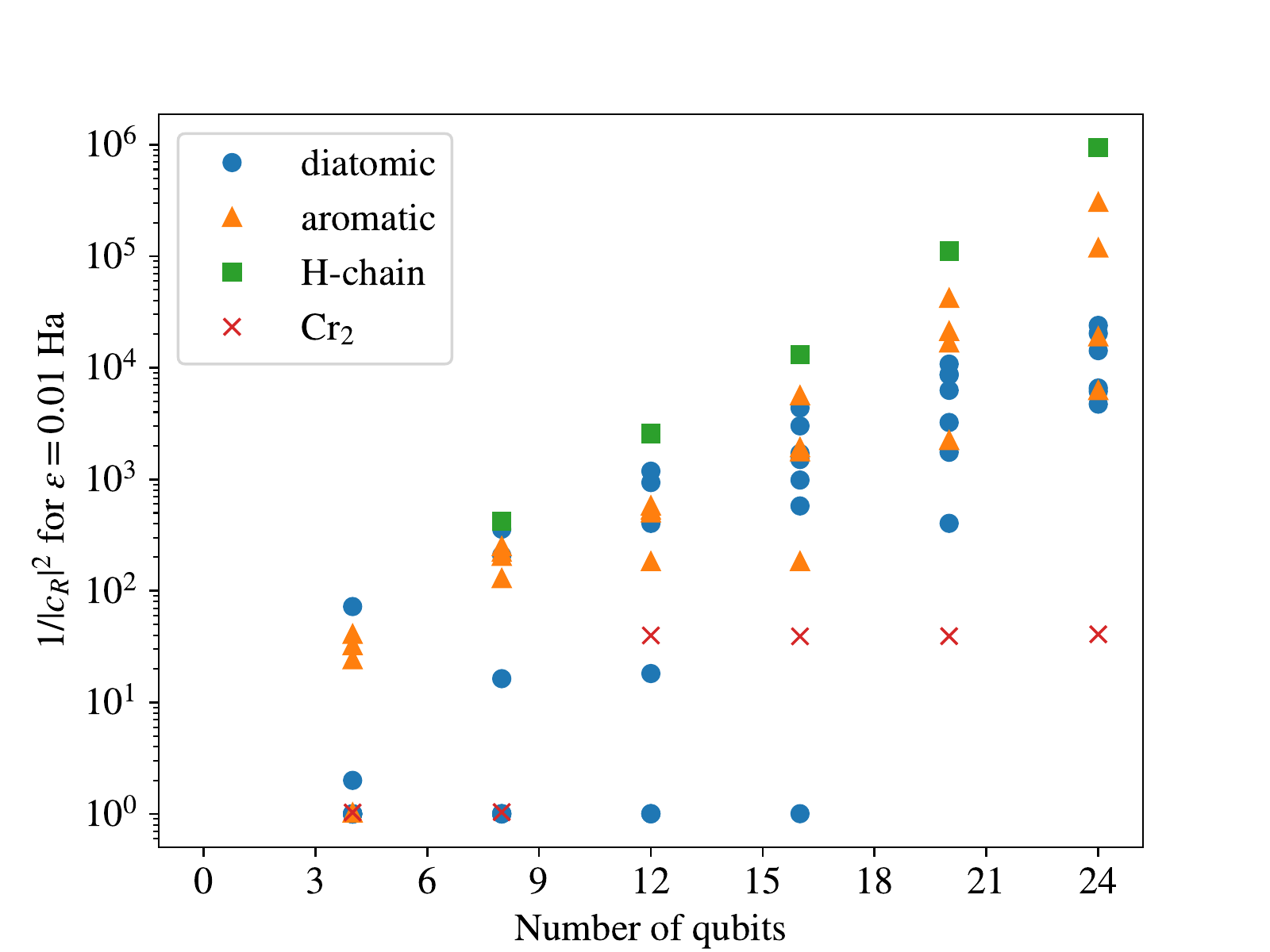}
     }
     
     \subfloat[][$R$ required for energy error $\epsilon=\SI{0.1}{Hartree}$.]{
         \includegraphics[width=.45\textwidth]{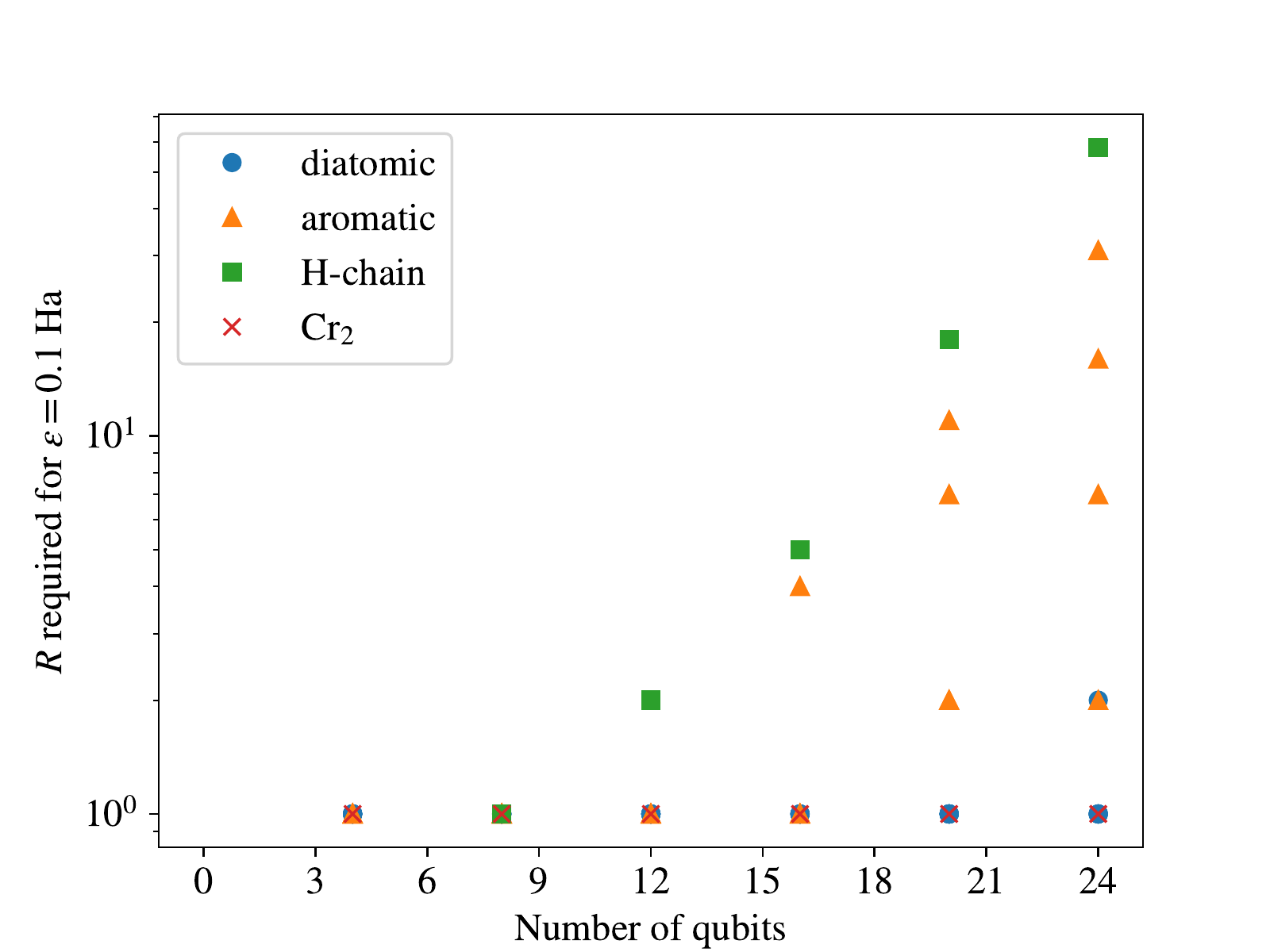}
     }
     \subfloat[][$1/\abs{c_R}^2$ for energy error $\epsilon=\SI{0.1}{Hartree}$.]{
         \includegraphics[width=.45\textwidth]{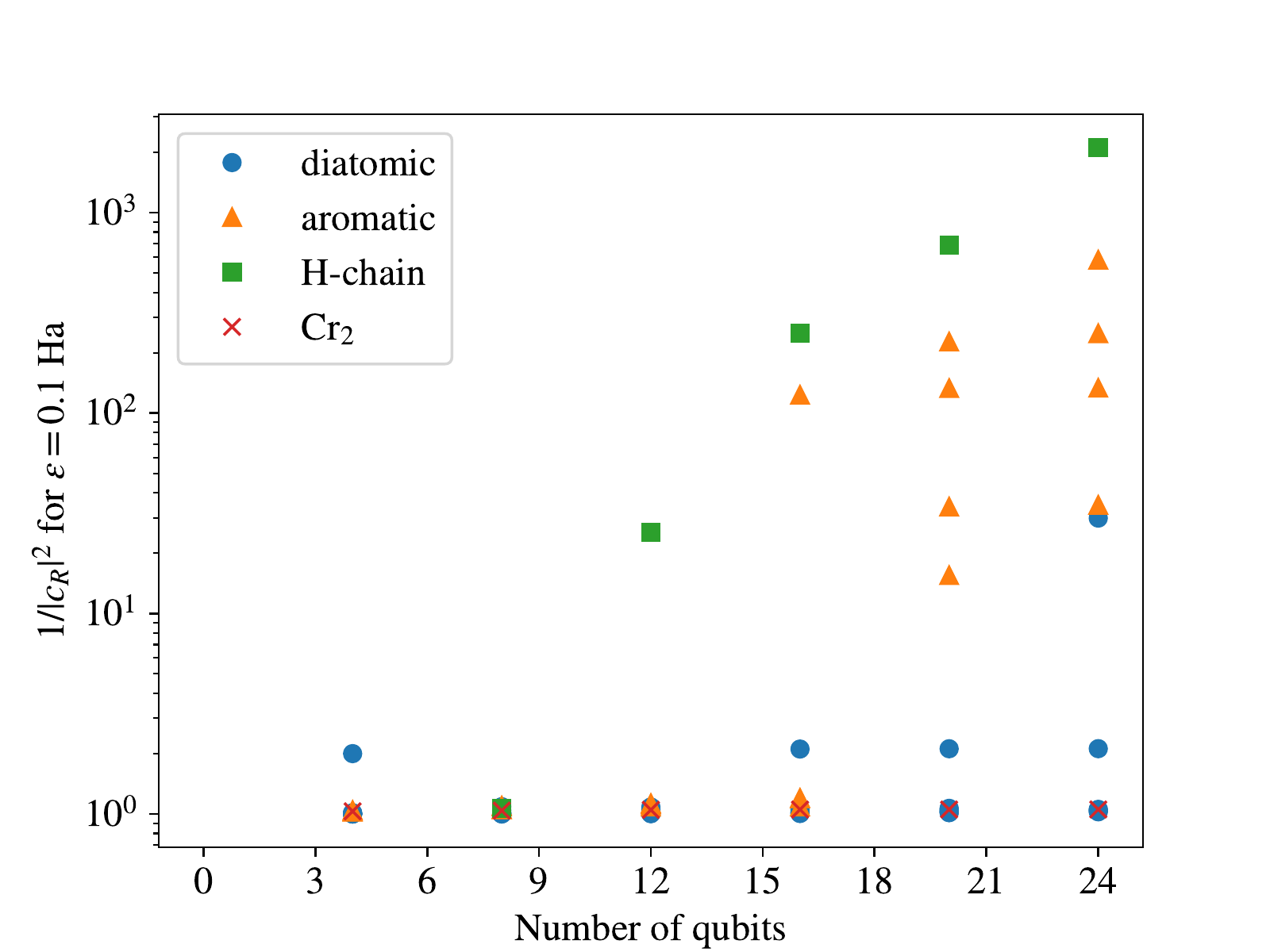}
         }
        \end{minipage}
    \caption{Estimated $R$ and $1/\abs{c_R}^2$ for various molecules with the same setup as Figs. \ref{fig:scaling-a} and \ref{fig:scaling-b}. The number of qubits is varied by changing the number of atoms for hydrogen chains, and by changing the active space for the other molecules.}
    \label{fig:all-scaling}
\end{figure*}

Figure~\ref{fig:all-scaling} shows the scaling of the classical and quantum computational costs, discussed in Sec.~\ref{subsec:scaling}, for different types of molecules. Here, we test three kinds of molecules: hydrogen chains, diatomic molecules, and aromatic molecules. The data for hydrogen chains are exactly the same as in the main text. Diatomic molecules are \ce{N2}, \ce{O2}, \ce{F2}, \ce{Cl2}, \ce{HCl},  \ce{CO}, and \ce{Cr2} with cc-pVQZ basis, and we tested them with various active spaces just as in the main text for \ce{Cr2}.
To test with larger molecules, four aromatic molecules are chosen: benzene, naphthalene, anthracene, and tetracene.
The Hamiltonian is generated by using the Hartree-Fock orbitals with STO-3G basis.
The active space of $n$ orbitals and $n$ electrons with varying $n$ was employed for the diatomic and aromatic molecules.
The geometries of these tested molecules are summarized in Table~\ref{tab: geometries}.

As can be seen in Fig.~\ref{fig:all-scaling}, hydrogen chains with various numbers of atoms show the worst scalings, while \ce{Cr2} is one of the least expensive systems among others.

\subsection{Sampling simulations with various molecules}
\label{ssec:appendix-sampling}
\begin{figure*}
     \begin{minipage}{0.9\textwidth}
     \subfloat[][\ce{H4} (8 qubits)]{
         \includegraphics[width=.45\textwidth]{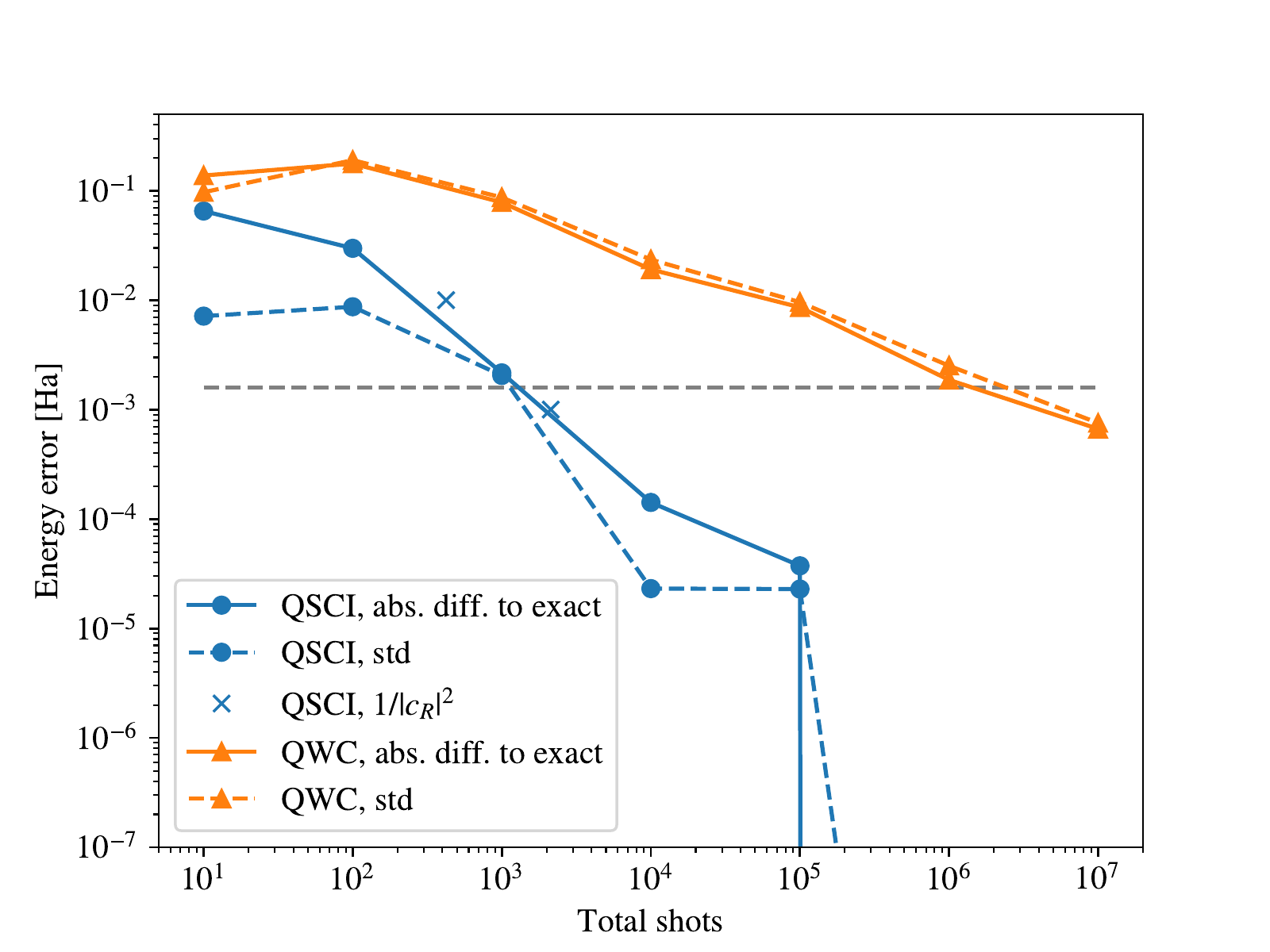}
     }
     \subfloat[][\ce{H6} (12 qubits)]{
         \includegraphics[width=.45\textwidth]{figures/sampling/H6_diff.pdf}
     }
     
     \subfloat[][\ce{H8} (16 qubits)]{
         \includegraphics[width=.45\textwidth]{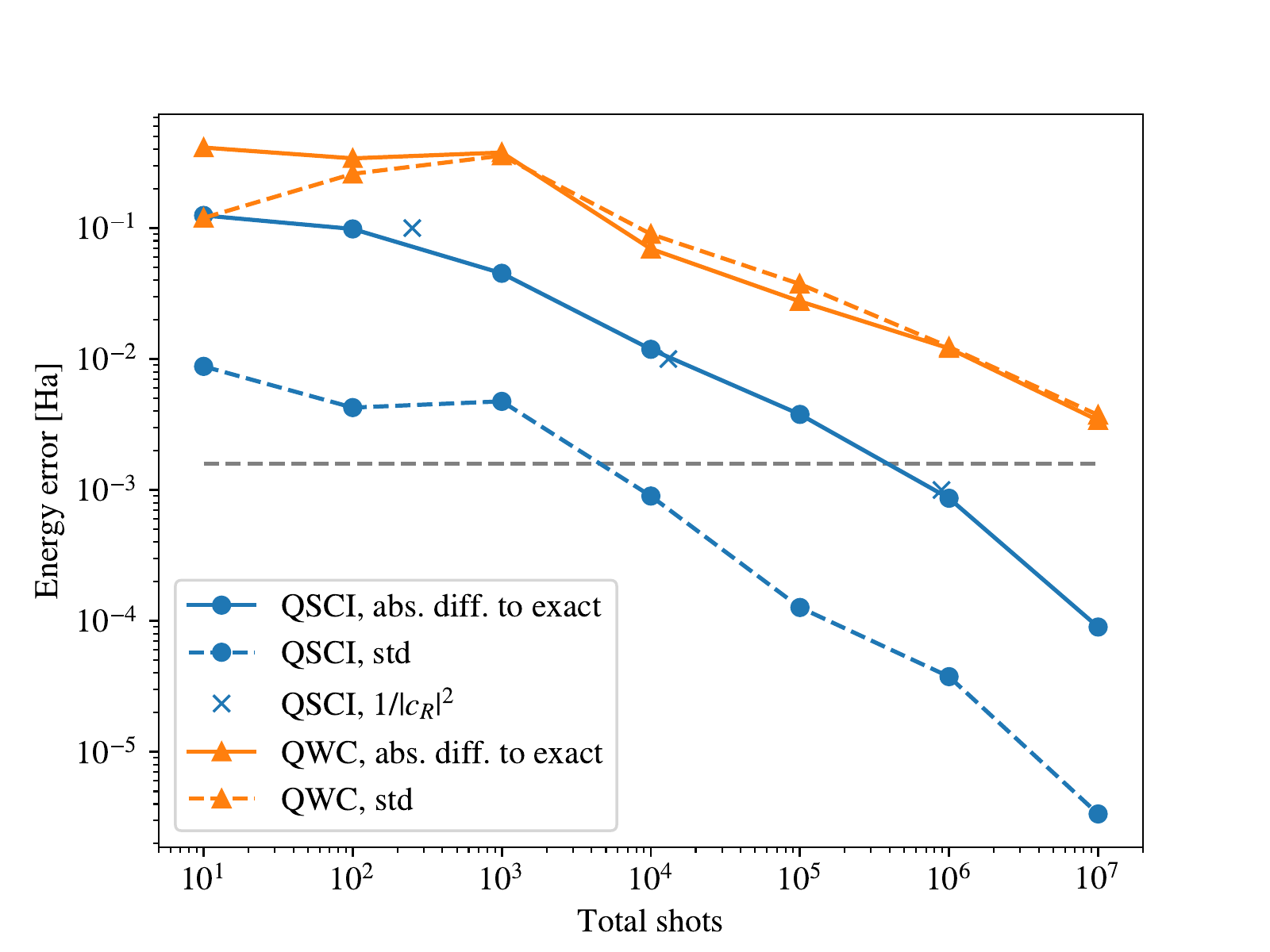}
     }
     \subfloat[][\ce{H10} (20 qubits)]{
         \includegraphics[width=.45\textwidth]{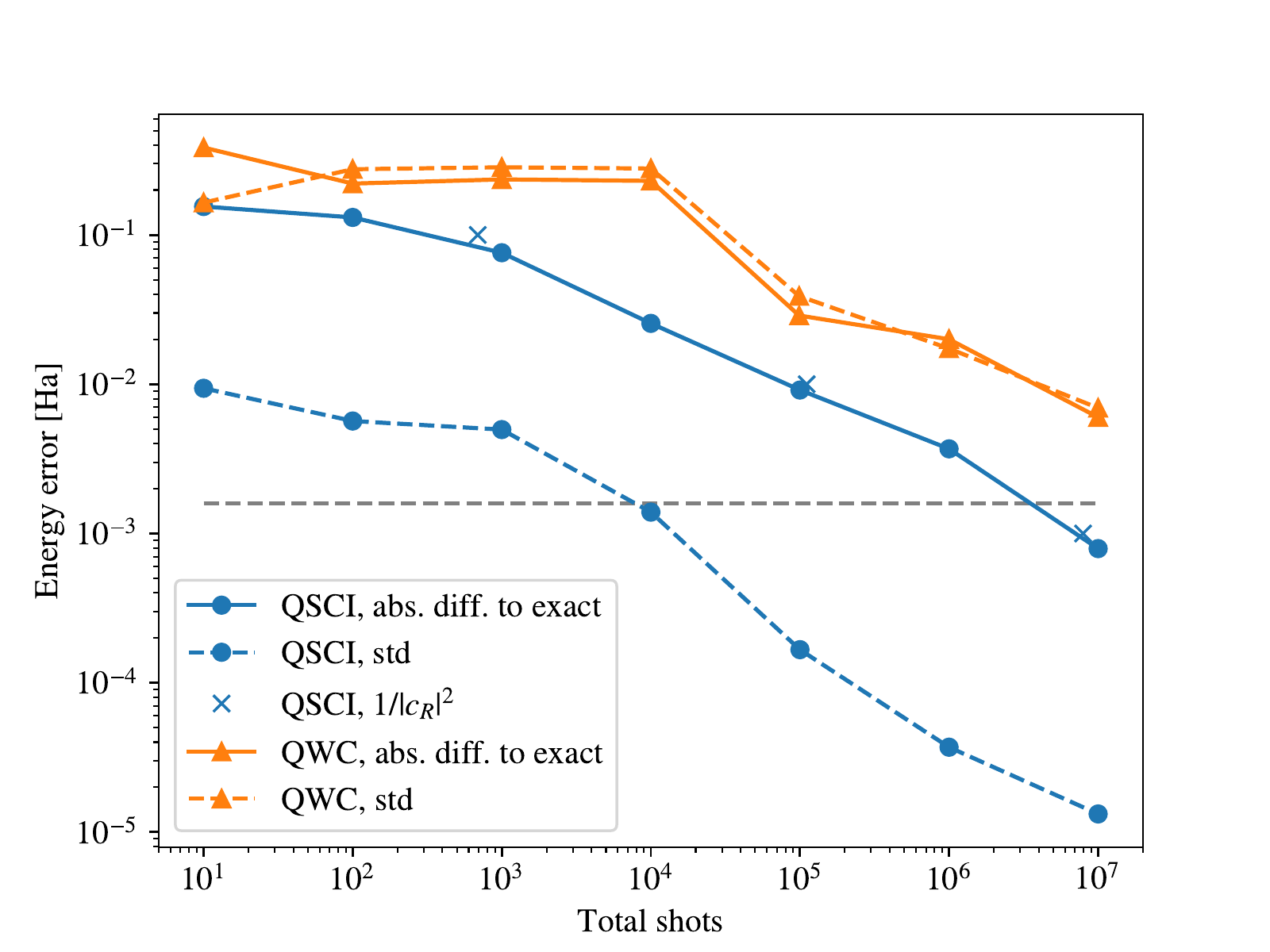}
     }
     
     \subfloat[][\ce{H2O} (12 qubits)]{
         \includegraphics[width=.45\textwidth]{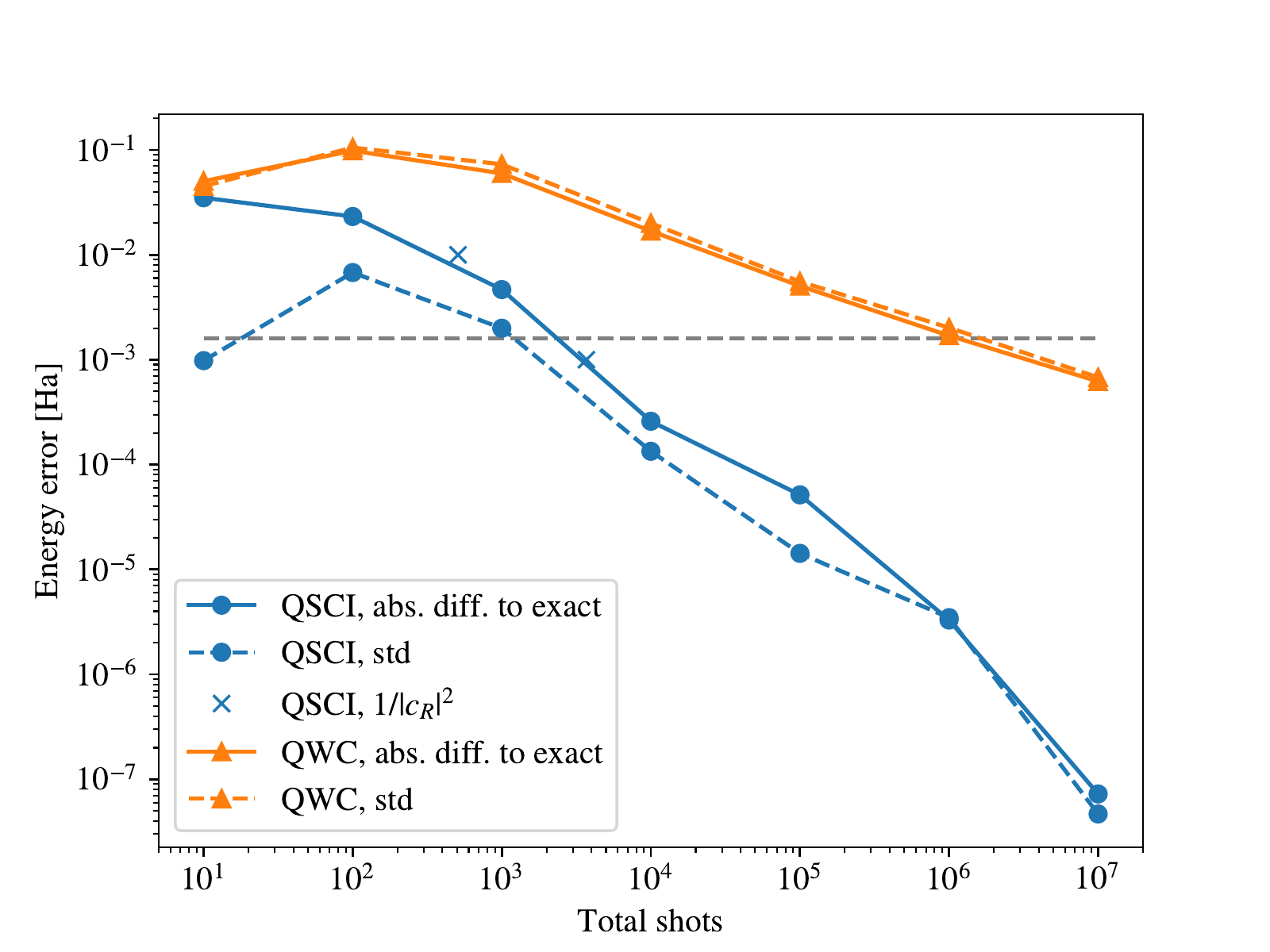}
     }
     \subfloat[][\ce{LiH} (12 qubits)]{
         \includegraphics[width=.45\textwidth]{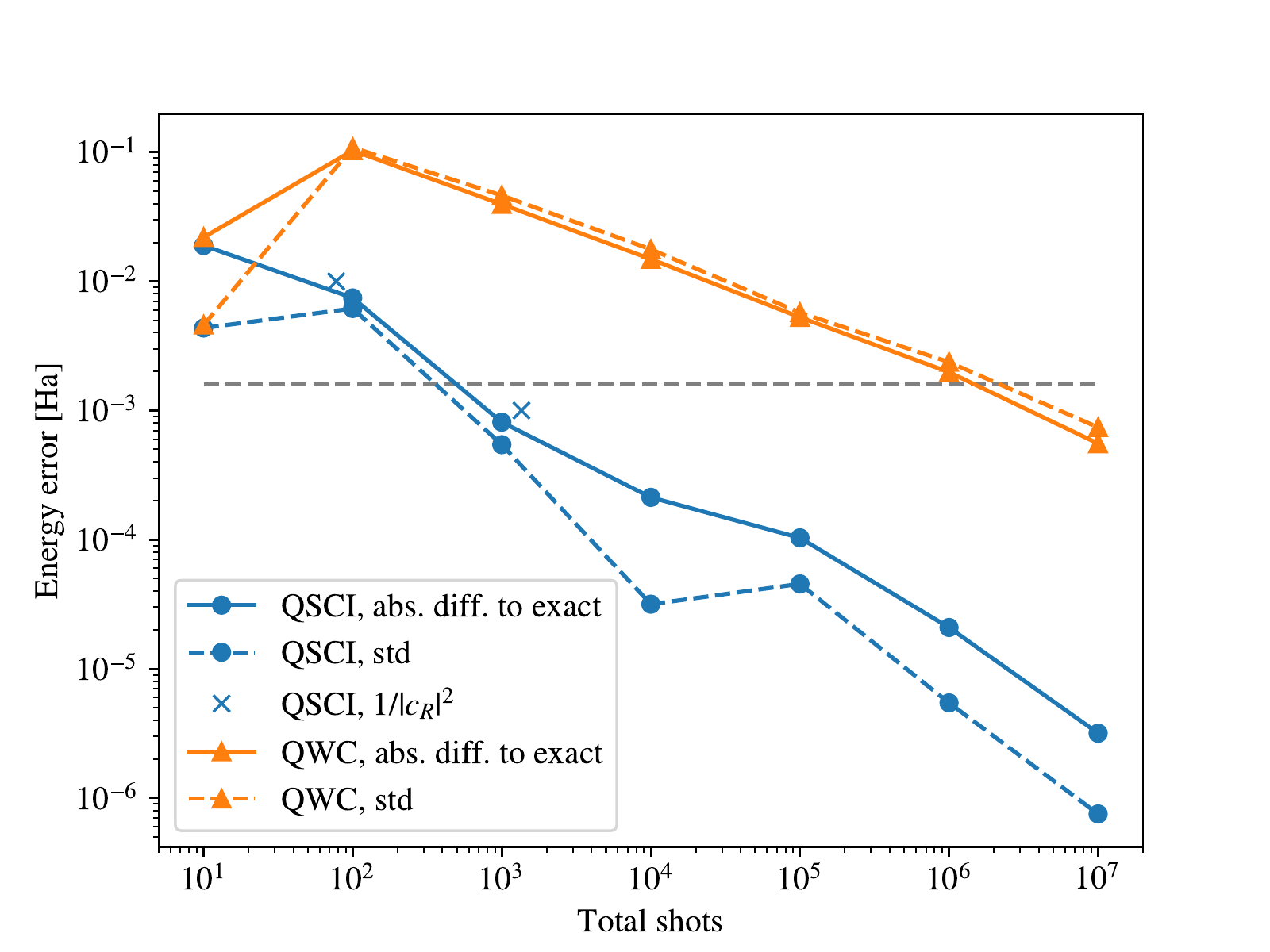}
         }
        \end{minipage}
    \caption{Sampling simulation with various molecules. See Fig.~\ref{fig:conventional} and the main text for details.}
    \label{fig:all-sampling}
\end{figure*}

In this subsection, we present similar results as Fig.~\ref{fig:conventional} but with various other molecules. The results, shown in Fig.~\ref{fig:all-sampling}, show the same features as \ce{H6} in the main text, such as the small standard deviation for QSCI and $1/\abs{c_R}^2$ giving an accurate estimation of the number of shots for given accuracy $\epsilon$. One can also see that the standard deviation is almost constant for hydrogen chains with various numbers of atoms, while the absolute error is highly dependent on the number of atoms. Comparing the three 12-qubit systems, it can be seen that the difference between the standard deviation and the absolute error depends on the system.

\subsection{Accuracy of expectation values of observables other than the Hamiltonian in QSCI}
\label{ssec:appendix-multiple-observable-accuracy}
\begin{figure}[h!]
  \begin{minipage}{0.45\textwidth}
     \subfloat[][$\epsilon=0.001$]{
         \includegraphics[width=\textwidth]{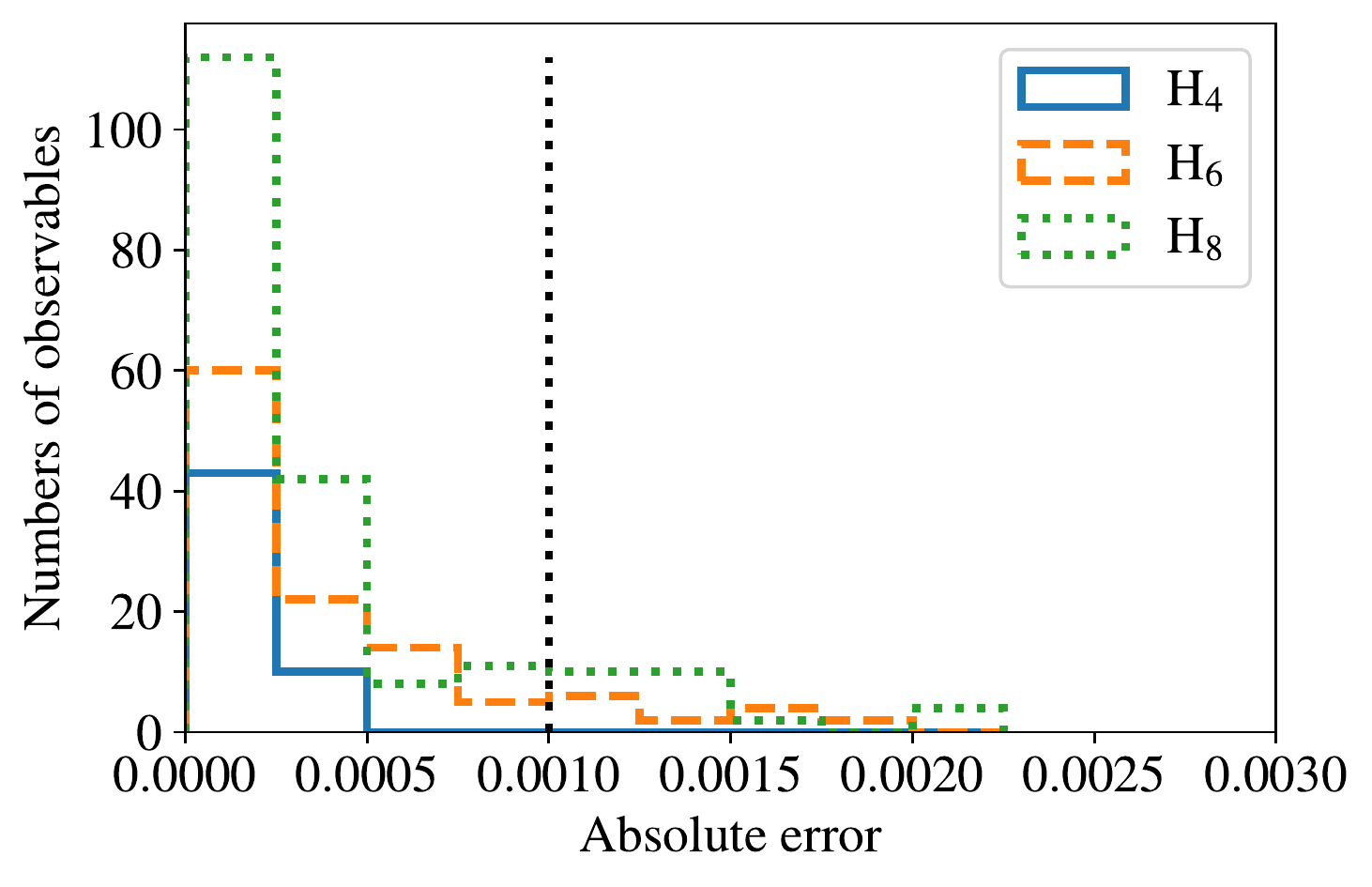}}
         
     \subfloat[][$\epsilon=0.01$]{
         \includegraphics[width=\textwidth]{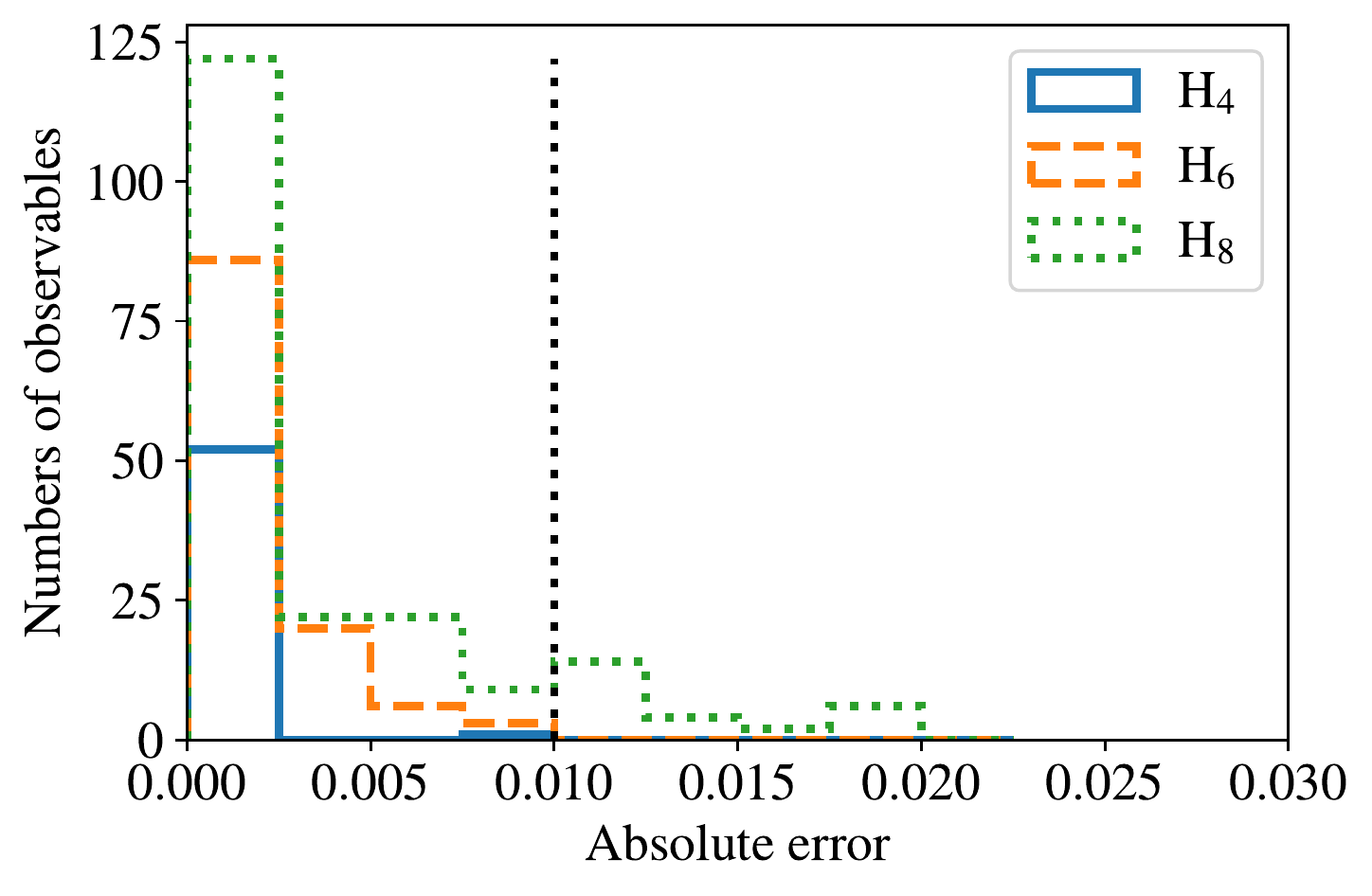}}
         
     \subfloat[][$\epsilon=0.1$]{
         \includegraphics[width=\textwidth]{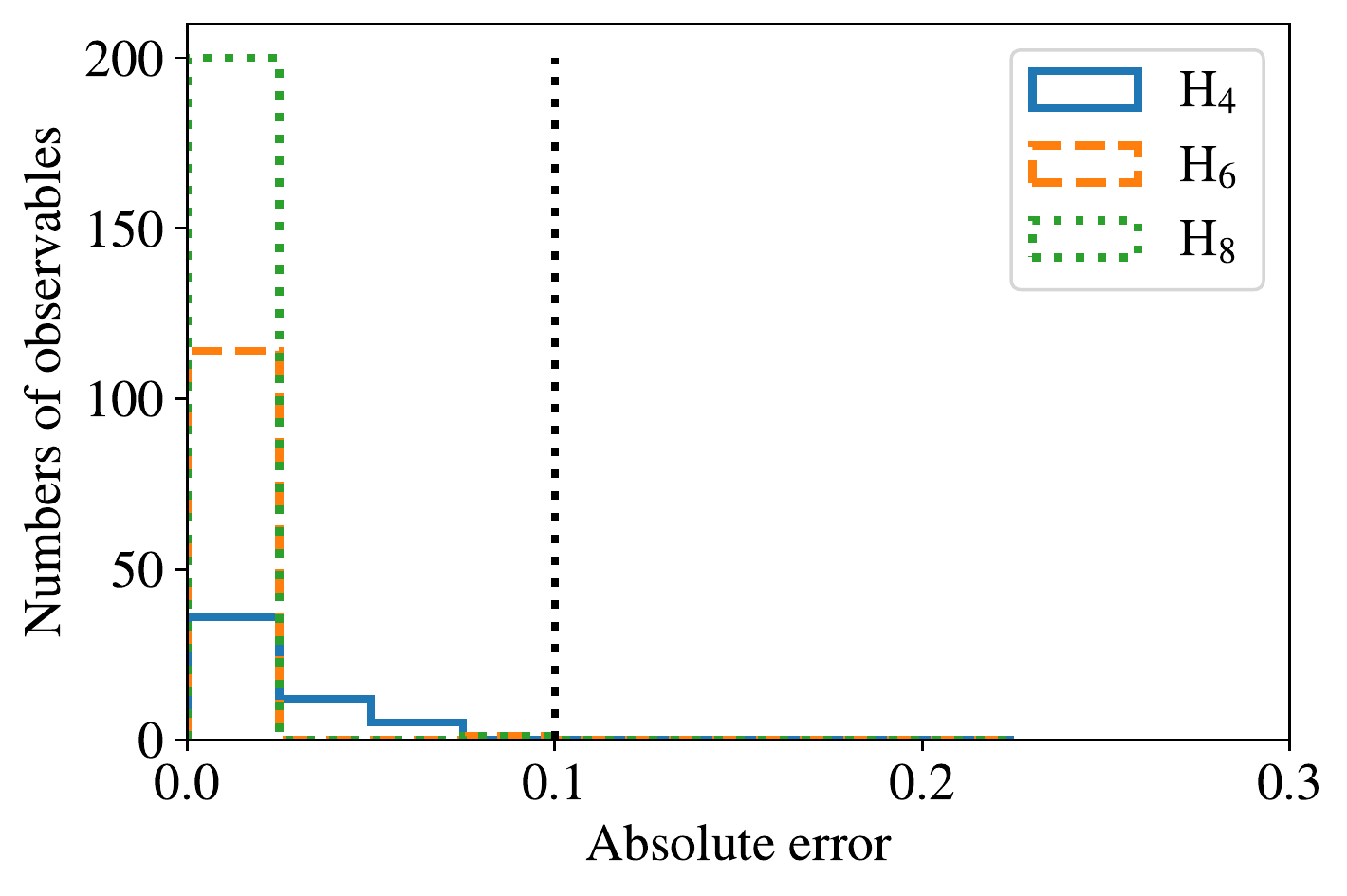}}
    \end{minipage}
    \caption{Histograms of absolute errors of nuclear gradients and Hessians for \ce{H4}, \ce{H6}, and \ce{H8} molecules in QSCI, compared to the exact, CASCI value. Each plot corresponds to QSCI calculations with an energy error $\epsilon$. Absolute errors are in units of Hartree, Hartree/\AA, or $\text{Hartree}/\text{\AA}^2$, depending on the observables. }
    \label{fig:appendix-multiple-observables}
\end{figure}
Here, we examine the accuracy of the expectation values of observables other than the Hamiltonian, estimated for the output state obtained by QSCI calculation.
Figure~\ref{fig:appendix-multiple-observables} shows the histograms for absolute errors of the expectation values for the gradient and Hessian, where the absolute error for an observable $\hat{O}$ is defined by
\begin{equation}
    \abs{\ev{\hat{O}}{\psiout}-\ev{\hat{O}}{\psi_\text{exact}}}.
\end{equation}
Here, $\ket{\psiout}$ is the output state of QSCI calculation with the idealized sampling from the exact ground state with $R$ given in Fig.~\ref{fig:scaling-a} for each error tolerance $\epsilon$ for energy, and $\ket{\psi_\text{exact}}$ is the exact ground state.
The observables $\hat{O}$ are set to be the nuclear gradient $\pdv{\hat{H}}{x_i}$ $(i=1,\dots,3N_\text{atom})$ and the Hessian $\pdv{\hat{H}}{x_i}{x_j}$ $(i,j=1,\dots,3N_\text{atom})$, where $N_\text{atom}$ is the number of atoms in the molecule and $x_i$ are coordinates of the nuclei.
The absolute error is shown in the unit of Hartree, Hartree/\AA, or $\text{Hartree}/\text{\AA}^2$, depending on the observables. Although there are some observables (i.e., components of the gradient or Hessian) whose expectation values exhibit larger absolute errors than that of the energy, the expectation values of the majority of the observables have similar accuracy as the energy.

\subsection{Bond length dependence}
\label{ssec:appendix-bond-length}
\begin{figure}[h!]
    \includegraphics[width=.45\textwidth]{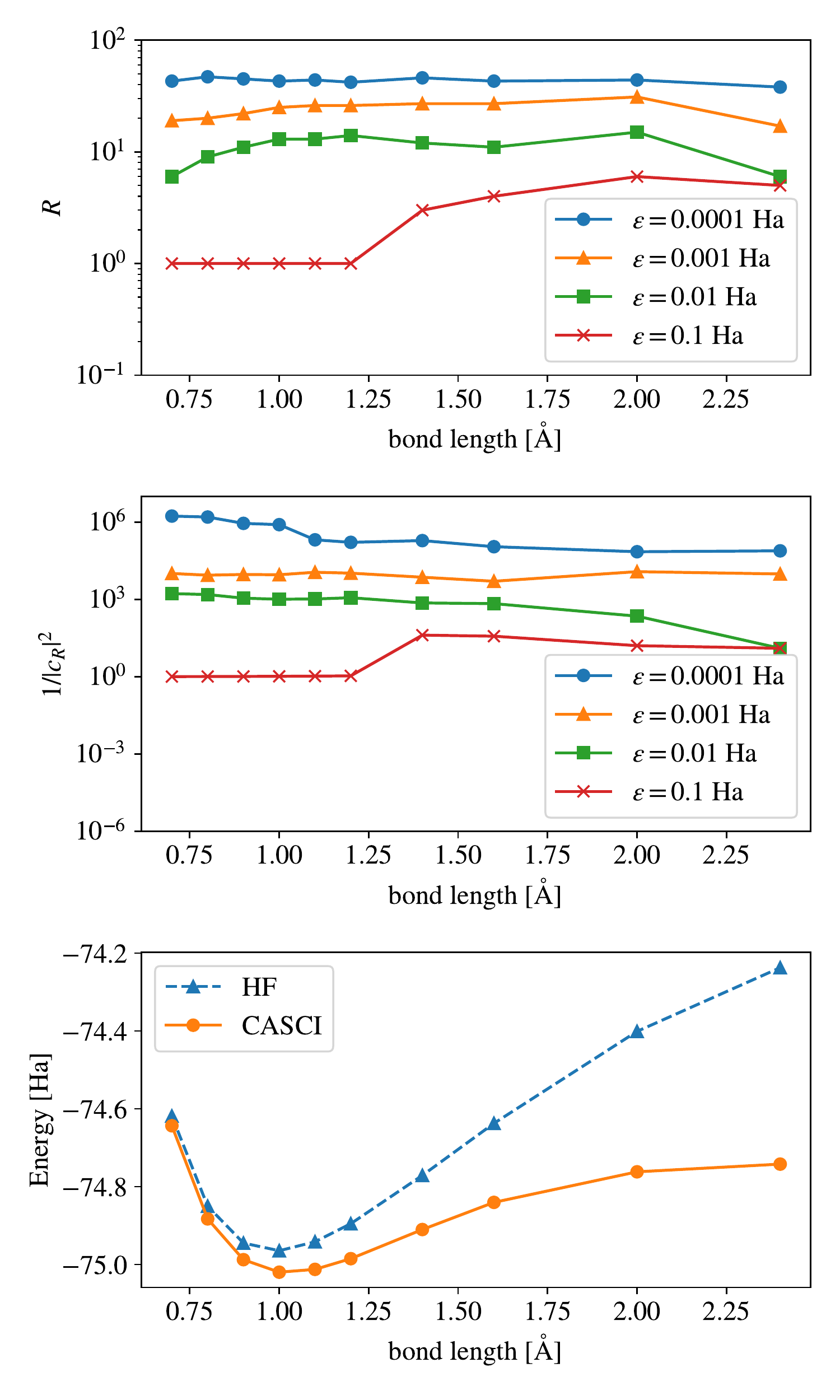}
    \caption{Estimated $R$ and $1/\abs{c_R}^2$ for \ce{H2O} molecule (14 qubits) with various bond lengths. Potential energy curves are also shown for reference. The estimation method of $R$ and $1/\abs{c_R}^2$ are the same as in Figs. \ref{fig:scaling-a} and \ref{fig:scaling-b}.}
    \label{fig:pec}
\end{figure}
The Hartree-Fock calculation is known to perform better for a stable geometry of a molecule than for the dissociation limit, so it is worth studying if QSCI also performs worse in the dissociation limit. Figure~\ref{fig:pec} shows the result of the same numerical analysis as Fig.~\ref{fig:scaling-a}, but for various bond lengths of \ce{H2O} molecules. The Hamiltonian is generated by the Hartree-Fock orbitals using STO-3G basis without specifying the active space, and is of 14-qubit after the Jordan-Wigner mapping. The bond lengths of two H-O bonds are taken to be equal, and the H-O-H angle is fixed to \ang{104.45}. The result implies that, although there is some dependency on the bond length for larger $\epsilon$, the dependency disappears for smaller $\epsilon$. It can be expected from the result that the potential energy surface calculated by QSCI has a relatively constant accuracy, at least compared to the Hartree-Fock result, when the error tolerance is not very large.

\subsection{Comparison to ASCI}
\label{ssec:comparison-to-asci}

\begin{figure}[h!]
    \includegraphics[width=.45\textwidth]{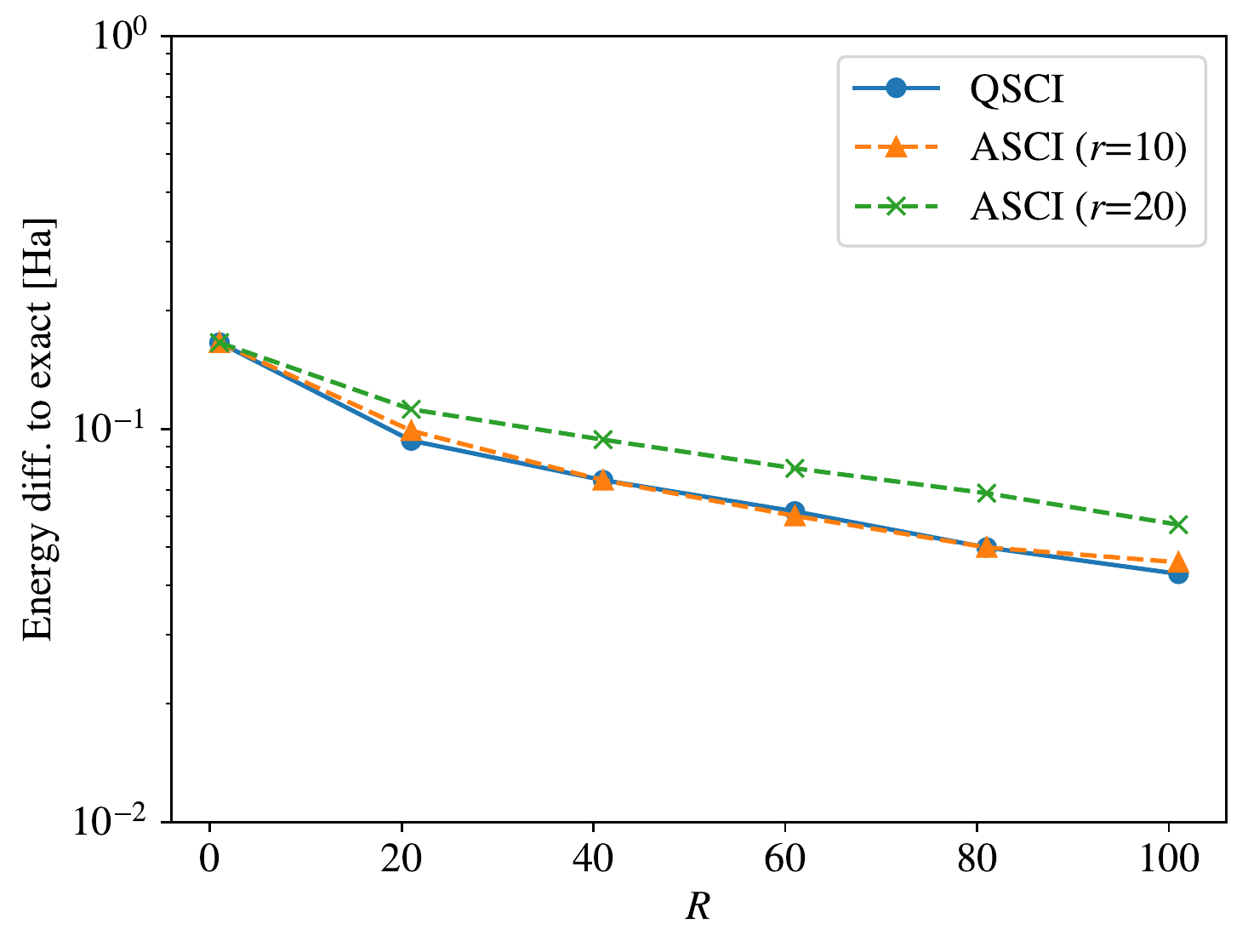}
\caption{Comparison of ASCI and QSCI for \ce{H10} molecule (20 qubits). The ASCI and QSCI energies as the difference to the exact CASCI energy are plotted against the size $R$ of the selected subspace. The parameter $r$ determines the size of the search space for ASCI.}
\label{fig:asci}
\end{figure}

Here, we investigate if there is 
a possibility that QSCI outperforms the state-of-the-art selected CI methods 
by taking ASCI for illustration.
ASCI is a
selected CI method
solely based on classical computation, which adaptively searches for the optimal subspace of the Fock space for the diagonalization.
In Fig.~\ref{fig:asci}, we compare QSCI with ASCI, for which we follow the description in Ref.~\cite{tubman2020modern}.
Here, we use the QSCI method with the idealized sampling from the ground state obtained by the exact diagonalization (full-CI) calculation.

The target molecule is the linear hydrogen chain \ce{H10} 
with the equal separation of 1.0 \AA.
The basis set is STO-3G and the Hamiltonian with the Hartree-Fock orbitals is mapped to the 20-qubit one by the Jordan-Wigner mapping.
For ASCI, in addition to the parameter $R$ (called as $N_{tdets}$ in Ref.~\cite{tubman2020modern}), there are two additional parameters: they are denoted by $\epsilon$ and $N_{cdets}$ in that paper, and are denoted by $\delta$ and $R_{\text{core}}$, respectively, in the following. The parameters $\delta$ and $R_{\text{core}}$ determine the size of the search space for the iterative search for the new determinants, while the cost for the generation and diagonalization of the Hamiltonian, which is common for both ASCI and QSCI, are determined solely by $R$. We fixed $\delta=\SI{0.05}{Hartree}$ and $r:=R/R_{\text{core}}=10$ or $20$ for ASCI, and run QSCI and ASCI calculations with various $R$.

While in the case of $r=10$ the two methods perform similarly, QSCI performs better for $r=20$, where less computational cost is required for searching for a better set of configurations in ASCI. The result shows that, depending on the hyperparameters for ASCI, there is a possibility that QSCI performs better, at least in the case of the idealized sampling from the exact ground state.

\bibliography{bib}

\end{document}